\documentclass[a4paper,11pt]{article}
\usepackage{jheppub} 
\usepackage{lineno}
\usepackage[utf8]{inputenc}
\usepackage[english]{babel}
\usepackage{csquotes}
\usepackage[T1]{fontenc}
\usepackage{graphicx}
\usepackage{chngcntr}
\usepackage{subcaption}
\usepackage{tabularx}
\usepackage{caption}
\renewcommand{\figurename}{Figure}
\usepackage{hyperref}
\usepackage[colorinlistoftodos]{todonotes}
\usepackage{epigraph}
\usepackage[export]{adjustbox}
\usepackage{amsmath}
\allowdisplaybreaks
\usepackage{amsthm}
\usepackage{bbold}
\usepackage{amssymb}
\usepackage{amscdx}
\usepackage{nccmath}
\usepackage{mathtools}
\usepackage{mathrsfs}
\usepackage{blindtext}
\usepackage{titlesec}
\usepackage{natbib}
\usepackage{comment}

\usepackage{subfiles}

\preprint{UUITP-03/25}

\title{The entropy of radiation for local quenches in higher dimensions}

\author[a,b]{Lorenzo Bianchi,}
\author[a,b]{Andrea Mattiello,}
\author[c]{Jacopo Sisti}
 \affiliation[a]{Dipartimento di Fisica, Università di Torino,\\Via P. Giuria 1, 10125 Torino, Italy}
\affiliation[b]{INFN - Sezione di Torino,\\Via P. Giuria 1, 10125 Torino, Italy}
\affiliation[c]{Department of Physics and Astronomy, Uppsala University,\\Box 516, SE-75120 Uppsala, Sweden}

\emailAdd{lorenzo.bianchi@unito.it, andrea.mattiello@unito.it, jacopo.sisti@physics.uu.se}

\abstract{We investigate the real time dynamics of the radiation produced by a local quench in a $d$-dimensional conformal field theory (CFT) with $d>2$. Using the interpretation of the higher-dimensional twist operator as a conformal defect, we study the time evolution of the entanglement entropy of the radiation across a spherical entangling surface. We provide an analytic estimate for the early- and late-time behavior of the entanglement entropy and derive an upper bound valid at all times. We extend our analysis to the case of a boundary CFT (BCFT) and derive similar results through a detailed discussion of the setup with two conformal defects (the boundary and the twist operator). We conclude with a holographic analysis of the process, computing the time evolution of the holographic entanglement entropy (HEE) as the area of the Ryu-Takayanagi surface in a backreacted geometry. This gives a Page-like curve in agreement with the early- and late-time results obtained with CFT methods. The extension to a holographic BCFT setup is generically hard and we consider the case of a tensionless end-of-the-world brane.}


\begin{document}
\maketitle
\flushbottom

\newpage

\section{Introduction and discussion}
\label{section:Introduction}

Understanding the properties of the radiation emitted by a physical subsystem is a crucial goal in physics. In collider physics, for instance, one is interested in measuring the energy and momentum of the emitted particles. In the context of the Hawking radiation, instead, an important observable is the entropy of the radiation emitted by an evaporating black hole. In particular, understanding the time evolution of this entropy would provide a crucial insight into the celebrated black hole information paradox \cite{Hawking:1976ra}. 

In this paper, we analyze the radiation produced by a local quench in a conformal field theory (CFT) and a boundary conformal field theory (BCFT) in dimension $d>2$. Quantum quenches are useful tools to study out-of-equilibrium physics and thermalization \cite{Calabrese:2005in,Calabrese:2006rx,Calabrese:2007rg,Calabrese:2007mtj}. A local quench is a sudden excitation of the CFT localized in space and producing a non-trivial real time dynamics. The quantum system jumps abruptly to an excited state, and the radiation produced by the quench carries information about this state. We want to understand how this information is encoded in the radiation collected infinitely far away. In particular, we will be interested in studying the real time evolution of the entropy of the radiation.

\begin{figure}[htbp!]
\centering
\includegraphics[scale=0.4]{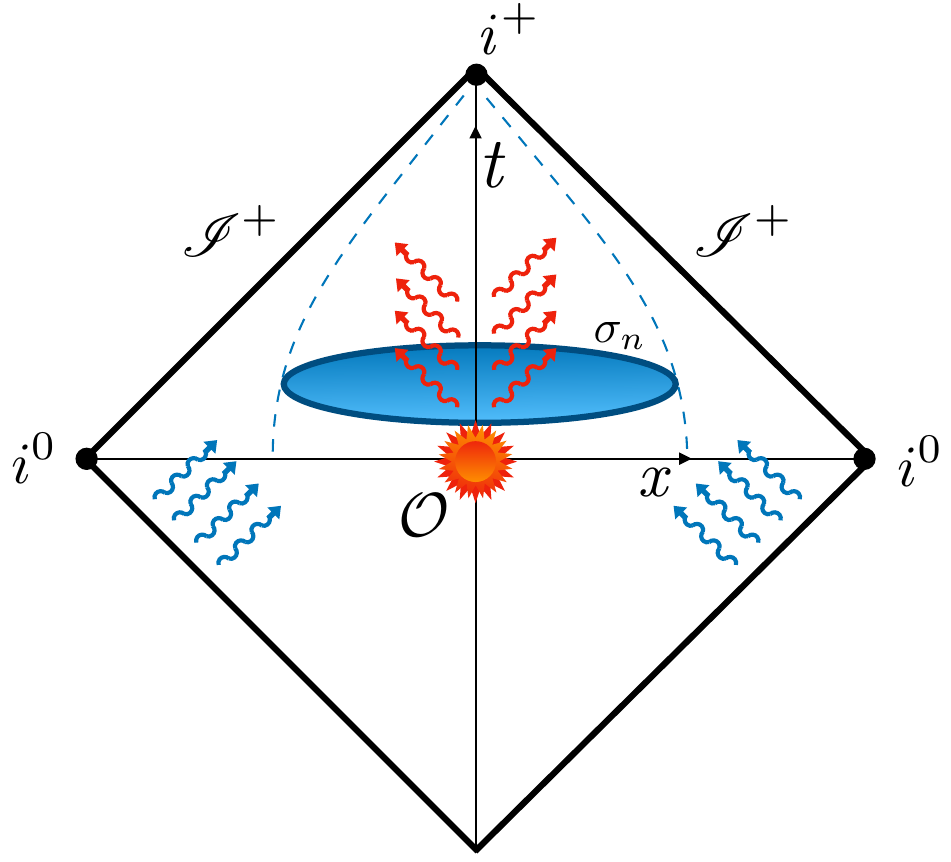}
\caption[Penrose diagram]{Three-dimensional schematic representation of our setup.}
\label{fig:Penrose}
\end{figure}

For a homogeneous CFT, a schematic representation of our setup is shown in figure \ref{fig:Penrose}. At a given time, we divide the space into two regions (the blue circle and the outside) and we measure the entanglement entropy of the radiation across the spherical entangling surface. Similar setups have been already considered in the literature \cite{Nozaki:2013wia,Asplund:2014coa,Jahn:2017xsg,Belin:2018juv,Agon:2020fqs,Belin:2021htw,David:2022czg}, but mostly in the context of two-dimensional CFTs and $\mathrm{AdS}_{3}/\mathrm{CFT}_{2}$ \footnote{The work \cite{Jahn:2017xsg} deals with a higher dimensional setup and we will comment on the relation with our work in section \ref{section:LocalQuench}}. Here, we show that some of the universal features of this radiation survive also in higher dimensions.  Specifically, using the operator product expansion (OPE) of the CFT and some assumptions on the spectrum, we provide a universal estimate for the early- and late-time behavior of the entropy in any dimension. We also find a universal bound for the entropy that is valid at any time.

 A holographic perspective, where the entanglement entropy is computed via the Ryu-Takayanagi prescription \cite{Ryu:2006bv,Ryu:2006ef}, provides a powerful approach to studying local quenches in strongly coupled field theories with gravitational duals and, conversely, any result on the CFT side gives us some insight into dynamical processes in quantum gravity. We model the local quench as the insertion of a massive particle in an asymptotically Anti-de Sitter (AdS) spacetime. The evolution of the holographic entanglement entropy is then related to the minimal surface area in the bulk geometry. Through numerical analysis, we obtain a Page-like curve for the entanglement entropy, which matches the early- and late-time predictions from the CFT analysis.  For a small black hole, we can also perform a perturbative expansion in the mass and compute analytically the leading order result for the area. Quite surprisingly, we find that this result precisely saturates the entropy bound at all times. 

\begin{figure}[htbp!]
    \centering
    \includegraphics[scale=0.4]{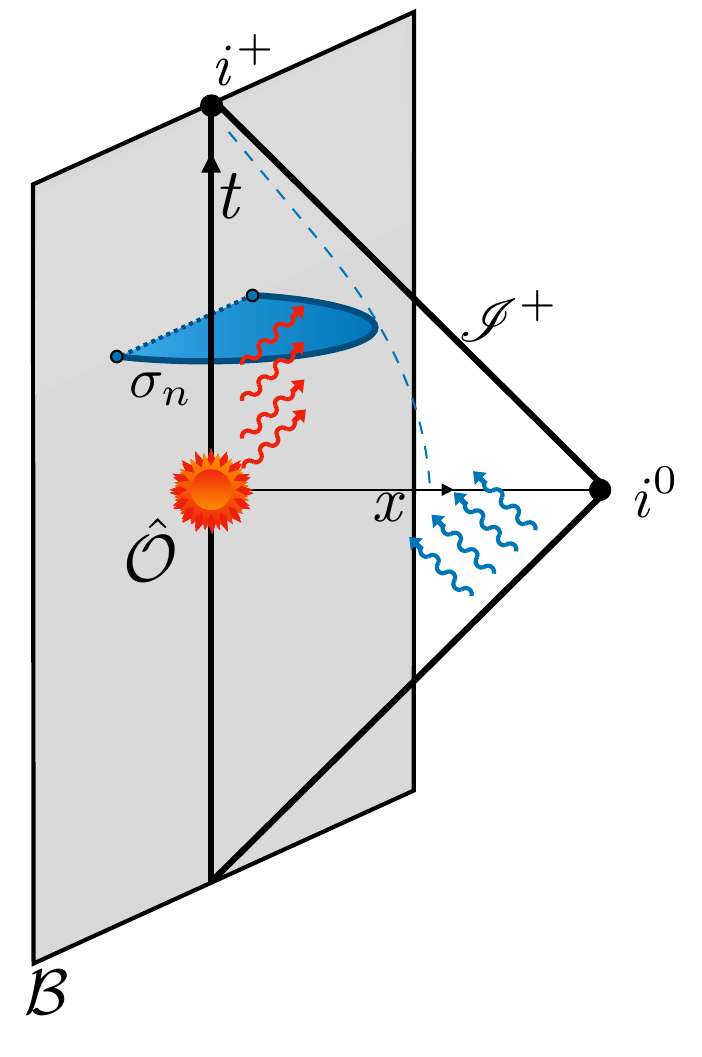}
    \caption[Penrose diagram]
    {Three-dimensional schematic representation of our boundary setup.}
    \label{fig:Penrose2}
\end{figure}

For a BCFT, we consider a boundary excited state, as represented in figure \ref{fig:Penrose2}. The entangling surface in this case is a hemisphere on a time slice ending on the boundary. This setup is interesting for at least two reasons. On the CFT side, it provides an interesting and poorly studied setup involving the interaction of two conformal defects (the boundary and the twist operator) and local operators. We study the kinematics of the problem and, also in this case, we provide an estimate for the early- and late-time behavior of the radiation, which extends previous results in two dimensions \cite{Kawamoto:2022etl,Bianchi:2022ulu}. 

\begin{figure}[htbp!]
\centering
\includegraphics[scale=0.3]{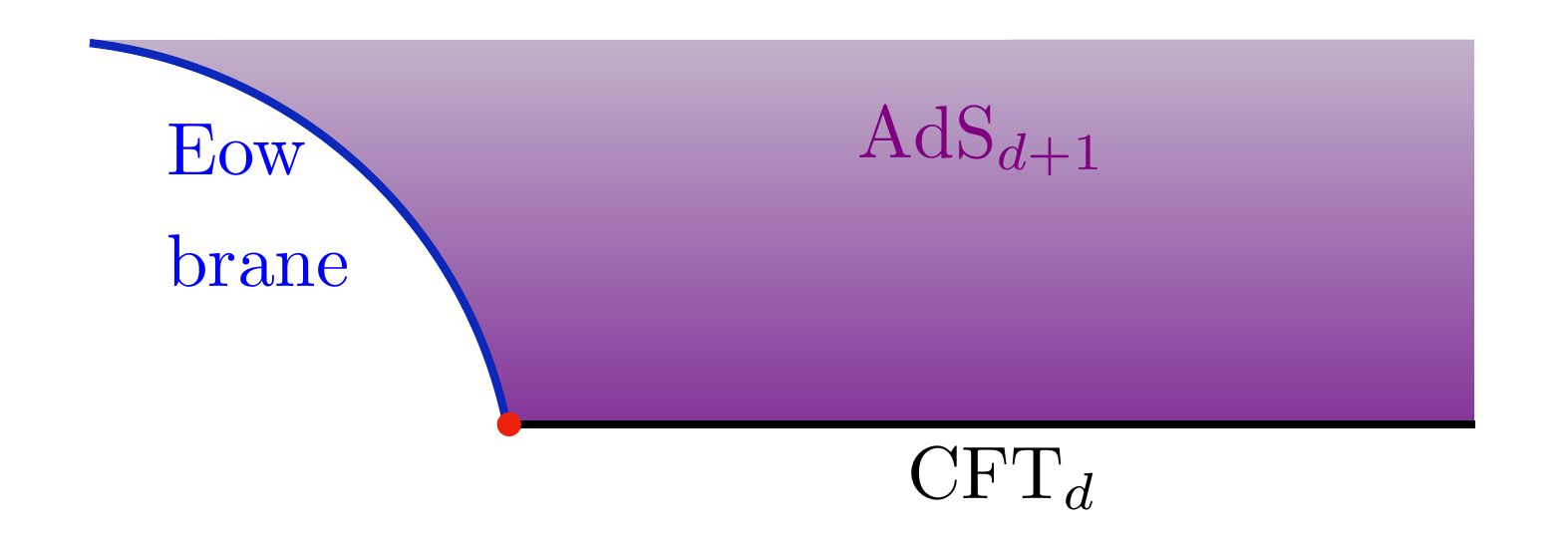}
\caption[Double holography setup]
    {The doubly holographic setup: the $d$-dimensional $\mathrm{CFT}_{d}$ is a rigid boundary to a $(d+1)$-dimensional $\mathrm{AdS}_{d+1}$ spacetime and it has a $(d-1)$-dimensional conformal boundary whose dual description is an end-of-the-world (EoW) brane, \textit{i.e.}\ a dynamical boundary for $\mathrm{AdS}_{d+1}$.}
\label{fig:DoubleHolography}
\end{figure}

On the gravity side, this BCFT setup is particularly relevant in light of the recent developments in the semiclassical understanding of the black hole information paradox. These developments, started by \cite{Penington:2019npb,Almheiri:2019psf,Almheiri:2019hni} (see \cite{Almheiri:2020cfm} for a review), have clarified that many features of the puzzle can be understood by looking at toy models on which we have a great control. One example is the doubly-holographic setup shown in figure \ref{fig:DoubleHolography} which can be interpreted either as a gravitational region -- the end-of-the-world (EOW) brane \cite{Randall:1999ee,Karch:2000gx,Takayanagi:2011zk,Fujita:2011fp} -- coupled to a reservoir or as a CFT where the radiation is collected with a (possibly non-conformal) boundary encoding all the information on the gravitational region. In this respect, the collected radiation could provide an insight into the gravitational process happening on the brane \cite{Geng:2020qvw,Geng:2020fxl,Rozali:2019day,Sully:2020pza,Chen:2020uac,Chen:2020hmv,Hernandez:2020nem,Grimaldi:2022suv,Geng:2021iyq,Anous:2022wqh,Bianchi:2022ulu,Izumi:2022opi,Geng:2024xpj} (see also \cite{Geng:2023qwm} for a review), especially if we manage to construct a boundary state which is dual to an evaporating black hole \cite{Anous:2016kss}. Here we consider a different excited state of the BCFT, which is dual to a black hole in the bulk $\mathrm{AdS}_{d+1}$ ending on the EOW brane. This black hole does not evaporate, but in the Poincar\'e section it has a non-trivial dynamics producing an interesting behavior for the entropy of the radiation.

The paper is structured as follows: In section 2, we introduce the local quench and we discuss the early- and late-time behavior as well as the derivation of the entropy bound. Section 3 extends the analysis to a BCFT framework. Section 4 explores the holographic dual description, including numerical results for the entanglement entropy evolution.

\section{The CFT Setup}
\label{section:CFTsetup}

\subsection{The local quench}
\label{subsection:ExState}
We begin our analysis by constructing a local quench in a conformal field theory (CFT), \textit{i.e.}\ an excited state produced by the insertion of a local operator. Let us consider a CFT in Lorentzian signature \( g_{\mu \nu} = \text{diag}(-1, 1, \dots, 1) \) with coordinates $x^{\mu} = (t, \vec{x})$. We create an excited state at the origin of the CFT by acting on the vacuum with a local operator at \( \vec{x} = 0 \) and \( t = 0 \), 

\begin{equation}
    |\mathcal{O}\rangle=\mathcal{O}(t=0,\vec{x}=0)|0\rangle\, ,
\end{equation}
where we take \( \mathcal{O} \) to be a primary scalar operator with scaling dimension \( \Delta \).
Such a state has infinite norm and, to regularize this divergence,  we evolve it for some Euclidean time $\epsilon$. This gives

\begin{equation}
    | \mathcal{O} \rangle_{{\epsilon}} = \sqrt{\mathcal{N}}e^{-\epsilon P_{0}}\mathcal{O}(t=0,\vec{x}=0)|0\rangle=\frac{1}{\sqrt{\langle \mathcal{O}(t = -i\epsilon) \mathcal{O}(t = i\epsilon) \rangle}} \mathcal{O}(t = i\epsilon,\vec{x} = 0) | 0 \rangle\, ,
    \label{eq:ExcitedState}
\end{equation}
where we introduced a proper normalization $\sqrt{\mathcal{N}}$ to make the operator $|\mathcal{O}\rangle_{\epsilon}$ unit normalized. Hereafter, we will denote the excited state in \eqref{eq:ExcitedState} simply as $|\mathcal{O}\rangle$ for ease of notation. The state in the origin produces a radiation in the CFT. The $\epsilon$ regulator makes the energy density of the radiation a smooth function that is picked on the lightcone centered in the origin.
 The setup is illustrated in figure \ref{fig:Penrose}.

We are interested in the time evolution of the entanglement entropy of this radiation across a spherical region of the CFT, depicted in blue in figure \ref{fig:Penrose}.  This entropy measures the entanglement between the radiation crossing the sphere at a given time slice and that crossing the exterior of the sphere. As time passes, the radiation will propagate towards lightlike infinity while the spherical entangling surface evolves on a timelike trajectory leading to a vanishing entropy at late time. In the following, under certain conditions, we will be able to quantify this late time behavior.

Note that this is a dynamical process, since the state in \eqref{eq:ExcitedState} is not an eigenstate of the Hamiltonian. However, since we are working in a CFT it is useful to notice that it is an eigenstate of a specific conformal generator (often denoted as conformal Hamiltonian) with eigenvalue \( \Delta \)

\begin{equation}
    \frac{1}{2} \left( \frac{K^0}{\epsilon} + \epsilon P^0 \right) | \mathcal{O} \rangle = \Delta | \mathcal{O} \rangle\, .
    \label{eq:eigenstate}
\end{equation}
In particular, mapping the problem to the Lorentzian cylinder $\mathbb{R}\times S^{2}$ this generator maps to the Hamiltonian, making the problem invariant under time translations. 

\subsection{The entanglement of radiation}

Consider a pure state \( |\psi\rangle \) in a quantum field theory. For a given time slice, we can split the system into two subsystems, \( A \) and its complement \( \bar{A} \), and we assume that the total Hilbert space decomposes as a direct product of the two subsystems, \textit{i.e.}\ \( \mathcal{H} = \mathcal{H}_{A} \otimes \mathcal{H}_{\bar{A}} \)\,. The reduced density matrix \( \rho_A \) for subsystem \( A \) is obtained by tracing out the degrees of freedom of subsystem \( \bar{A} \), namely $\rho_A = \mathrm{tr}_{\bar{A}}(\rho)$.
The entanglement entropy \( S_A \) for the subsystem \( A \), which in our case corresponds to the region inside the spherical entangling surface, is given by the von Neumann entropy

\begin{equation}
    S_A = -\mathrm{tr}_A \left( \rho_A \log{\rho_A} \right)\,,
    \label{eq:VonNeumannEntropy}
\end{equation}
where \( S_A = S_{\bar{A}} \) for a pure state. It is worth noting that, in general, the entanglement entropy is subject to ultraviolet (UV) divergences, due to high frequency modes close to the entangling surface \( \partial A \).

To compute the entanglement entropy it is useful to define a  one-parameter generalization, the Rényi entropy

\begin{equation}
    S_A^{(n)} = \frac{1}{1-n} \log \left( \mathrm{tr}_A \left( \rho_A^n \right) \right)\,,
    \label{eq:Renyi}
\end{equation}
where \( n \) is an integer. In particular, the \( n \to 1 \) limit recovers the usual entanglement entropy.
Computing \( \mathrm{tr}_A \rho_A^n \) for \( n \in \mathbb{R} \) in a general CFT is a challenging task. Therefore, one usually employs the replica trick \cite{Holzhey:1994we,Calabrese:2004eu} computing \( \mathrm{tr}_A \rho_A^n \) only for \( n \in \mathbb{Z}^+ \), and then analytically continues to general values of \( n \) \footnote{The analytic continuation of a function from a discrete set of integer values to a continuum domain is not unique and this has been discussed at length in the literature \cite{Calabrese:2004eu,Calabrese:2009qy,Calabrese:2009ez}. We will have nothing to add on this topic here.}. This computation reduces to that of the partition function \( Z_n \) of the \( n \)-fold cover \( \mathcal{M}_n \) of the original space

\begin{equation}
    \mathrm{tr}_A \left( \rho_A^n \right) = \frac{1}{Z^n} \quad \includegraphics[scale=0.23, valign=c]{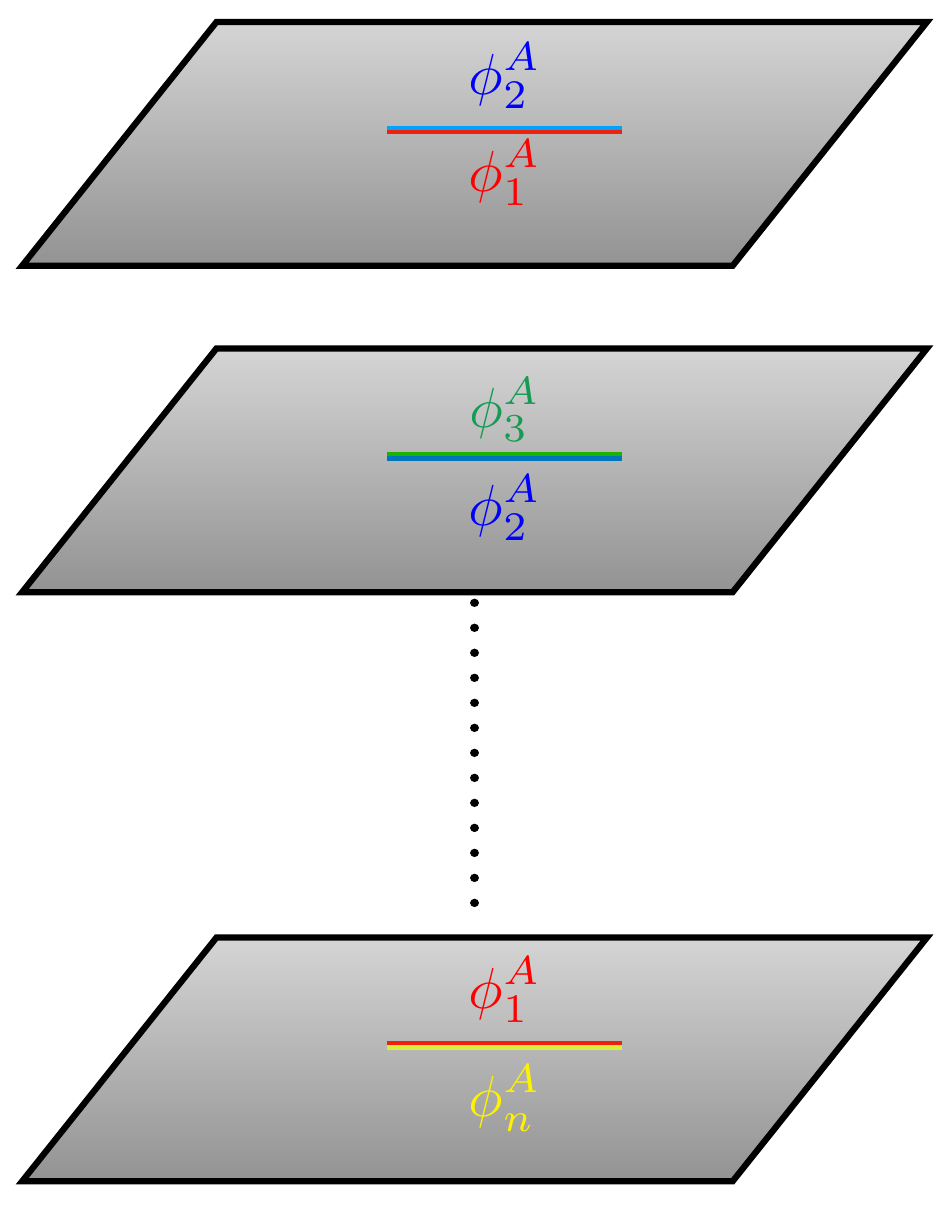} = \frac{Z_n}{Z^n}\,.
    \label{eq:pathintegralrho}
\end{equation}
This computation is carried out by cyclically sewing together the $n$ copies of the manifold along the region \( A \).  

From this construction, the vacuum Rényi entropy \eqref{eq:Renyi} is given by
\begin{equation}
    S_A^{(n)} = \frac{1}{1 - n} \log \left( \frac{Z_n}{Z^n} \right)\,.
\end{equation}
An equivalent description is obtained by introducing an extended non-local operator $\sigma_{n}$, called \textit{twist operator} \cite{Hung:2014npa,Bianchi:2015liz}.
The twist operator behaves as a conformal defect in the orbifolded theory $\mathrm{CFT}^{n}/\mathbb{Z}_{n}$ defined on a single flat spacetime.
In this framework, the Rényi entropy in the vacuum state of the CFT is
\begin{equation}
    S_{A}^{(n)}=\frac{1}{1-n}\log{\langle\sigma_{n}\rangle}\,.
\end{equation}
For a CFT excited by a local quench, we extend this definition to an excited state $|\mathcal{O}\rangle$ \cite{Alcaraz:2011tn}, yielding
\begin{equation}
    S_{A}^{\mathcal{O},(n)}=\frac{1}{1-n}\log{\langle\mathcal{O}^{\otimes n}|\sigma_{n}|\mathcal{O}^{\otimes n}\rangle}\,,
\end{equation}
where $|\mathcal{O}^{\otimes n}\rangle$ represents the state associated with a multiple-copy operator $\mathcal{O}^{\otimes n}$, which corresponds to a local scalar operator inserted at the same point in each CFT replica. We are interested in the excess of entropy, \textit{i.e.}\ the variation of the entropy relative to the vacuum state
\begin{equation} \label{mainobs}
    \Delta S_{A}^{(n)}=S_{A}^{\mathcal{O},(n)}-S_{A}^{(n)}=\frac{1}{1-n}\log{\frac{\langle\mathcal{O}^{\otimes n}|\sigma_{n}|\mathcal{O}^{\otimes n}\rangle}{\langle\sigma_{n}\rangle}}\,.
\end{equation}
This is the main observable we are going to consider in the rest of the paper.

\subsection{Kinematics of the correlation function} \label{subsection:kinem}

Following the setup outlined in section \ref{subsection:ExState}, we are interested in computing the time evolution of the Rényi entropy for a spherical entangling surface. According to \eqref{mainobs}, this is equivalent to computing the 2$n$-point function of scalar primary operators in the presence of a spherical defect on the orbifolded theory. 
To understand the kinematics of the problem, we need to take into account that all the multiple-copy operators are inserted at the same point on the different replicas. Therefore, the kinematics of the problem is identical to that of a two-point function in the presence of a defect. Specifically, we will show in the following that the observable \eqref{mainobs} depends on a single conformal cross-ratio.  

\begin{figure}[htbp!]
\centering
\includegraphics[scale=0.4]{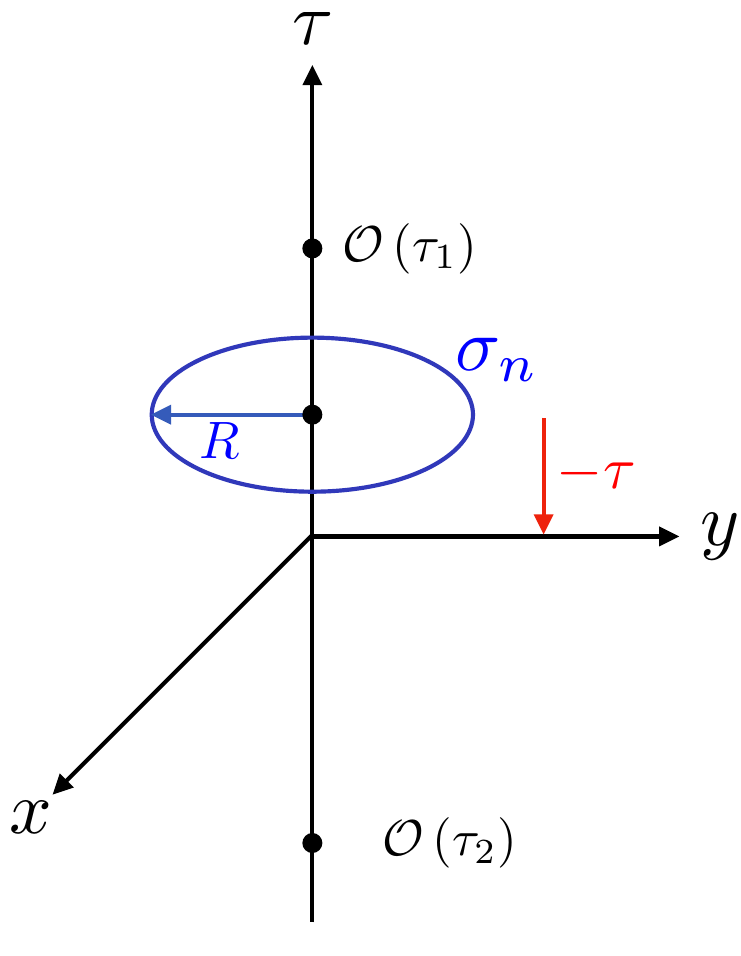}
\caption[Entangling surface]{
    Three-dimensional Euclidean setup of the CFT with two bulk insertions \( \mathcal{O} \) and a circular defect $\sigma_n$ in blue. We can use translation invariance to put the circle at $\tau=0$.}
\label{fig:EntanglingSurface}
\end{figure}

To show this, let us work in Euclidean signature with $\tau=i t$. We define the conformal defect $\sigma_n$, as a codimension-two spacelike spherical defect of radius \( R \) centered at the origin in space and at a given $\tau$. The bulk insertions \( \mathcal{O}_{1,2} \), in a given replica, are located at \( \tau_{1} = -\tau_{2} = \epsilon \), as described in \eqref{eq:ExcitedState}. This configuration is depicted in figure \ref{fig:EntanglingSurface}.
For simplicity, we perform a translation along the \( \tau \)-coordinate such that the defect $\sigma_n$ is located at $\tau=0$, while \(\mathcal{O}_{1,2} \) are located at points $x_{1}^{\mu}= (\tau_{1}-\tau, \vec{0})$ and $x_{2}^{\mu}= (\tau_{2}-\tau,\vec{0})$, respectively. 

In general, the correlation function of two bulk scalar insertions \( \mathcal{O} \) in the presence of a codimension-two defect $\sigma_n$
\begin{equation}
    \langle \mathcal{O}(x_{1})\mathcal{O}(x_{2})\rangle_{\sigma_n}=\frac{F_n(\cos{\phi},\xi)}{|r_{1}|^{\Delta}|r_{2}|^{\Delta}}\, 
    \label{eq:three-point}
\end{equation}
is a function of two conformally invariant cross-ratios  \cite{Billo:2016cpy}. Here, $r_{1,2}$ are the radial distances of the operators from the defect \footnote{For a flat defect, these distances are just the orthogonal distances from the defect. For a generic spherical defect, the explicit expressions can be found in \cite{Billo:2016cpy} and, for our particular kinematics, they reduce to $r_{i}=\frac{R^{2}+(\tau-\tau_{i})^{2}}{2R}$.}. In this setup, we choose the following two cross-ratios
\begin{equation}
\begin{split}
& \cos{\phi}=\frac{ \left[ \left( \tau_{1}-\tau\right)\left( \tau_{2}-\tau\right)+R^{2}\right]^{2}-R^{2}\left( \tau_{1}-\tau_{2}\right)^{2}}{\left[ R^{2}+\left( \tau-\tau_{1}\right)^{2}\right]\left[ R^{2}+\left( \tau-\tau_{2}\right)^{2}\right]}\,,\\
& \xi=\frac{R^{2}\left( \tau_{1}-\tau_{2}\right)^{2}}{\left[ R^{2}+\left( \tau-\tau_{1}\right)^{2}\right]\left[ R^{2}+\left( \tau-\tau_{2}\right)^{2}\right]}\, . 
  \end{split}
    \label{eq:CrossRatioXi}
\end{equation}
Notice that these two cross-ratios, in the constrained kinematics $\tau_{1}=-\tau_{2}=\epsilon$, are related to each other. Specifically, we obtain
\begin{equation}
 \cos{\phi}=1-2\xi\, .
    \label{eq:CrossRatioZeta}
\end{equation}

Consequently, in our specific kinematics, the correlation function \eqref{eq:three-point} depends only on one conformal cross-ratio, namely \( \xi \) in \eqref{eq:CrossRatioXi}. For what we said above, the same applies to the multiple-copy correlator
\begin{equation}
\langle \mathcal{O}^{\otimes n}(x_{1})\mathcal{O}^{\otimes n}(x_{2})\rangle_{\sigma_n}=\langle\sigma_{n}(R)\mathcal{O}^{\otimes n}(x_{1})\mathcal{O}^{\otimes n}(x_{2})\rangle=\frac{(2 R)^{2n\Delta}G_n (\xi)}{\left[ R^{2}+\left( \tau-\tau_{1}\right)^{2}\right]^{n\Delta}\left[ R^{2}+\left( \tau-\tau_{2}\right)^{2}\right]^{n\Delta}}  \,,  
\end{equation}
where $G_{n}(\xi)$ is a function of the cross-ratio $\xi$, and to the quantity \eqref{mainobs}, which we rewrite as
\begin{equation}
    \Delta S_{A}^{(n)}=\frac{1}{1-n}\log{\frac{\langle\sigma_{n}(R)\mathcal{O}^{\otimes n}(x_{1})\mathcal{O}^{\otimes n}(x_{2})\rangle}{\langle\sigma_{n}(R)\rangle\langle\mathcal{O}(x_{1})\mathcal{O}(x_{2})\rangle^{n}}}=\frac{1}{1-n}\log\left[(4\xi)^{n\Delta} G_n(\xi)\right]\,.
    \label{eq:DeltaSn}
\end{equation}
In higher-dimensional setups, obtaining explicit results for this quantity is hard, yet we will explore its universal behavior at early and late  time using the OPE. To do this, we need to analyze the analytic properties of the correlator in restricted kinematics.

\subsection{Analyticity of the correlator}
A generic bulk two-point function with a conformal defect in Euclidean signature is characterized by two OPE channels, which are conveniently analyzed using the alternative cross-ratios $z$ and $\bar z$ defined by
\begin{equation}
    \xi = \frac{(1 - z)(1 - \bar{z})}{4\sqrt{z \bar{z}}}\,, \quad \cos{\phi} = \frac{z + \bar{z}}{2\sqrt{z \bar{z}}}\,.
    \label{eq:xiZ}
\end{equation}
When the correlator is Euclidean $z$ and $\bar z$ are related by complex conjugation. The bulk channel is controlled by the regime $z,\bar z \to 1$ and the defect channel by $z,\bar z \to 0$. In Lorentzian kinematics $z$ and $\bar z$ are independent real variables and the analytic continuation can result in various operators orderings in the correlation function \eqref{eq:three-point}, depending on the chosen contour. Specifically, when the operators \( \mathcal{O} \) become timelike separated, a branch cut arises when one operator crosses the lightcone of the other, and the path chosen around this discontinuity dictates the ordering. Here, we want to understand the analytic structure of the correlator in \eqref{eq:DeltaSn} during the Lorentzian time evolution. In particular, the correlator in \eqref{eq:DeltaSn}, as we mentioned, corresponds to a bulk two-point function in restricted kinematics and the analytic structure in the $\xi$ variable is affected by this restriction. Quite surprisingly, we find that the Lorentzian time evolution does not correspond to a Lorentzian configuration for the two-point function \eqref{eq:three-point} in restricted kinematics. 

To see this, let us perform a Wick rotation in real time $t$ and consider the real-time evolution of the cross-ratio $\xi$. In real-time, we find that the cross-ratio \eqref{eq:CrossRatioXi} becomes

\begin{equation}
    \xi = \frac{4 R^{2} \epsilon^{2}}{\left( R^{2} + (i t - \epsilon)^{2} \right)\left( R^{2} + (i t + \epsilon)^{2} \right)}\,.
    \label{eq:xicrossratiofinale}
\end{equation}

\begin{figure}[htbp!]
\centering
\includegraphics[scale=0.8]{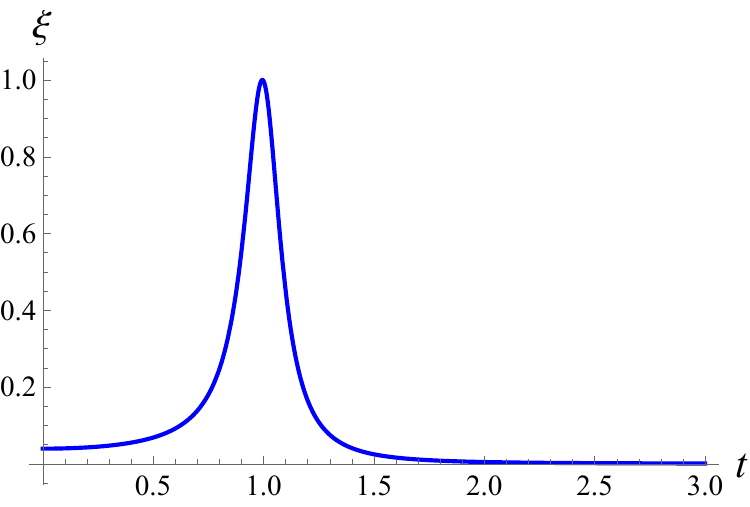}
\caption[cross-ratio evolution]
    {The cross-ratio $\xi$ as a function of time $t$ at fixed $\epsilon=0.1$, $R=1$.}
\label{fig:crossratioevolution}
\end{figure}

\noindent
We start from $t=0$ where the bulk OPE converges. Then, as time evolves, the cross-ratio \( \xi \) remains real, reaches its maximum \( \xi = 1 \) at \( t = \sqrt{R^{2} - \epsilon^{2}} \), and varies within \( \left[ 0, 1 \right] \), as shown in figure \ref{fig:crossratioevolution}.

\begin{figure}[htbp!]
\centering
\includegraphics[scale=0.23]{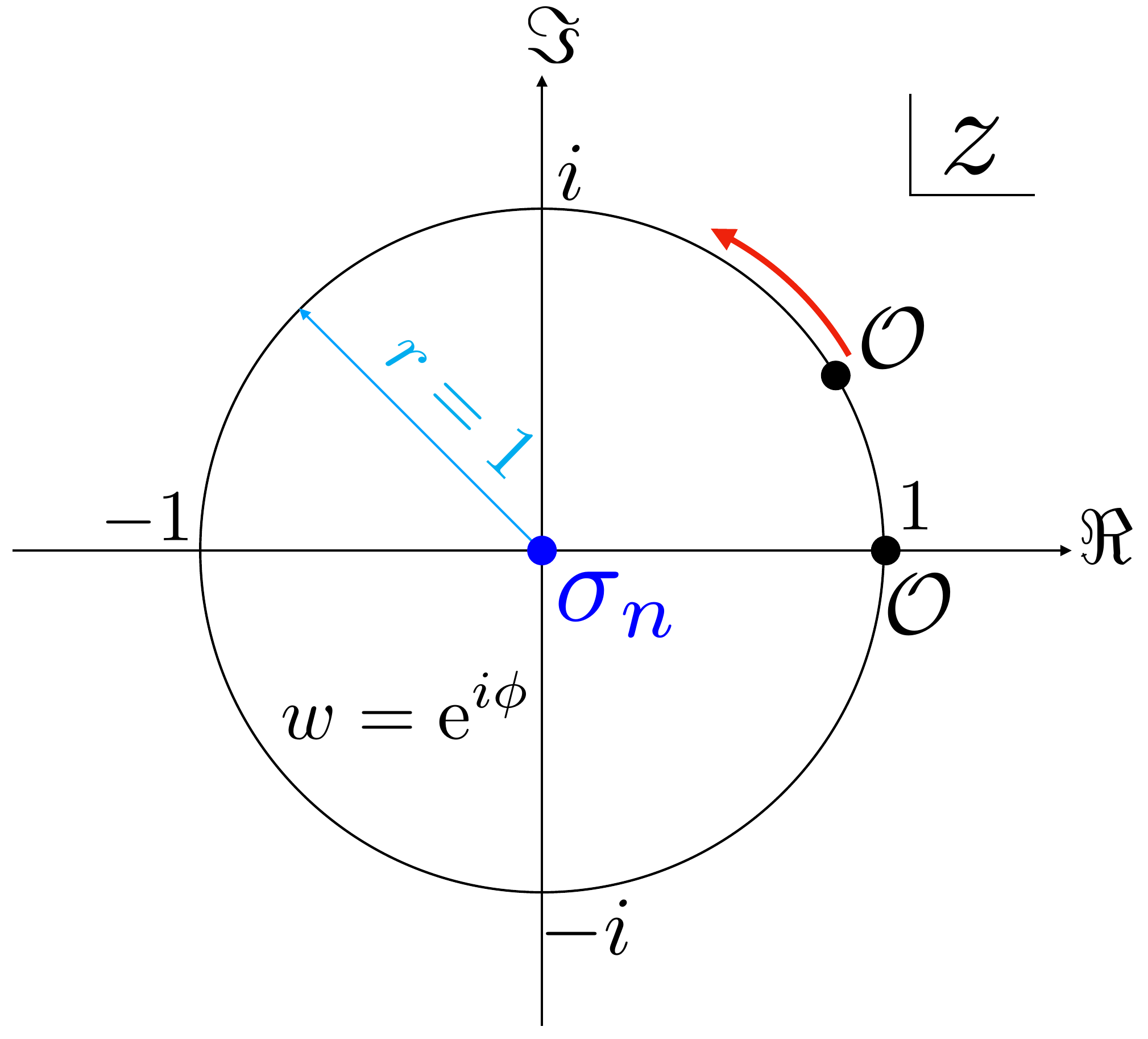}
\caption[z-complex]
    {In the $z$ complex plane the two bulk operators $\mathcal{O}$ are confined on the unit circle.}
\label{fig:z-complex}
\end{figure}

Given the time evolution of the cross-ratio $\xi$, we are now able to show that the Lorentzian time evolution of \eqref{eq:DeltaSn} corresponds to a Euclidean configuration of \eqref{eq:three-point} in our restricted kinematics. Let us first introduce the new variables $r$ and $w$, defined by
$z=r w$ and $\bar{z}=\frac{r}{w}$. In Euclidean kinematics $r$ is a positive real number and $w$ is a phase, while in Lorentzian kinematics $r$ and $w$ are independent real variables.
Since the constrained kinematics \eqref{eq:CrossRatioZeta} requires \( r = 1 \), we can write the cross-ratio $\xi$ as 
\begin{equation}
    \xi=\frac{(1-w)(1-\frac{1}{w})}{4} \,.
\end{equation}
Comparing this expression with \eqref{eq:xicrossratiofinale}, it is straightforward to check that $w$ is just a phase.
This tells us that, during the time evolution of the correlator, the bulk two-point function \eqref{eq:three-point}  is always Euclidean, as  \( z \) and \( \bar{z} \) are complex conjugates of each other.
Moreover, it is easy to see that along the contour dictated by the time evolution, the correlator does not cross any singularity. Indeed, by a conformal transformation, one can map the defect to a plane and the complex $z$ plane corresponds to two orthogonal directions with one operator fixed at $z=\bar z=1$ and the other operator in $z,\bar z$. The constrained kinematics \eqref{eq:CrossRatioZeta} forces the operator to move around the unit circle, as shown in figure \ref{fig:z-complex}. The singularity at $z=0$ is therefore never reached, as one should expect since the operators are fixed at the center of the spherical entangling surface. The singularity at $z=1$, \textit{i.e.}\ $\xi=0$, is relevant both for the early- and late-time behavior of the entanglement entropy. Along the contour, no other singularity is present and this also ensures the convergence of the OPE at late time.

\label{subsection:anal}

\subsection{OPE at early and late  times}
\label{subsection:OPE}
We have seen that the bulk OPE limit controls the early and late time behavior of the entanglement entropy. In the following, we will use this fact to compute the leading order term in the late time asymptotics of the correlator in \eqref{eq:DeltaSn}. For the moment, we assume that the lightest exchanged operator in the bulk channel is the stress-energy tensor. At the end of this section, we will comment on this assumption and analyze its domain of validity.

The leading contribution for $\xi\to 0$ is determined by the lightest operator exchanged in the OPE of the two scalar operators $\mathcal{O}$. The identity operator, that is certainly present for two identical operators, contributes with a disconnected term
\begin{equation}
    \frac{\langle\sigma_{n}(R)\mathcal{O}^{\otimes n}(x_{1})\mathcal{O}^{\otimes n}(x_{2})\rangle}{\langle\sigma_{n}(R)\rangle\langle\mathcal{O}(x_{1})\mathcal{O}(x_{2})\rangle^{n}} \sim \frac{\langle\sigma_{n}(R)\rangle\langle\mathcal{O}(x_{1})\mathcal{O}(x_{2})\rangle^{n}}{\langle\sigma_{n}(R)\rangle\langle\mathcal{O}(x_{1})\mathcal{O}(x_{2})\rangle^{n}}=1\,,
\end{equation}
which is precisely canceled by the vacuum contribution in \eqref{eq:DeltaSn}, yielding $\Delta S_{A}^{(n)}|_{\text{id}}=0$. Consequently, under our assumptions, the leading term for $\xi \rightarrow 0$ is provided by the stress-energy tensor $\mathcal{T}$, a single-copy operator symmetrized over all $n$ replicas
\begin{equation}
    \mathcal{T}_{\mu\nu}=\sum_{m=0}^{n-1}\mathbb{1}^{\otimes m}\otimes T_{\mu\nu}\otimes\mathbb{1}^{\otimes (n-m-1)}\,,
    \label{eq:singlecopy}
\end{equation}
with $T_{\mu\nu}$ inserted at the same point in every replica. 

When we take the operators close to each other, they get close in every replica, but only one replica contains the stress tensor in \eqref{eq:singlecopy}. All other replicas contain the identity, giving a disconnected term $\langle\mathcal{O}(x_{1})\mathcal{O}(x_{2})\rangle^{n-1}$. Furthermore, each term in the sum in \eqref{eq:singlecopy}  contributes the same way, yielding an overall factor of $n$ . This gives 
\begin{equation}
    \langle\sigma_{n}(R)\mathcal{O}^{\otimes n}(x_{1})\mathcal{O}^{\otimes n}(x_{2})\rangle\overset{\xi\rightarrow0}{\sim}n\epsilon_{\mu\nu}(x_{12})\langle\sigma_{n}(R)T^{\mu\nu}(x_{2})\rangle\langle\mathcal{O}(x_{1})\mathcal{O}(x_{2})\rangle^{n-1}\,,
    \label{eq:correlationfunctionPartial}
\end{equation}
where $\epsilon_{\mu\nu}(x_{12})$ is a tensorial structure depending on $x_{12}=x_{1}-x_{2}$ and containing also the OPE coefficient. The explicit expression of this tensorial structure is

\begin{equation}
    \epsilon_{\mu\nu}(x_{12}) = \frac{c_{\mathcal{O} \mathcal{O} T}}{C_{T} |x_{12}|^{2\Delta - d + 2}} x_{12\mu} x_{12\nu}\,.
    \label{eq:epsilontensor}
\end{equation}
Inserting it and dividing by the vacuum contribution we get

\begin{equation}
\begin{split}
    \frac{\langle \sigma_{n}(R) \mathcal{O}^{\otimes n}(x_{1}) \mathcal{O}^{\otimes n}(x_{2}) \rangle}{\langle \sigma_{n}(R) \rangle \langle \mathcal{O}(x_{1}) \mathcal{O}(x_{2}) \rangle^{n}} & \overset{\xi \rightarrow 0}{\sim} 1 + \frac{n\,  c_{\mathcal{O} \mathcal{O} T}}{C_{T} |x_{12}|^{-d+2}} x_{12 \mu} x_{12 \nu} \frac{\langle \sigma_{n}(R) T^{\mu \nu}(x_{2}) \rangle}{\langle \sigma_{n}(R) \rangle}\,.
    \label{eq:correlationfunctionPartial4}
\end{split}
\end{equation}
The one-point function of the stress tensor was computed in \cite{Hung:2014npa} and we adapt their result  to our configuration in figure \ref{fig:EntanglingSurface}. In particular, since the entangling surface lies on a constant time slice at \( \tau=0 \), we label the coordinates as \( x^{\mu} = (\tau, x^{i}) \), with \( i \) being directions orthogonal to \( \tau \). The stress tensor is inserted at $x_2^i=0$ and \( x^1_{2} = \tau_{2} - \tau \), leading to
\begin{equation}
    \frac{\langle \sigma_{n}(R) T_{\tau\tau} \rangle}{\langle\sigma_{n}(R)\rangle} = \frac{d-1}{2 \pi n} \left( \frac{2 R}{R^{2} + (\tau_{2} - \tau)^{2}} \right)^{d} h_{n}\,, \quad \frac{\langle \sigma_{n}(R) T_{i j} \rangle}{\langle\sigma_{n}(R)\rangle} = -\frac{\delta_{i j}}{2 \pi n} \left( \frac{2 R}{R^{2} + (\tau_{2} - \tau)^{2}} \right)^{d} h_{n}\,.
    \label{eq:1pointtwistsphericalFINAL}
\end{equation}
Using these expressions and noting \( x_{12}^{i} = 0 \), we substitute these results into \eqref{eq:correlationfunctionPartial4} to find
\begin{equation}
    \frac{\langle \sigma_{n}(R) \mathcal{O}^{\otimes n}(x_{1}) \mathcal{O}^{\otimes n}(x_{2}) \rangle}{\langle \sigma_{n}(R) \rangle \langle \mathcal{O}(x_{1}) \mathcal{O}(x_{2}) \rangle^{n}} \overset{\xi \rightarrow 0}{\sim} 1 + \frac{2^{d-1} (d-1)}{\pi} \frac{c_{\mathcal{O} \mathcal{O} T}\, h_{n} }{ C_{T}} \frac{R^{d} |\tau_{1} - \tau_{2}|^{d}}{(R^{2} + (\tau_{2} - \tau)^{2})^{d}}\,.
\end{equation}
We recognize a power of the cross-ratio \( \xi \) \eqref{eq:CrossRatioXi}, yielding the excess of Rényi entropy in the OPE limit
\begin{equation}
    \Delta S_{A}^{(n)} \overset{\xi \rightarrow 0}{\sim} \frac{   2^{d-2} d \, \Gamma(d/2)}{(n-1) \pi^{d/2+1}} \frac{ h_{n} \, \Delta }{ C_{T}} \, \xi^{d/2}\,,
    \label{eq:partialDeltaS}
\end{equation}
where we fixed the coefficient $c_{\mathcal{O}\mathcal{O}T}$ with a conformal Ward identity \cite{Osborn:1993cr}
\begin{equation}
    c_{\mathcal{O}\mathcal{O}T}=-\frac{d\Delta}{(d-1)S_{d}}\,,\quad S_{d}=\frac{2\pi^{d/2}}{\Gamma(d/2)}\,.
    \label{eq:coot}
\end{equation}
It is also useful to show the late time behavior of the excess of Rényi entropy exploiting the real time expression of $\xi$ \eqref{eq:xicrossratiofinale} for large $t$ into \eqref{eq:partialDeltaS}

\begin{equation}
    \Delta S_{A}^{(n)} \overset{t \rightarrow \infty}{\sim} \frac{   2^{d-2} d \, \Gamma(d/2)}{(n-1) \pi^{d/2+1}} \frac{ h_{n} \, \Delta }{ C_{T}} \, \frac{(2R)^{d} \epsilon^{d}}{t^{2d}}\,.
    \label{eq:partialDeltaSrealtime}
\end{equation}

To compute the entanglement entropy, we take the \( n \rightarrow 1 \) limit and \( h_{n} \) reduces to the coefficient \( C_{T} \) of the stress-energy tensor two-point function \cite{Hung:2014npa}
\begin{equation}
    \partial_{n}h_{n}|_{n=1}=2\pi^{\frac{d}{2}+1}\frac{\Gamma(d/2)}{\Gamma(d+2)}C_{T}\,.
    \label{eq:confdimh}
\end{equation}
Expanding \( h_{n} \) in \eqref{eq:partialDeltaS} around \( n \rightarrow 1 \), we obtain the excess of entanglement entropy in the early and late time limit \( \xi \rightarrow 0 \) 
\begin{equation}
    \Delta S_{A}^{EE} = \lim_{n \rightarrow 1} \Delta S_{A}^{(n)} = \frac{  2^{d-1} d\, \Gamma(d/2)^{2}}{\Gamma(d+2)} \,\Delta \, \xi^{d/2} + \mathcal{O}(\xi^{d})\,,
    \label{eq:DeltaSEEFInale}
\end{equation}
which displays a universal behavior depending only on the dimension $\Delta$ of the operator creating the state.

For general $n$ and in $d=2$, our result \eqref{eq:partialDeltaS} reduces to 
\begin{equation}
    \Delta S_{A}^{(n)} \overset{\xi \rightarrow 0}{\sim} \frac{2\Delta  h_{n}}{\pi^{2}(n-1)\, C_{T}} \, \xi\,,
    \label{eq:partialDeltaSd=2}
\end{equation}
matching the one obtained in \cite{Bianchi:2022ulu}\footnote{Their result was obtained in the presence of a conformal boundary, but the final result is identical to the homogeneous case, and we can compare it with our result. A careful reader should be aware of the different convention between our work and \cite{Bianchi:2022ulu}. In $2d$ they used complex coordinates $z, \bar{z}=x\pm i\tau$ to define the stress-energy tensor $T(z)=-2\pi T_{zz}$, $\bar{T}(\bar{z})=-2\pi T_{\bar{z}\bar{z}}(\bar{z})$. This led to a different notation in the CFT data as $C_{D}\rightarrow4\pi^{2}C_{T}$, $c_{\mathcal{O}\mathcal{O}D}\rightarrow\pi c_{\mathcal{O}\mathcal{O}T}$, $c_{\mathcal{O}\mathcal{O}T}\rightarrow\frac{2\Delta}{\pi}$ and $a_{D,n}=-4\Delta_{n}$. Moreover, in \cite{Hung:2014npa} it has been proved that the conformal dimension $h_{n}$ of the twist operator in higher dimensions is the same as the conformal dimension $\Delta_{n}$ in $2d$ when restricting to the two-dimensional case. In two dimensions, the twist operator becomes a pair of local operators, which in our case are located at $\tau=0$ and $x=\pm R$.}.
Also, from \eqref{eq:DeltaSEEFInale} in $d=2$ we find

\begin{equation}
    \Delta S_{A}^{EE} = \frac{2}{3} \Delta \xi + \mathcal{O}(\xi^{2})\,,
    \label{eq:DeltaSee2d}
\end{equation}

\noindent
again, matching the result in \cite{Bianchi:2022ulu}.

Let us comment now on our initial assumption on the spectrum.
Since the operators $\mathcal{O}^{\otimes n}$ are multiple-copy insertions, their OPE can include both single- and multiple-copy operators, respecting the $\mathbb{Z}_{n}$ symmetry.
However, only operators with a non-vanishing one-point function in the presence of the twist operator will contribute to the OPE described above. The first order for $n\to 1$ of the one-point function is computed by the vacuum modular Hamiltonian $H_{A}^{0}$ according to the recipe \cite{Smolkin:2014hba,Hung:2014npa}
\begin{equation}
    \lim_{n\to 1} \frac{\partial}{\partial n} \langle \mathcal{O}(x)\rangle_{\sigma_n} = -\langle \mathcal{O}(x) H_A^{0}\rangle\,.
\end{equation}
In our setup, where the modular Hamiltonian is local and proportional to the integral of the stress tensor, this correspondence implies that the only single-copy operator relevant in the OPE is the stress tensor itself since all the other operators have a vanishing two-point function with the stress tensor. This holds in the $n\rightarrow1$ limit, but since we are mostly interested in computing the entanglement entropy, we can safely use this assumption. 

The next class of operators that might play a role in the OPE are double-copy operators. Consider, for instance, the lightest scalar operator $\mathcal{O}_{L}$ of dimension $\Delta_{L}$ that is exchanged in the OPE of the two $\mathcal{O}$ operators (excluding the identity). We can construct $\left[ \frac{n}{2}\right]$ double-copy scalar operators,
\begin{equation}
    \mathcal{O}^{(2)}_{i} = \mathbb{1}^{\otimes m} \otimes \mathcal{O}_{L}(0) \otimes \mathbb{1}^{\otimes i} \otimes \mathcal{O}_{L}(0) \otimes \mathbb{1}^{\otimes(n-m-i-2)} + \mathbb{Z}_{n}\text{-symm}\,,
\end{equation}
with dimension $2\Delta_{L}$. If $\Delta_L>\frac{d}{2}$, the stress tensor is lighter than these operators and our assumption holds. Otherwise, similar considerations can be carried out for $\mathcal{O}_L$ leading to a different late-time behavior.

In holographic theories, characterized by a large central charge and a sparse spectrum, single-trace operators other than the stress tensor in the OPE of $\mathcal{O} \times \mathcal{O}$ are suppressed, and double-trace operators dominate. The lightest of these has dimension $2\Delta$, and with double-copy insertions, the dominant contribution to the excess of entanglement entropy comes from operators with dimension $4\Delta$. Therefore, the stress-energy tensor remains the lightest operator in holographic theories as long as we create the state with an operator of dimension $\Delta > \frac{d}{4}$.

\subsection{Universal bound on the entanglement entropy in higher dimensions}
\label{subsection:bound}

As pointed out in \cite{Bianchi:2022ulu}, it is possible to establish a universal bound on the early- and late-time behavior of the excess of entanglement entropy as a consequence of the relative entropy bound \cite{Casini:2008cr,Blanco:2013joa,Sarosi:2016oks,Sarosi:2016atx}. Here, we adapt the argument of \cite{Bianchi:2022ulu}, which was carried out for a boundary CFT, to the homogeneous case, and we show that the (early-) late-time (grow) decay is bounded by a power-law in the cross-ratio $\xi$ \eqref{eq:CrossRatioXi}. 

The relative entropy \cite{Casini:2008cr} is an entropic quantity which quantifies the distance between two different states in the Hilbert space. It can be written as
\begin{equation}
S(\rho_A | \rho_A^0) = \mathrm{Tr}(\rho_A \log \rho_A) - \mathrm{Tr}(\rho_A \log \rho_A^0) \, ,
\end{equation}
and it satisfies the positivity property
\begin{equation}
\label{eq:pos_RE}
S(\rho_A | \rho_A^0) \ge 0 \, .
\end{equation}
Using the definition of entanglement entropy \eqref{eq:VonNeumannEntropy} and introducing the modular Hamiltonian $H_A^0$ as
\begin{equation}
    \rho_A^0 = \frac{e^{-H_A^0}}{\mathrm{Tr}(e^{-H_A^0})} \, ,
    \label{eq:rhohammodu}
\end{equation}
we can rewrite the equation \eqref{eq:pos_RE} as
\begin{equation}
    \Delta S_A^{EE} = S(\rho_A) - S(\rho_A^0) \le \mathrm{Tr}(H_A^0 \rho_A) - \mathrm{Tr}(H_A^0 \rho_A^0)\,,
    \label{eq:boundHam}
\end{equation}
which, as anticipated, provides a bound on the excess of entropy. Notably, in our setup, the r.h.s.\ can be computed explicitly. Indeed, the modular Hamiltonian \( H_A^0 \) associated with a vacuum density matrix reduced on a $(d-1)$-dimensional sphere \( S^{d-1} \) of radius \( R \) takes the local form \cite{Casini:2011kv}

\begin{equation}
    H_A^0(R, t) = 2\pi \int_A \mathrm{d}^{d-1}\vec{x} \, \frac{R^2 - \vec{x}^2}{2R} T_{tt}(t, \vec{x})\,,
    \label{eq:modularHam}
\end{equation}
where \( A \) is the spacelike sphere \( S^{d-1} \), \( x^\mu = (t, \vec{x}) \), with \( \mu = 0, \ldots, d-1 \), and \( \vec{x}^2 \le R^2 \). 

\noindent
Thus, the modular Hamiltonian is given by the spatial integral over the region \( A \) of the stress-energy tensor component \( T_{tt} \).
Consequently, the first term in the bound \eqref{eq:boundHam} is the spatial integral of the stress tensor expectation value in the excited state

\begin{equation}
\begin{split}
    \mathrm{Tr}\left( H_{A}^{0}\rho_{A}\right)& = 2\pi \int_{A} \mathrm{d}^{d-1} \vec{x} \, \frac{R^{2}-\vec{x}^{2}}{2R} \langle \mathcal{O} | T_{tt}(t, \vec{x}) | \mathcal{O} \rangle \\
    & = 2\pi \int_{A} \mathrm{d}^{d-1} \vec{x} \, \frac{R^{2}-\vec{x}^{2}}{2R} \frac{\langle \mathcal{O}(\tau_{1}) T_{tt}(t, \vec{x}) \mathcal{O}(\tau_{2}) \rangle}{\langle \mathcal{O}(\tau_{1}) \mathcal{O}(\tau_{2}) \rangle}\,,
\end{split}
\label{eq:integralbound}
\end{equation}
where \( \tau_{1} = -\tau_{2} = \epsilon \). The second term in \eqref{eq:boundHam} is the vacuum expectation value of the stress-energy tensor, and it vanishes. Therefore, the bound is determined by the integral in \eqref{eq:integralbound}, where the two-point function \( \langle \mathcal{O} \mathcal{O} \rangle \) and the three-point function \( \langle \mathcal{O} \mathcal{O} T \rangle \) are fully determined by conformal invariance

\begin{equation}
\begin{split}
     & \langle\mathcal{O}(x_{1})\mathcal{O}(x_{2})\rangle=\frac{1}{|x_{12}|^{2\Delta}}\,, \quad \langle \mathcal{O}(x_{1}) \mathcal{O}(x_{2}) T_{\alpha\beta}(x_{3}) \rangle = \frac{c_{\mathcal{O} \mathcal{O} T} H_{\alpha\beta}(x_{1}, x_{2}, x_{3})}{|x_{12}|^{2\Delta - d + 2} |x_{13}|^{d-2} |x_{23}|^{d-2}}\,, \\
    & H_{\alpha\beta} = V_{\alpha} V_{\beta} - \frac{1}{d} V_{\mu} V^{\mu} \delta_{\alpha\beta}\,, \quad V_{\alpha} = \frac{x_{13\alpha}}{x_{13}^{2}} - \frac{x_{23\alpha}}{x_{23}^{2}}\,.
     \end{split}
     \label{eq:two-three point function}
\end{equation}
To do the computation, we analytically continue to Euclidean time \( \tau \) and consider the Euclidean time-time component of the stress-energy tensor \( T_{\tau \tau}(\tau, r) \), which introduces an overall minus sign. We switch to spherical coordinates \( x^{\mu} = (\tau, \vec{x}) = (\tau, x, y^{i}) = (\tau, r \sin \theta, \Omega^{i} r \cos \theta) \), with \( \theta \in [0, \pi] \) and \( \Omega^{i} \Omega_{i} = 1 \). Then, the integral in \eqref{eq:integralbound} becomes

\begin{equation}
\begin{split}
    \mathrm{Tr}\left( H_{A}^{0} \rho_{A} \right) & = -\frac{2\pi \Omega_{d-3}}{2R} \int_{0}^{\pi} \mathrm{d}\theta \int_{0}^{R} \mathrm{d}r \, r^{d-2} (\sin \theta)^{d-3} (R^{2} - r^{2}) \frac{\langle \mathcal{O}(\tau_{1}) T_{\tau \tau}(\tau, r) \mathcal{O}(\tau_{2}) \rangle}{\langle \mathcal{O}(\tau_{1}) \mathcal{O}(\tau_{2}) \rangle}\,,
\end{split}
\label{eq:integralbound2}
\end{equation}
with \( \Omega_{d-3} = \frac{2\pi^{\frac{d}{2} - 1}}{\Gamma\left[ \frac{d}{2} - 1 \right]} \). The integral gives 
\begin{equation}
    \Delta S_{A}^{EE} \le \frac{1}{2} \sqrt{\pi} \Delta \, \xi^{d/2} \, \Gamma\left(\frac{d}{2}\right) \, {}_2\tilde{F}_1\left(1, \frac{d}{2}, \frac{d + 3}{2}; \xi\right)\,,
    \label{eq:boundDdim}
\end{equation}
where ${}_2\tilde{F}_{1}(a, b, c; z) = {}_2 F_{1}(a, b, c; z) / \Gamma(c)$ is the regularized hypergeometric function. Notice that the bound in \eqref{eq:boundDdim} is valid for any dimension $d$ and at all times $t$, contrary to what could be done in \cite{Bianchi:2022ulu} for the boundary case.

We can expand the result for $\xi\to 0$, \textit{i.e.}\ in the early- and late-time limit, finding
\begin{equation}\label{boundOPElatet}
    \Delta S_{A}^{EE} \lesssim \frac{  2^{d-1} d\, \Gamma(d/2)^2}{\Gamma(d+2)} \Delta \,\xi^{d/2}\,,
\end{equation}
which precisely matches the behavior derived in the OPE limit \eqref{eq:DeltaSEEFInale}. This implies that the contribution of the stress tensor in the OPE saturates the bound in any dimension. Consequently, excluding the stress-energy tensor, the lightest operator exchanged in the OPE of $\mathcal{O}\times\mathcal{O}$, whether it is lighter or heavier than the stress-energy tensor itself, must provide a negative contribution to the excess of entanglement entropy.

We can also compare our early and late time bound \eqref{boundOPElatet} with the result in \cite{Bianchi:2022ulu}, where the authors found, in the OPE limit $\xi \rightarrow 0$,

\begin{equation}
    \Delta S_{A}^{EE} \lesssim c_{\mathcal{OO}D} \frac{\pi^{\frac{d+1}{2}}}{2(d-1)\Gamma\left( \frac{d+3}{2} \right)} \xi^{\frac{d}{2}}\,,
    \label{eq:boundDLorenzo}
\end{equation}

\noindent
where $c_{\mathcal{O}\mathcal{O}D}$ in $d>2$ cannot be determined, as there are no known Ward identities to fix the OPE coefficient solely in terms of the conformal dimension $\Delta$. However, we note that the power of $\xi$ in the bound remains the same for configurations with and without a conformal boundary.

Let us focus on the specific case \( d = 3 \), which will be relevant for the holographic computations in section \ref{section:LocalQuench}. Here, the bound in \eqref{eq:boundDdim} simplifies to
\begin{equation}
    \Delta S_{A}^{EE} \leq \frac{\pi \Delta \left( 2 - \xi - 2 \sqrt{1 - \xi} \right)}{2 \sqrt{\xi}}\,, \quad (d=3)\,,
    \label{eq:bound3xi}
\end{equation}
being the r.h.s.\ always non-negative in the range of validity $\xi = [0,1]$.
When expanding the bound for $\xi\rightarrow 0$ in \eqref{eq:bound3xi}, the first non-zero term is completely saturated by the early and late time limits of the CFT result in $d=3$ dimensions \eqref{eq:DeltaSEEFInale},

\begin{equation}
    \Delta S_{A}^{EE} = \frac{\pi \Delta}{8}\xi^{\frac{3}{2}} + \mathcal{O}(\xi^{2})\,, \quad (d=3)\,.
    \label{eq:boundOPE3}
\end{equation}
\noindent

Let us now examine the simpler two-dimensional case and compare it with the result obtained in \cite{Bianchi:2022ulu}.
In two dimensions, the bound in \eqref{eq:boundDdim} simplifies to

\begin{equation}
    \Delta S_{A}^{EE} \leq 2 \Delta \left[1-\sqrt{\frac{1}{\xi} - 1} \arcsin{\left(\sqrt{\xi}\right)} \right] \simeq \frac{2}{3}\Delta \, \xi + \mathcal{O}(\xi^{2}) \quad (d=2)\,,
    \label{eq:bound2xi}
\end{equation}
which reproduces the result obtained in \cite{Bianchi:2022ulu} for the two-dimensional bound. A reoccurring feature of this result is that the bound is again saturated in the early and late time limit $\xi \rightarrow 0$ by the CFT computation in \eqref{eq:DeltaSee2d}.

\begin{figure}[htbp!]
\centering
\includegraphics[scale=0.5]{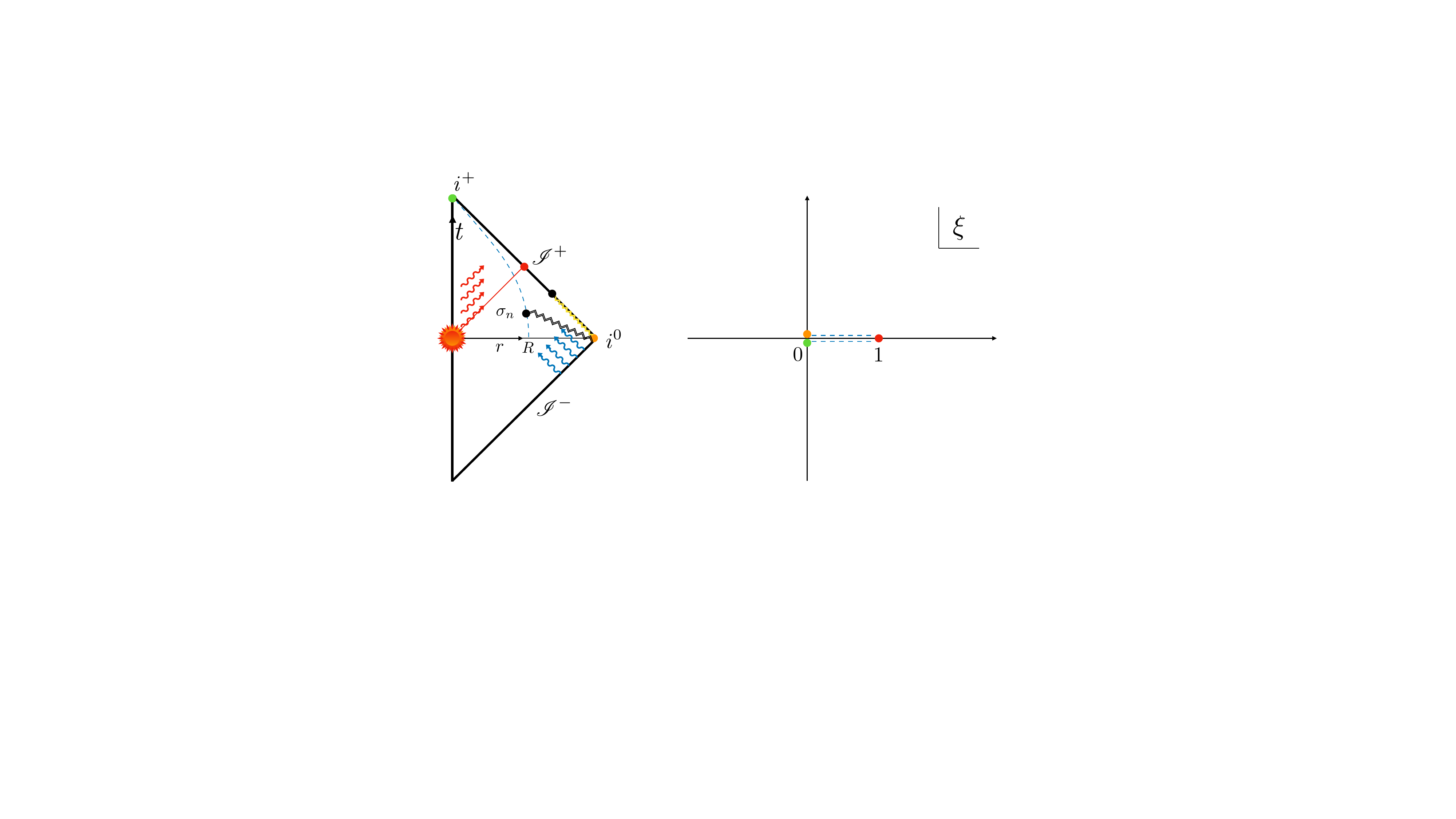}
\caption[Evolution of Cross-Ratio]
    {Sketch of the time evolution of the observable (black dot) in radial coordinates within a Penrose diagram (left) and in cross-ratio space $\xi$ (right). As the observable moves along the future null infinity $\mathscr{I}^{+}$, the cross-ratio $\xi$ exhibits a perfect $\mathbb{Z}_{2}$ symmetry. In lightcone coordinates, starting at $t=0$, corresponding to spatial infinity $i^{0}$, the cross-ratio is $\xi=0$. On the lightcone emanating from the origin (red), $\xi=1$, and at future infinity $i^{+}$, the cross-ratio is back to $\xi=0$. }
\label{fig:EvolCrossRatio}
\end{figure}

Lastly, we would like to plot the bound for several spacetime dimensions. 
To do so, we will take the limit where the spherical twist operator $\sigma_{n}$ goes to the asymptotic region $\mathscr{I}^{+}$ as shown in figure \ref{fig:EvolCrossRatio}. Although this is slightly unconventional, it is necessary to obtain an early-time behaviour that is not affected by the radiation coming from the past (see also \cite{Meineri:2019ycm,Bianchi:2022ulu}). Indeed, the observable in \eqref{eq:DeltaSn} is clearly time-reversal invariant, while the physical process we have in mind is not. We want to create a burst of energy at $t=0$ and study the emitted radiation, but the way this happens in the observable \eqref{eq:DeltaSn} is through a radiation coming from the past, converging at the origin at $t=0$ and spreading again for $t>0$. This produces a non-vanishing entropy at $t=0$ unless we find a way to exclude the past radiation. We do so by taking the twist operator to null infinity.

\begin{figure}[htbp!]
\centering
\includegraphics[scale=0.65]{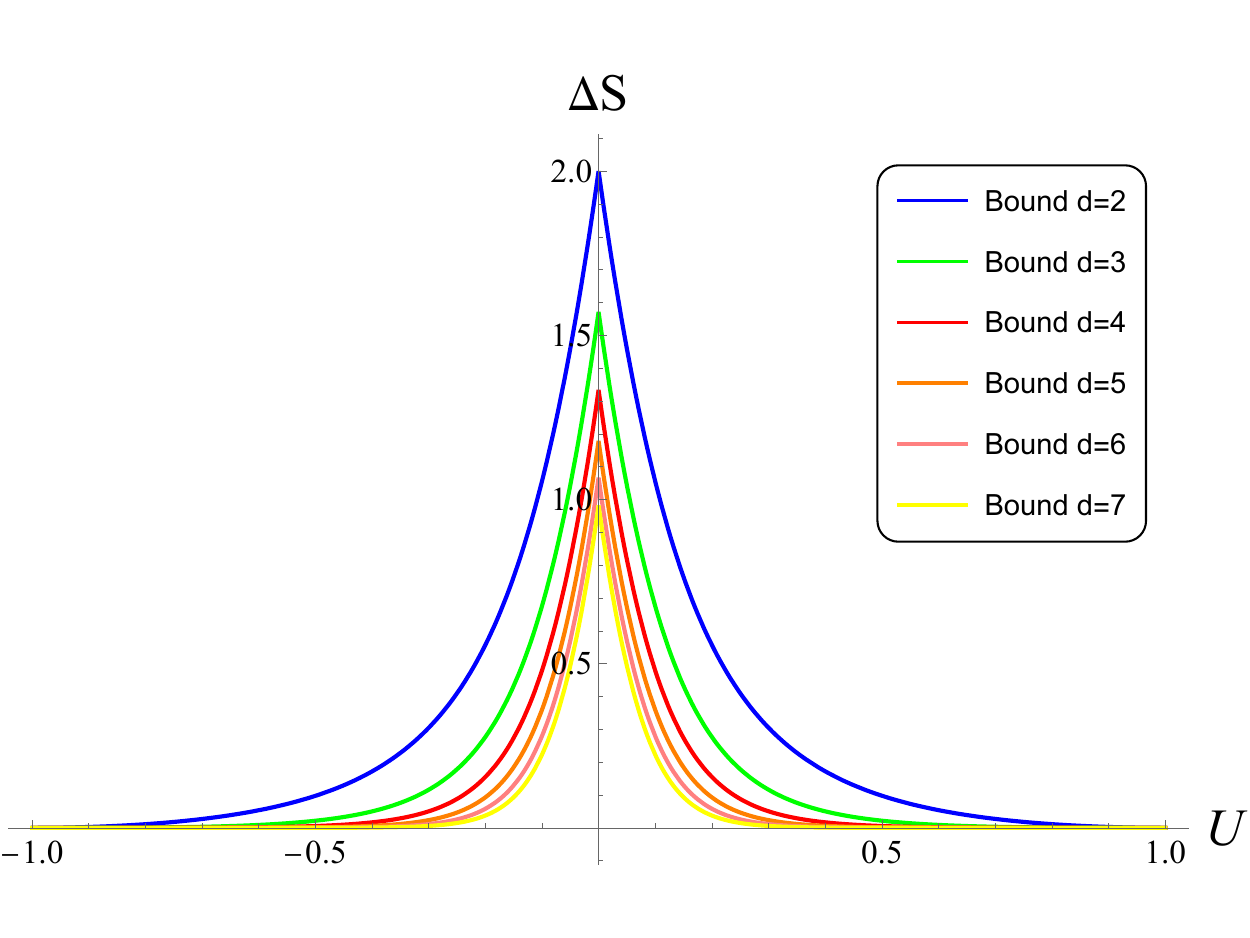}
\caption[Delta SEE]
    {Plot of the entanglement entropy bound $\Delta S$ as a function of the lightcone coordinate $U$ on $\mathscr{I}^{+}$, at fixed $\epsilon=0.4$ and $\Delta=1$. The bound is shown for dimensions $d=2$ (blue), $d=3$ (green), $d=4$ (red), $d=5$ (orange), $d=6$ (pink), and $d=7$ (yellow).}
\label{fig:PlotBounds}
\end{figure}

Consider $d$-dimensional Minkowski space in spherical coordinates:
\[
\mathrm{d}s^{2} = -\mathrm{d}t^{2} + \mathrm{d}r^{2} + r^{2} \mathrm{d}\Omega_{d-2}^{2}\,,
\]
with the twist operator located at constant $t$ and $r=R$.
To simplify the analysis, we parametrize the position of the twist operator with compact lightcone coordinates:
\begin{equation}
    \begin{split}
        U &= \frac{2}{\pi} \arctan\left( \frac{R-t}{l} \right)\,, \\
        V &= \frac{2}{\pi} \arctan\left( \frac{R+t}{l} \right)\,,
    \end{split}
    \label{eq:lightcone coord}
\end{equation}
where $U, V \in (-1, 1)$ and $l$ is an arbitrary scale. In these coordinates, the cross-ratio $\xi$ becomes:
\begin{equation}
    \xi = \frac{\left(\frac{\epsilon}{l}\right)^{2}\left(\tan{\frac{\pi U}{2}} + \tan{\frac{\pi V}{2}}\right)^{2}}{\left[\tan^{2}\frac{\pi U}{2} + \left(\frac{\epsilon}{l}\right)^{2}\right]\left[\tan^{2}\frac{\pi V}{2} + \left(\frac{\epsilon}{l}\right)^{2}\right]}.
\end{equation}
Notice that this expression is identical to the two-dimensional case of \cite{Bianchi:2022ulu}, although in that case the defect is the boundary and the twist operator is local, while in this case there is not boundary and the defect is the twist operator.
For simplicity, setting $l = 1$ and taking the future null infinity limit $\mathscr{I}^{+}$ ($V \to 1$), we find:
\begin{equation}
    \xi = \frac{\epsilon^{2}}{\tan^{2}\left( \frac{\pi U}{2} \right) + \epsilon^{2}} + \mathcal{O}(V - 1)\,.
\end{equation}
This expression reveals that the evolution of $\xi$ is symmetric under $U \to -U$, as depicted in figures \ref{fig:EvolCrossRatio} and \ref{fig:PlotBounds}. In particular, both the early-time ($U \to 1$) and late-time ($U \to -1$) limits are clearly governed by the same OPE $\xi \to 0$.
Using these lightcone coordinates in the $V \to 1$ limit, we plot the evolution of the bound for different spacetime dimensions in figure \ref{fig:PlotBounds}. As $d$ increases, the bound decreases, approaching zero as $d \to \infty$.

\noindent
\section{The BCFT Setup, a tale of two defects}
\label{section:BCFTsetup}

In this section, we generalize our construction along the lines of \cite{Bianchi:2022ulu} by introducing a flat, timelike conformal boundary, which partially breaks the original conformal symmetry of the system. This setup is particularly insightful due to its holographic dual, which includes an End-of-the-World (EoW) brane. Such brane configuration enables black hole evaporation, allowing radiation to be collected in a reservoir represented by the BCFT itself.

\subsection{Excitation in the BCFT}
\label{subsection:BCFTexcitation}

We construct an excited state on the BCFT along the lines of what we did in the homogeneous case.  Consider $d$-dimensional Minkowski spacetime with coordinates $x^{\mu}=(t, x, \vec{y})$, where $\vec{y}=(y_{1}, \ldots, y_{d-2})$. The spacetime has a conformal boundary located on the hyperplane $(t, \vec{y})$ at $x=0$, such that the CFT is defined on the half-space $x\ge0$.
The presence of this boundary reduces the original conformal group $SO(d,2)$ of the homogeneous CFT to the subgroup $SO(d-1,2)$, preserving conformal symmetry along the boundary. 

The local operators in the spectrum of a BCFT can be divided into two main classes: bulk operators which are defined for $x>0$ and organized in irreducible representations of the bulk conformal algebra and boundary operators, denoted as $\hat{\mathcal{O}}$, which are localized on the boundary $x=0$ and organized in irreducible representations of the preserved subalgebra.
We excite the BCFT by inserting a local boundary operator $\hat{\mathcal{O}}$ at $x=0$, $t=0$, and $\vec{y}=0$. As in section \ref{subsection:ExState}, regularization requires displacing the operator in Euclidean time $\tau=i t$ by an amount $\epsilon\in\mathbb{R}^{+}$
\begin{equation}
|\hat{\mathcal{O}}\rangle \equiv \frac{1}{\sqrt{\langle \hat{\mathcal{O}}(t=-i \epsilon, \vec{y}=0) \, \hat{\mathcal{O}}(t=i \epsilon, \vec{y}=0) \rangle}} \, \hat{\mathcal{O}}(t=i \epsilon, \vec{y}=0) |0\rangle\,.
\label{eq:BoundaryExcState}
\end{equation}
We consider $\hat{\mathcal{O}}$ to be a scalar primary boundary operator with scaling dimension $\hat{\Delta}$. The excited state is characterized by radiation reflecting from the conformal boundary, propagating outward to infinity, as depicted in figure \ref{fig:Penrose2}.

\noindent
\subsection{Constrained kinematics in correlation functions}
\label{subsection:ConstrainedKinematics}

In dimension $d>2$, as we discussed above, the twist operator is an extended operator on the replicated manifold. In this case, there are $n$ replicas of the BCFT and the relevant observable for computing the excess of entropy is a $2n$-point function of boundary local operators $\hat{\mathcal{O}}$, in the presence of two distinct defects: the conformal boundary and the twist operator
\begin{equation}
    \langle \hat{\mathcal{O}}^{\otimes n}(\tau_{1}, \vec{y}=0) \, \hat{\mathcal{O}}^{\otimes n}(\tau_{2}, \vec{y}=0) \rangle_{\sigma_n, \mathcal{B}} = 
    \frac{\langle \hat{\mathcal{O}}^{\otimes n}(\tau_{1}, \vec{y}=0) \, \hat{\mathcal{O}}^{\otimes n}(\tau_{2}, \vec{y}=0) \, \sigma_n(\tau) \, \mathcal{B} \rangle}{\langle \sigma_n(\tau) \, \mathcal{B} \rangle}\,.
    \label{eq:2point2defects}
\end{equation}
As discussed in section \ref{subsection:kinem}, this quantity, at the kinematical level, is equivalent to the two-point function of boundary operators in the presence of two defects. We then move to the discussion of this particular kinematics, with the restrictions dictated by our setup.

\subsubsection{Correlation functions of two defects: embedding space formalism}
\label{subsubsection:2defects}

We are interested in considering correlators involving two defects with different dimensions in a vacuum CFT, and we will use the embedding space formalism to analyze their kinematics \cite{Gadde:2016fbj}. We first review the method before applying it to our configuration.

In general, a $p$-dimensional conformal defect (with codimension $q$ such that $d=p+q$) in Euclidean signature preserves the symmetry subgroup $SO(p+1,1) \times SO(q)$, retaining the conformal group on the defect and transverse rotational symmetry. The embedding formalism consists in linearizing the conformal transformations by considering its action in a $(d+1,1)$-dimensional embedding space $\mathbb{R}^{d+1,1}$. Points on $\mathbb{R}^d$ are mapped to points $X$ on the projective lightcone in $\mathbb{R}^{d+1,1}$, such that $X^2=0$ and the equivalence relation $X \sim \lambda X$ holds. Specifically, we denote the lightcone coordinates in embedding space by $X^{M} = (X^{+}, X^{-}, X^{\mu})$ with $\mu=1, \ldots, d$. One way to satisfy the null condition $X^{2}=0$ is to take $X^{M} = (a, x^{2}/a, x^{\mu})$. We can also fix the $GL(1)$ redundancy by setting $a=1$, \textit{i.e.}\ $X^{+}=1$, known as the Poincaré section, thus recovering the original $d$-dimensional coordinates $x^{\mu}$ with the identification $X^{M}=(1, x^{2}, x^{\mu})$. Different projections are associated to different conformal frames in the original space.

\begin{figure}[htbp!]
    \centering
    \includegraphics[scale=0.45]{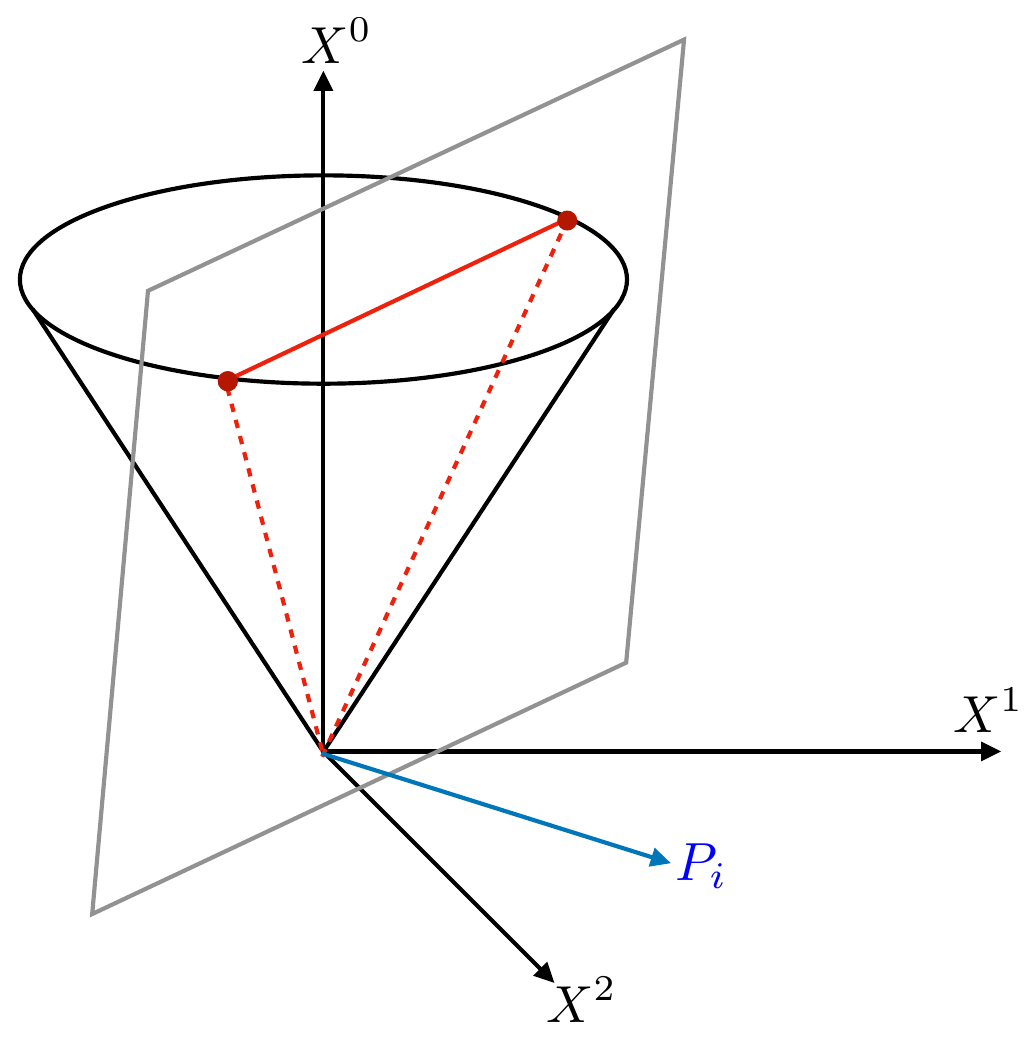}
    \caption[Embedding defect]
        {Example of a 3-dimensional embedding space $(X^{0}, X^{1}, X^{2})$. The two-dimensional hyperplane, parametrized by its orthogonal vector $P_{i}$, intersects the null cone, yielding a pair of points, corresponding to a spherical 0-dimensional defect.}
    \label{fig:EmbDef}
\end{figure}

In $\mathbb{R}^{d+1,1}$, a codimension-$q$ timelike hyperplane preserves the same subgroup as a conformal defect of codimension-$q$. The intersection between this hyperplane and the null cone yields a $p=d-q$ dimensional space, identifying the defect in the original space. Figure \ref{fig:EmbDef} illustrates this for a $0$-dimensional defect.

Thus, we identify a conformal defect in embedding space by defining a codimension-$q$ hyperplane specified by $q$ orthogonal vectors $P_i$, with $i = 1, \ldots, q$, with the defect locus given by the embedding coordinates $X^{M}$ satisfying
\begin{equation}
    X \cdot X = 0\,, \quad X \cdot P_i = 0\,,
    \label{eq:CondTransv}
\end{equation}
up to projective identification, leaving a $GL(q)$ redundancy. The metric defining the scalar product in embedding space is given by $\eta_{+-} = \eta_{-+} = -\frac{1}{2}$ and $\eta_{\mu \nu} = \delta_{\mu\nu}$ for $\mu,\nu = 1, \ldots, d$. Imposing \eqref{eq:CondTransv} on the null vector $X$ in the Poincaré section yields $q$ equations corresponding to codimension-one spheres for each of the $q$ orthogonal vectors $P_i$. For a timelike codimension-$q$ hyperplane, these spheres intersect in a codimension-$q$ sphere, forming a spherical conformal defect. As suggested in \cite{Gadde:2016fbj}, it is convenient to construct $GL(q)$-invariant expressions involving $P_i$ and the point at infinity, $\Omega=(0,1,0^{\mu})$ (absent in the Poincaré section), by choosing an orthonormal basis for $P_i$ with $P_i \cdot P_j = \delta_{ij}$, leaving an $O(q)$ gauge redundancy.

With these ingredients, we can now determine the orthonormal vectors $P_i$ associated with a spherical or flat defect in the original space. By definition, the $q$ vectors $P_i$ are orthogonal to the $(d+2-q)$-dimensional hyperplane in the embedding space $\mathbb{R}^{d+1,1}$. Fixing this hyperplane requires identifying $d+2-q$ points. Therefore, for a general codimension-$q$ defect $\mathscr{D}^{(q)}$, we select $d+2-q$ points in real space, uplift them to points on the projective lightcone $X_{A}$ ($A=1,\ldots,d+2-q$) and solve for the vectors $P_i$ such that $X_{A} \cdot P_i = 0$.

Let us first consider a planar defect of codimension $q_1$. The codimension-$q_{1}$ flat defect in real space is defined on a hyperplane spanned by $d-q_{1}$ vectors $e_{a}$ forming an orthonormal basis. We uplift them to $X_{a} = (1,1,e_{a})$ and add the origin $\tilde{X}=(1,0,0^{\mu})$ and the point at infinity $\Omega = (0,1,0^{\mu})$, thus identifying a $(d-q_{1}+2)$-hyperplane in embedding space. Solving \eqref{eq:CondTransv} gives us the intuitive result
\begin{equation}
    P_i= (0,0,e_{d-q_{1}+i})\,, \quad i = 1,\ldots,q_{1}\,.
    \label{eq:orthovecboundaryq1}
\end{equation}

Then, we consider a codimension-$q_{2}$ spherical defect of radius $R$ centered in the origin. This defect is aligned to a $d-q_{2}+1$ dimensional hyperplane, spanned by $d-q_{2}+1$ vectors $\hat{e}_{k}$ forming once again an orthonormal basis. As suggested in \cite{Gadde:2016fbj}, we can pick as the $d-q_{2}+2$ points satisfying \eqref{eq:CondTransv}, the points $\hat{X}_{j}=\left(1,R^{2},R\hat{e}_{k}\right)$, with $k=1,\ldots,d-q_{2}+1$, and the point $\hat{X}_{d-q_{2}+2}=\left( 1,R^{2},-R\hat{e}_{1}\right)$. With this choice, a convenient orthonormal basis satisfying \eqref{eq:CondTransv} is

\begin{equation}
    \hat{P}_{i}=\left( 0,0,\hat{e}_{d-q_{2}+1+i}\right)\,, \quad i=1,\ldots,q_{2}-1\,, \qquad \hat{P}_{q_{2}}=\left( \frac{1}{R},-R,0^{\mu} \right)\,.
\end{equation}

With this general setup in mind, we examine the specific case of a codimension-one flat defect (the boundary) and a codimension-two spherical defect (the twist operator). Hereafter, we will use the following coordinates in real space $(\tau,x,\vec{y})$, where $\tau$ and the $(d-2)$ coordinates $\vec{y}$ are parallel to the boundary, which sits at $x=0$. For the boundary $\mathcal{B}$, the orthogonal vector is obviously
\begin{equation}
    P^{\mathcal{B}} = (0,0,0,e_{x},\vec{0})\,. 
    \label{eq:orthovecboundary}
\end{equation}
The twist operator, instead, is a  codimension-two spherical defect. We start with a spherical defect of radius $R$ on the hyperplane $x=0$, centered at the origin,
\begin{equation}
    \sigma_n = (R \cos{\phi}, 0, \vec{\Omega} R \sin{\phi})\,,
    \label{eq:sphericaldefectbeginning}
\end{equation}
where $\vec{\Omega}\cdot \vec{\Omega}=1$ parametrizes a $(d-3)$-sphere in the $(d-2)$-dimensional hyperplane spanned by $\vec{y}$. 
Since the boundary is located at $x=0$ this is certainly not the most general configuration of the two defects. To reach it, we translate the center of the sphere by an amount $l_x$ away from  the origin in the $x$ direction, and we perform a rotation $M_{\tau x}(\theta)$ in the $(\tau,x)$ directions by an angle $\theta$ (rotations in other planes involving the $\vec{y}$ directions are equivalent due to the symmetry preserved by the boundary). 

Afterwards, we uplift $d$ points on the spherical defect to the Poincaré section, choosing for simplicity the points identified by

\begin{equation}
\begin{split}
& \hat{X}_{1}=(1,R^{2},R\cos{\theta}\hat{e}_{\tau}+(l_{x}-R\sin{\theta})\hat{e}_{x})\,,\\
& \hat{X}_{k}=(1,R^{2}, l_{x}\hat{e}_{x}+R\hat{e}_{k})\,,\quad k=2,\ldots,d-1\,,\\
&\hat{X}_{d}=(1,R^{2},l_{x}\hat{e}_{x}-R\hat{e}_{y_{1}})\,.
\end{split}
\end{equation}

\noindent
Finally, applying \eqref{eq:CondTransv}, we find two orthonormal vectors $\hat{P}_i^{\sigma_{n}}$ that specify the spherical defect

\begin{equation}
\begin{split}
    & \hat{P}_{1}^{\sigma_{n}} = \left(\frac{1}{R}, -R, 0,0,\vec{0}\right)\,, \\
    & \hat{P}_{2}^{\sigma_{n}} = \frac{1}{\sqrt{-l_{x}^{2} + R^{2} \sec{\theta}^{2}}}\left( \frac{l_{x}}{R}, l_{x} R,  R \tan{\theta}, R, \vec{0}\right)\,.
\end{split}
\label{eq:2orthovecsphere}
\end{equation}

\noindent
These vectors satisfy \eqref{eq:CondTransv} and form a two-dimensional orthonormal basis.

We now have all the necessary ingredients to study the correlation function of two conformal defects. A word of caution is needed. Correlation functions of defects generically involve UV divergences. The famous area term in the computation of the entanglement entropy, for instance, can be seen as the UV-divergent contribution in the expectation value of the twist operator $\sigma_n$. These power-law divergent contributions are generically not universal (they depend on the regulator), but one can always isolate a universal term, either a logarithm or a finite part, depending on the dimensions of the defects and of the bulk. This will not be an issue for our observable \eqref{eq:2point2defects} because the normalization makes it UV finite. For the following paragraph, though, we are going to analyze the kinematics of two defects without local operator insertions and, therefore, our considerations apply to the universal part of the result.

In general, as found in \cite{Gadde:2016fbj}, the correlation function of two defects with different codimensions 
\begin{equation}
\langle\mathscr{D}^{(q_{1})}(P_i)\mathscr{D}^{(q_{2})}(\hat{P}_j)\rangle\,,
\end{equation}
depends on a set of conformal cross-ratios determined by the codimensions $q_{1}, q_{2}$ of the defects and the spacetime dimension $d$. The exact number is
\begin{equation}
    \#\text{cross-ratios}=\mathrm{min}(q_{1}, q_{2}, d+2-q_{1}, d+2-q_{2})\,.
\end{equation}
These cross-ratios are specific combinations of the orthogonal vectors $P_i$ and $\hat{P}_j$ that remain invariant under conformal transformations. A possible conformally invariant cross-ratio is given by
\begin{equation}
    \eta=\mathrm{Tr}M\,, \quad M_{ij}=(P_i\cdot \hat{P}_k)(\hat{P}_k\cdot P_j)\,,
    \label{eq:crossratio2defects}
\end{equation}
which is also manifestly gauge invariant under the $O(q_{1})$ and $O(q_{2})$ redundancies. In general, we can build more cross-ratios as

\begin{equation}
    \eta_{n}\equiv\mathrm{Tr}M^{n}\,, \quad n=1,\ldots,\#\text{cross-ratios}\,.
\end{equation}
For our case, the correlation function involving a flat boundary $q_{1}=1$ and a codimension-two spherical defect $q_{2}=2$ will depend on a single cross-ratio
\begin{equation}
    \langle\mathscr{D}^{(1)}(P^{\mathcal{B}})\mathscr{D}^{(2)}(\hat{P}_i^{\sigma_{n}})\rangle=F(\eta)\,.
    \label{eq:2defectfunction}
\end{equation}
Using these results, along with \eqref{eq:orthovecboundary}, \eqref{eq:2orthovecsphere} and \eqref{eq:crossratio2defects}, we find that
\begin{equation}
    \eta^{-1}=\sec{\theta}^{2}-\frac{l_{x}^{2}}{R^{2}}\,,
    \label{eq:crossratioetadefects}
\end{equation}
which depends only on the dimensionless ratio between the orthogonal distance $l_{x}$ of the sphere center from the boundary and the radius $R$ of the sphere itself, as well as on the angle $\theta$, which relates to the orientation of the entangling surface in the real spacetime relative to the conformal boundary.

Our configuration, however, is not general. We consider a sphere centered on the boundary and orthogonal to the $\tau$ direction. This means that $\theta=\frac{\pi}{2}$ and $l_{x}=0$ leading to
\begin{equation}
    \eta\underset{\theta\rightarrow\frac{\pi}{2}}{\longrightarrow}0\,.
\end{equation}
This calculation confirms that in our constrained kinematics, the expression in the denominator of \eqref{eq:2point2defects} does not depend on any cross-ratio.

\subsubsection{Residual symmetry}
\label{subsubsection:PresTransf}

\begin{figure}[h!]
\centering
\begin{subfigure}{.45\textwidth}
\centering
\includegraphics[width=0.9\linewidth]{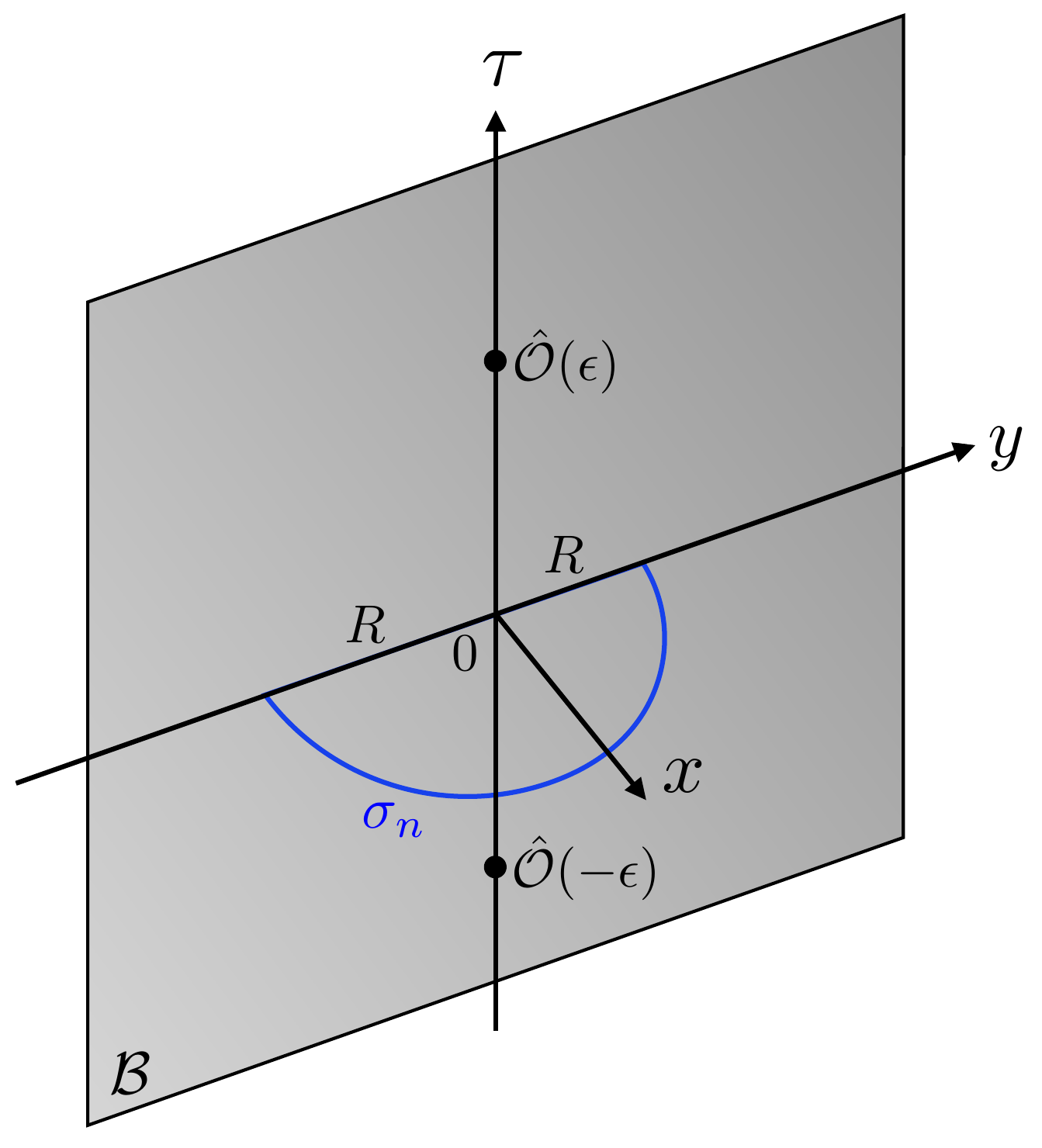}
\caption{}
\label{fig:setupboundarytwista}
\end{subfigure}%
\hspace{0.02\textwidth} 
\begin{subfigure}{.45\textwidth}
\centering
\includegraphics[width=0.9\linewidth]{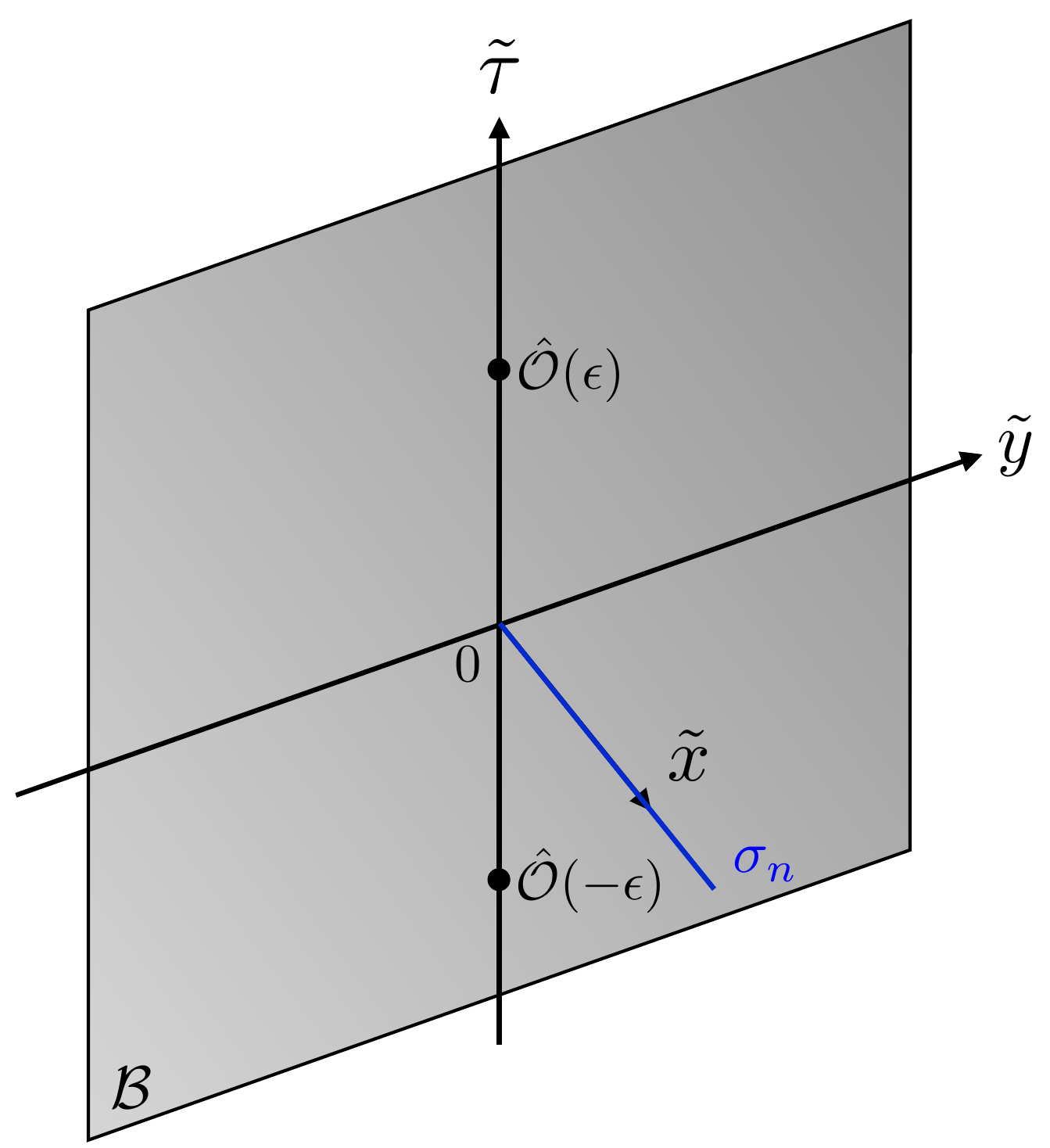}
\caption{}
\label{fig:setupboundarytwistb}
\end{subfigure}
\caption{(a) Illustration of the configuration in $d=3$ dimensions with a conformal boundary and a spherical defect. (b) The entangling surface is mapped via \eqref{eq:mappingspheretohalplane} to a half-line in the positive $\tilde{x}$ direction at $\tilde{y}=\tilde{\tau}=0$.}
\label{fig:setupboundarytwist}
\end{figure}

We have established that the correlator of the two defects in our specific configuration does not produce any conformal invariant quantity. Furthermore, the symmetry of the configuration is such that there is a residual symmetry. In other words, a subgroup of the original conformal group is preserved. It is easy to see this by realizing that the two defects intersect on a codimension-three sphere on the boundary (in the 3d case depicted in figure \ref{fig:setupboundarytwista} the intersection is just a pair of points) and we can use the conformal transformations preserved by the boundary to map this hypersphere to a codimension-three plane. 
Therefore, our configuration is conformally equivalent to a codimension-one and a codimension-two planar defects intersecting on a codimension-three plane. This particular configuration clearly preserves conformal transformations in the $(d-3)$ common directions and rotations around the codimension-two defect. Let us make this explicit.

Consider Euclidean space $\mathbb{R}^{d}$ with coordinates $x^{\mu}=(\tau, x, y_{1},\ldots, y_{d-2})$. The conformal boundary lies at $x=0$, while the entangling surface is a half codimension-two hypersphere located at $\tau=0$ on the $(d-1)$-dimensional hyperplane spanned by $x$ and $y_{1},\ldots,y_{d-2}$. The mapping to two planar defects illustrated in figure \ref{fig:setupboundarytwist} is achieved through the following transformation:
\begin{equation}
\begin{split}
    & \tilde{\tau} = \frac{R^{2} \tau}{\tau^{2} + x^{2} + y^{\hat{a}}y_{\hat{a}}+ (R - y_{d-2})^{2}}\,,  \\
    & \tilde{x} = \frac{R^{2} x}{\tau^{2} + x^{2} + y^{\hat{a}}y_{\hat{a}} + (R - y_{d-2})^{2}}\,, \\
    & \tilde{y}_{\hat{a}}=\frac{R^{2} y_{\hat{a}}}{\tau^{2} + x^{2} + y^{\hat{a}}y_{\hat{a}} + (R - y_{d-2})^{2}}\,, \quad \hat{a}=1,\ldots,d-3\,, \\
    & \tilde{y}_{d-2} = -\frac{R}{2} + \frac{R^{2} (R - y_{d-2})}{\tau^{2} + x^{2} + y^{\hat{a}}y_{\hat{a}} + (R - y_{d-2})^{2}}\,.
\end{split}
\label{eq:mappingspheretohalplane}
\end{equation}
This transformation places the codimension-two halfplane at $\tilde{\tau}=0$ and $\tilde{y}_{d-2}=0$.

As previously mentioned, this new setup preserves a conformal subgroup $\tilde{SO}(d-2,1)$, which is localized on the codimension-three intersection of the two defects, spanned by $\tilde{y}_{\hat{a}}$, with $\hat{a}=1,\ldots,d-3$. Additionally, this configuration is invariant under the rotation $\tilde{M}_{\tilde{\tau} \tilde{y}_{d-2}}$ on the plane orthogonal to the codimension-two defect. Thus, the symmetry group in this new configuration is $\tilde{SO}(d-2,1)\times \tilde{SO}(2)$. The rotation $\tilde{M}_{\tilde{\tau}\tilde{y}_{d-2}}$, associated with the $\tilde{SO}(2)$ symmetry group, commutes with the generators of the conformal subalgebra $\tilde{so}(d-2,1)$. Using the mapping \eqref{eq:mappingspheretohalplane}, these generators can be expressed in terms of generators in the original coordinates as
\begin{equation}
\begin{split}
    & \tilde{M}_{\tilde{\tau} \tilde{y}_{d-2}}=\frac{1}{2R}\left( R^{2} P_{\tau} + K_{\tau}\right)\,,\\
    & \tilde{D} = \frac{1}{2R}\left( R^{2} P_{y_{d-2}} - K_{y_{d-2}} \right)\,,\\
    & \tilde{P}_{\tilde{y}_{\hat{a}}}=\frac{1}{R^{2}}\left( R^{2}P_{y_{\hat{a}}}-K_{y_{\hat{a}}}-2R M_{y_{\hat{a}}y_{d-2}}\right)\,, \quad \hat{a}=1,\ldots,d-3\,,\\
    & \tilde{K}_{\tilde{y}_{\hat{a}}}=\frac{1}{4}\left( -R^{2}P_{y_{\hat{a}}}+K_{y_{\hat{a}}}-2R M_{y_{\hat{a}}y_{d-2}}\right)\,, \quad \hat{a}=1,\ldots,d-3\,,\\
    & \tilde{M}_{\tilde{y}_{\hat{a}}\tilde{y}_{\hat{b}}}=M_{y_{\hat{a}}y_{\hat{b}}}\,, \quad \hat{a},\hat{b}=1,\ldots,d-3\,.
\end{split}
\end{equation}

We recognize that the generators $M_{y_{\hat{a}}y_{d-2}}$ and $M_{y_{\hat{a}}y_{\hat{b}}}$, with $\hat{a},\hat{b}=1,\ldots,d-3$, forming the $SO(d-2)$ group of rotations, leave the codimension-two sphere invariant. This group is accompanied by $(d-1)$ generators written as combinations of translations and SCTs along the conformal boundary, specifically $R^{2}P_{y_{m}}-K_{y_{m}}$, with $m=1,\ldots,d-2$, and $R^{2}P_{\tau}+K_{\tau}$. These generators are mutually commuting, but they don't commute with the aforementioned rotations.

\subsubsection{Inserting boundary operators}

To understand the kinematics of \eqref{eq:2point2defects} we need to consider the correlator of two local operators and two defects in the specific configuration depicted in figure \ref{fig:setupboundarytwista}
\begin{equation}
    \langle\hat{\mathcal{O}}(\tau_{1}, \vec{y}=0)\hat{\mathcal{O}}(\tau_{2}, \vec{y}=0)\sigma_n(\tau)\mathcal{B}\rangle\,.
    \label{eq:corrfunct2defects}
\end{equation}
We can use conformal invariance to constrain the kinematics of the correlator and count how many coordinates are needed to specify the positions of the operators. This will correspond to the number of conformal cross ratios. Since both operators live on the boundary, we can first use the $(d-1)$ generators $R^{2}P_{y_{m}}-K_{y_{m}}$ and $R^{2}P_{\tau}+K_{\tau}$ to fix one of the two operators at the origin. A consequence of this is that the one-point function of a local operator with two defects in our specific configuration is completely fixed by conformal invariance (see appendix \ref{onepoint} for the explicit expression). The second local boundary operator is specified by its $(d-1)$ coordinates $(\tau,y_{1},\ldots,y_{d-2})$. Using the rotations $M_{y_{\hat{a}}{y_{d-2}}}$, with $\hat{a}=1,\ldots,d-3$, the operator can be placed in the plane $y_{1}=\ldots=y_{d-3}=0$ and it is specified by two coordinates $(\tau,y_{d-2})$. The other rotations $M_{y_{\hat{a}}y_{\hat{b}}}$, generating a $SO(d-3)\subset SO(d-2)$ do not act on these two coordinates, proving that the correlator \eqref{eq:corrfunct2defects} is fixed up to two conformal cross ratios. As a consistency check, we notice that, using the map in figure \ref{fig:setupboundarytwista}, the correlator can be mapped to a two-point function of boundary operators in the presence of another planar defect. The symmetry of this configuration is $SO(d-2,1)\times SO(2)$, \textit{i.e.}\ the same as a codimension-two conformal defect in $(d-1)$ dimensions. Also for that case, the correlators of two bulk operators with the defect is determined up to two cross-ratios.

Let us consider now the simple case of $d=3$ (for higher dimensions, it is possible to use additional preserved rotations to make the problem effectively three-dimensional).
One way to find the expressions of these cross-ratios is to start with complex coordinates on the boundary  and insert one operator at the origin and the other one at a generic point $(\tau,y)$ with $z=\tau+i y$ and $\bar z=\tau-i y$. Performing a conformal transformation, we bring the operator at the origin to a generic point and rewriting the parameters of the transformation in terms of the positions of the two operators $(\tau_1,y_1)$ and $(\tau_2,y_2)$ we get the two cross ratios
\begin{equation}
\begin{split}
    & z = \frac{R \left( -i y_1 + i y_2 - \tau_1 + \tau_2 \right)}{R^2 - (y_1 - i \tau_1)(y_2 - i \tau_2)}\,, \\
    & \bar{z} = \frac{R \left( i y_1 - i y_2 - \tau_1 + \tau_2 \right)}{R^2 - (y_1 + i \tau_1)(y_2 + i \tau_2)}\,.
\end{split}
\label{eq:zandbarz}
\end{equation}

For us, this is not the end of the story because the two boundary operators are not placed at a generic point, but they sit at $y_1=y_2=0$, $\tau_1 = \tau - \epsilon$ and $\tau_2 = \tau + \epsilon$. In this constrained kinematics the two cross-ratios are equal
\begin{equation}
    z = \bar{z} = \frac{2R \epsilon}{R^2 + \tau^2 - \epsilon^2}\,.
\label{eq:zconstrained}
\end{equation}
Therefore, also for the boundary case, our problem is effectively described by a single cross-ratio. In particular, we can use the same cross-ratio that we used in the homogeneous case \eqref{eq:xicrossratiofinale} through the mapping
\begin{equation}
    \xi =\frac{z^2}{1 + z^2} = \frac{4R^2 \epsilon^2}{\left( R^2 + (\tau - \epsilon)^2 \right) \left( R^2 + (\tau + \epsilon)^2 \right)}\,.
\label{eq:xiconstrained}
\end{equation}
The behavior of $\xi$ is the same as in section \ref{subsection:anal} and its time evolution is shown in figure \ref{fig:crossratioevolution}. 
In particular, the early- and late-time behavior are still controlled by the limit $\xi\to 0$.

\subsection{OPE at early and late times}
\label{subsection:OPE2defects}

The $\xi \to 0$ limit of the correlator \eqref{eq:2point2defects}, relevant for the early- and late-time behavior, is controlled by an OPE. Following the same strategy outlined in section \ref{subsection:OPE}, we extract the power-law behavior of the correlator at small $\xi$. Our main observable is still the excess of entanglement entropy, which is obtained in the $n\to 1$ limit of the excess of Rényi entropy
\begin{equation}
    \Delta S_{A}^{(n)} = \frac{1}{1-n} \log{\left( \frac{\langle \sigma_{n}(R)\mathcal{B} \hat{\mathcal{O}}^{\otimes n}(x^{a}_{1}) \hat{\mathcal{O}}^{\otimes n}(x^{a}_{2}) \rangle}{\langle \sigma_{n}(R) \mathcal{B} \rangle \langle \hat{\mathcal{O}}(x^{a}_{1}) \hat{\mathcal{O}}(x^{a}_{2}) \rangle^{n}} \right)}\,,
    \label{eq:deltaSn2defectsBCFT}
\end{equation} 
where the local operators $\hat{\mathcal{O}}$ sit on the boundary parametrized by the coordinates $x^a$. This expression involves a $2n$-boundary-point function with two defects: one is the boundary itself $\mathcal{B}$, and the other is the twist operator $\sigma_n(R)$. 

For $\xi \to 0$, as in the case without the boundary, we can perform the OPE of the two scalar operators (in this case though they are boundary operators). The leading contribution is given by the lightest exchanged local boundary operator, \textit{i.e.}\ the identity, which leads to a disconnected contribution canceling the contribution of the vacuum
\begin{equation}
    \frac{\langle \sigma_n(R)\mathcal{B} \hat{\mathcal{O}}^{\otimes n}(x^{a}_{1}) \hat{\mathcal{O}}^{\otimes n}(x^{a}_{2}) \rangle}{\langle \sigma_n(R) \mathcal{B} \rangle \langle \hat{\mathcal{O}}(x^{a}_{1}) \hat{\mathcal{O}}(x^{a}_{2}) \rangle^{n}} \sim \frac{\langle \sigma_n(R) \mathcal{B} \rangle \langle \hat{\mathcal{O}}(x^{a}_{1}) \hat{\mathcal{O}}(x^{a}_{2}) \rangle^{n}}{\langle \sigma_n(R) \mathcal{B} \rangle \langle \hat{\mathcal{O}}(x^{a}_{1}) \hat{\mathcal{O}}(x^{a}_{2}) \rangle^{n}} = 1\,,
\end{equation}
giving no contribution to the excess of entropy $\Delta S_{A}^{(n)}|_{\textit{id}} = 0$.

Moving to the lightest operator above the identity, contrary to the homogeneous example, we have two boundary operators and there is no conserved stress tensor on the boundary. Instead, we assume that the lightest exchanged single-copy operator is the displacement operator $\mathcal{D}$. This is a protected scalar operator of dimension $d$, that is associated to the broken translation in the orthogonal direction by the Ward identity $\partial_{\mu}\mathcal{T}^{\mu m} = \mathcal{D} \delta(x_{\perp})$, where $m$ labels the normal direction to the boundary and $\delta(x_{\perp})$ localizes the r.h.s.\ at the boundary. For the case of a conformal boundary, the orthogonal components of the stress tensor have a non-singular boundary OPE and the displacement can also be defined as $\mathcal{T}^{m m}|_{\text{bry}} = \mathcal{D}$. We recall that this lightest operator is assumed to be a single-copy boundary operator symmetrized over the $n$-copies of the CFT,
\begin{equation}
    \mathcal{D} = \sum_{j=0}^{n-1} \mathbb{1}^{\otimes j} \otimes D \otimes \mathbb{1}^{\otimes(n-j-1)}\,,
    \label{eq:displacementcopies}
\end{equation}
where the displacement operator is inserted at the same boundary point in each replica. 

Performing the OPE of the two boundary operators, the numerator of the logarithm in \eqref{eq:deltaSn2defectsBCFT} reduces to 
\begin{equation}
   \left.\langle \sigma_n(R)\mathcal{B} \hat{\mathcal{O}}^{\otimes n}(x_1^a) \hat{\mathcal{O}}^{\otimes n}(x_2^a) \rangle\right|_{\text{disp}} \overset{\xi \to 0}{\sim} \frac{n\, c_{\hat{\mathcal{O}} \hat{\mathcal{O}} D}}{C_D} |x_{12}^a|^{d-2 \hat{\Delta}} \langle \sigma_n(R) \mathcal{B} D(x_2^a) \rangle \langle \hat{\mathcal{O}}(x_1^a) \hat{\mathcal{O}}(x_2^a) \rangle^{n-1}\,,
   \label{eq:OPE2boundaryoperators}
\end{equation}
where we factorized a two-point function over the $(n-1)$-sheets, where no insertions are present. The OPE coefficient $c_{\hat{\mathcal{O}} \hat{\mathcal{O}} D}$ is divided by the Zamolodchikov norm of the displacement operator $C_D$, accounting for the fact that the normalization of the displacement is not arbitrary, but it is fixed by Ward identities for the broken translation.
Dividing by the vacuum contribution we get
\begin{equation}
    \frac{\langle \sigma_n(R) \mathcal{B} \hat{\mathcal{O}}^{\otimes n}(x_1^a) \hat{\mathcal{O}}^{\otimes n}(x_2^a) \rangle}{\langle \sigma_n(R) \mathcal{B} \rangle \langle \hat{\mathcal{O}}(x_1^a) \hat{\mathcal{O}}(x_2^a) \rangle^n} \overset{\xi \to 0}{\sim} 1 + \frac{n\, c_{\hat{\mathcal{O}} \hat{\mathcal{O}} D}|x_{12}^a|^{d}}{C_{D} } \frac{\langle \sigma_n(R) \mathcal{B} D(x_2^a) \rangle}{\langle \sigma_n(R) \mathcal{B} \rangle}\,,
    \label{eq:OPEtwist1boundary}
\end{equation}
where we exploited the trivial two-point function $\langle \hat{\mathcal{O}}(x_1^a) \hat{\mathcal{O}}(x_2^a) \rangle=|x_{12}^{a}|^{-2\hat{\Delta}}$\,.

The last term we need to consider is the one-point function of the single-copy displacement operator \( D \) in the presence of the boundary and the spherical defect
\begin{equation}
    \langle D(x^{a}) \rangle_{\sigma_{n}(R), \mathcal{B}} = \frac{\langle \sigma_{n}(R) \mathcal{B} D(x^{a}) \rangle}{\langle \sigma_{n}(R)\mathcal{B} \rangle}\,.
\end{equation}
As discussed in section \ref{subsubsection:PresTransf}, this one-point function does not depend on any cross-ratios. In particular, we show in appendix \ref{onepoint} that it takes the form \footnote{The factor of $n$ was introduced analogously to \eqref{eq:1pointtwistsphericalFINAL} to canonically normalize the operator symmetrized over the replicas.}

\begin{equation}
\begin{split}
     \langle D(x^{a})\rangle_{\sigma_{n}(R), \mathcal{B}} & =\frac{a_{D,n}}{n}\frac{(2R)^{d}}{\left[\left( (R-y_{d-2})^{2}+y^{\hat{a}}y_{\hat{a}}+\tau^{2}\right)\left( (R+y_{d-2})^{2}+y^{\hat{a}}y_{\hat{a}}+\tau^{2}\right)\right]^{\frac{d}{2}}}\\
     & \equiv \frac{a_{D,n}}{n}\frac{(2R)^{d}}{l_{min}^{d}l_{max}^{d}}\,,
     \end{split}
     \label{eq:1point2defects}
\end{equation}
where $\hat{a}=1,\ldots,d-3$. Here, \( x^{a} = (\tau, y^{\hat{a}}, y_{d-2}) \) represent the parallel coordinates labeling the position of the local operator \( D \), while \( l_{\text{max}} \) and \( l_{\text{min}} \) denote the maximum and minimum distances, respectively, between the displacement operator insertion and the codimension-three defect generated by the intersection of the entangling surface with the boundary.

Inserting \eqref{eq:1point2defects} in \eqref{eq:OPEtwist1boundary} we get
\begin{equation}
\begin{split}
   \frac{\langle \sigma_n(R) \mathcal{B} \hat{\mathcal{O}}^{\otimes n}(x_1^a) \hat{\mathcal{O}}^{\otimes n}(x_2^a) \rangle}{\langle\sigma_{n}(R)\mathcal{B}\rangle\langle\hat{\mathcal{O}}(x_{1}^{a})\hat{\mathcal{O}}(x_{2}^{a})\rangle^{n}} &\overset{\xi \to 0}{\sim} 1+ 2^{d}\frac{a_{D,n}c_{\hat{\mathcal{O}} \hat{\mathcal{O}}D}}{C_{D}}\left(\frac{2R\epsilon}{R^{2}+(\tau+\epsilon)^{2}}\right)^{d}\,,
   \label{eq:correlationfunctionBoundaryPartial1OPE}
\end{split}
\end{equation}
where we used the constrained kinematics \( x_{1}^{a}=(\tau-\epsilon,\vec{0}) \) and \( x_{2}^{a}=(\tau+\epsilon,\vec{0}) \).
In the OPE limit \( \xi\rightarrow0 \), corresponding to \( \epsilon, R \rightarrow 0 \), the kinematic term in \eqref{eq:correlationfunctionBoundaryPartial1OPE} reduces to
\begin{equation}
    \frac{2R\epsilon}{R^{2}+(\tau+\epsilon)^{2}}=\frac{2R\epsilon}{\tau^{2}}+\mathcal{O}(\epsilon^{2})\,,
\end{equation}
and we can reconstruct the power of the cross ratio $\xi$  \eqref{eq:xicrossratiofinale} at leading order for $\xi\to0$
\begin{equation}
    \frac{\langle \sigma_n(R) \mathcal{B} \hat{\mathcal{O}}^{\otimes n}(x_1^a) \hat{\mathcal{O}}^{\otimes n}(x_2^a) \rangle}{\langle\sigma_{n}(R)\mathcal{B}\rangle\langle\hat{\mathcal{O}}(x_{1}^{a})\hat{\mathcal{O}}(x_{2}^{a})\rangle^{n}} \overset{\xi \to 0}{\sim} 1+2^{d} \frac{a_{D,n}c_{\hat{\mathcal{O}} \hat{\mathcal{O}}D}}{C_{D}}\xi^{\frac{d}{2}}\,.
\end{equation}
Substituting this expression into \eqref{eq:deltaSn2defectsBCFT} yields
\begin{equation}
    \Delta S_{A}^{(n)} \overset{\xi\rightarrow0}{\sim} \frac{2^{d}}{1-n} \frac{a_{D,n} \, c_{\hat{\mathcal{O}}\hat{\mathcal{O}}D}}{C_{D}} \xi^{\frac{d}{2}}.
    \label{eq:deltaSn2defectsfinal3}
\end{equation}
Comparing this result with \eqref{eq:partialDeltaS} we notice that the late-time behavior with and without the boundary are characterized by the same power of the cross-ratio under the assumption that the stress tensor and the displacement operator are the lightest exchanged operators in the two cases.  What is different is that, while for the case without the boundary we could fix the coefficient $c_{\mathcal{O}\mathcal{O}T}$ in terms of the dimension $\Delta$, for the boundary case in $d>2$ there is no Ward identity, to our knowledge, that could fix $c_{\hat{\mathcal{O}}\hat{\mathcal{O}}D}$ in terms of other defect CFT data.

In the $n\to 1$ limit, we can relate the coefficient $a_{D,n}$ with the two-point function $C_D$. A full derivation is given in appendix \ref{sec:detailsnto1}. The final result is
\begin{equation}
     \partial_n a_{D,n} \Big|_{n=1} = - \frac{\pi^{(d+1)/2}}{2^{d+1} (d-1) \Gamma \left( \frac{d+3}{2} \right)} C_D \,.
    \label{eq:bsigmad}
\end{equation}
Expanding \eqref{eq:deltaSn2defectsfinal3} close to $n\to1$, the excess of the entanglement entropy reduces to
\begin{equation}
    \Delta S_{A}^{EE} \overset{\xi\rightarrow0}{\sim}  \frac{\pi^{\frac{d+1}{2}}}{2 (d-1) \Gamma \left( \frac{d+3}{2} \right)}c_{\hat{\mathcal{O}}\hat{\mathcal{O}}D} \, \xi^{\frac{d}{2}}\,.
    \label{eq:EEFinalBoundary}
\end{equation}
Interestingly, also in this case the result perfectly saturates the bound on the excess of entropy in presence of a boundary found in \cite{Bianchi:2022ulu}. Therefore, as discussed below \eqref{boundOPElatet}, after subtracting the contribution of the displacement operator, the lightest operator in the OPE must give a negative contribution to the excess of the entropy. This extends the two-dimensional result obtained in \cite{Bianchi:2022ulu}.

As shown in appendix \ref{section:2OPEBCFT}, we can alternatively compute the excess of entanglement entropy in the early and late time limits by performing a boundary OPE both for the boundary scalar multiple-copy operators \( \hat{\mathcal{O}}^{\otimes n} \) and for the twist operator $\sigma_{n}$.

\section{Holographic Study of Local Quenches}
\label{section:LocalQuench}

In this section, we construct the gravity dual of the setup introduced in section \ref{section:CFTsetup} and compute the holographic entanglement entropy (HEE) using the Ryu-Takayanagi prescription \cite{Ryu:2006bv,Ryu:2006ef}, which involves finding the area of the minimal surface anchored to the entangling surface at the conformal boundary of the spacetime.
Typically, for time-dependent states, we would need to consider more general and covariant prescriptions \cite{Hubeny:2007xt,Wall:2012uf}. On the other hand, analogously to the CFT calculation discussed in the sections above, it is possible to map the time-dependent problem to a static one, allowing us to employ the simpler RT formula.

In what follows, we obtain both  numerical and perturbative HEE results in $\mathrm{AdS}_{4}$ that we will compare to those obtained within the $\mathrm{CFT}_{3}$ framework.

\subsection{Holographic dictionary for a local quench}
\label{subsection:HeavyOperators}

A possible holographic description of a local quench has been studied in \cite{Nozaki:2013wia} where the authors followed an argument given in \cite{Horowitz:1999gf}. The idea is to consider a falling massive particle in AdS, whose metric in Poincaré coordinates reads
\begin{equation}
    \mathrm{d}s^{2}_{\mathrm{Poincar\acute{e}}} = \frac{L^{2}}{z^{2}} \left( -\mathrm{d}t^{2} + \mathrm{d}z^{2} + \mathrm{d}x^{2} + \mathrm{d}y^{2} \right)\,,
    \label{eq:PoincareVacuum4}
\end{equation}
where the conformal boundary at $z=0$ corresponds to the Minkowski spacetime in which the CFT is defined. 

In this bulk spacetime, the massive particle follows a timelike geodesic with trajectory \( z(t) = \sqrt{t^{2} + \epsilon^{2}} \). For small $\epsilon$, the particle is localized near the conformal boundary at \( t = 0 \) within a region of size $\epsilon$. In the CFT, this corresponds to preparing a state in the CFT at $t=0$ with a localized excitation of small radius $\epsilon$, as discussed in section \ref{subsection:ExState}. The falling massive particle generically induces a localized backreaction in the metric, as described in \cite{Horowitz:1999gf}. Under time evolution, this backreaction propagates towards the Poincaré horizon and spreads throughout spacetime.

Studying the backreaction is in general a difficult task that can be simplified if we focus on \textit{heavy} local excitations.
For this reason, we consider the state created by a heavy local operator $\mathcal{O}$ in a three-dimensional CFT, as described in \eqref{eq:ExcitedState}. The operator $\mathcal{O}$ has a conformal dimension $\Delta$ of the same order as the central charge $C_T$\footnote{In $d=2$ the Virasoro central charge $c$ quantifies the number of degrees of freedom of the theory. In $d=3$, such a quantity has been proposed to be the sphere free-energy $F$ \cite{Klebanov:2011gs}. However, for CFTs with an Einstein gravity holographic dual, $F$ is proportional to the stress-tensor central charge $C_T = 64/\pi^2 F$ \cite{Chester:2014fya}. Thus, $C_T$ can be used interchangeably with $F$ in this context.}, and we work in the large-$C_T$ limit using a semiclassical approximation. 

To construct the backreacted metric, it is convenient to consider the empty $\mathrm{AdS}_{4}$ in global coordinates

\begin{equation}
    \mathrm{d}s^{2} = -\left( r^{2} + L^{2} \right) \mathrm{d}\tau^{2} + \frac{L^{2}}{r^{2} + L^{2}} \mathrm{d}r^{2} + r^{2} \mathrm{d}\theta^{2} + r^{2} \sin^{2}\theta \, \mathrm{d}\phi^{2}\,,
    \label{eq:MetricEmptyAdS}
\end{equation}
where the conformal boundary is now located at $r \rightarrow +\infty$. In this background metric, we consider a static massive particle sitting at \( r = 0 \). The following coordinate transformation between global and Poincaré $\mathrm{AdS}$ patches

\begin{equation}
    \begin{aligned}
        & \sqrt{L^{2} + r^{2}} \cos{\tau} = \frac{\alpha L^{2} + \left( z^{2} + x^{2} + y^{2} - t^{2} \right) / \alpha}{2z}\,, \\
       & \sqrt{L^{2} + r^{2}} \sin{\tau} = \frac{Lt}{z}\,, \\
        & r \cos{\theta} = \frac{\alpha L^{2} + \left( -z^{2} - x^{2} - y^{2} + t^{2} \right) / \alpha}{2z}\,, \\
        & r \sin{\theta} \cos{\phi} = \frac{Lx}{z}\,, \\
        & r \sin{\theta} \sin{\phi} = \frac{Ly}{z}\,,
    \end{aligned}
    \label{eq:CoordinateTransformation}
\end{equation}
maps the static solution \( r=0 \) in global coordinates to \( z^{2} - t^{2} = L^{2} \alpha^{2} \) in Poincaré $\mathrm{AdS}_{4}$, where $\alpha$ is a positive boost parameter needed to map the $r=0$ trajectory to one that travels arbitrarily close to the boundary. Identifying \( \epsilon = \alpha L \), the trajectory of the massive particle in empty Poincaré corresponds to the static geodesic \( r=0 \) in global coordinates.

For sufficiently heavy states, the bulk geometry becomes equivalent to a black hole sitting at the center of $\mathrm{AdS}_{4}$, described by

\begin{equation}
    \mathrm{d}s^{2} = -L^2 f(r)  \mathrm{d}\tau^{2} + \frac{1}{f(r)} \mathrm{d}r^{2} + r^{2} \mathrm{d}\theta^{2} + r^{2} \sin^{2}\theta \, \mathrm{d}\phi^{2}\,,
    \label{eq:MetricBHAdS}
\end{equation}
with
\begin{equation}
\label{eq:f(r)}
    f(r) \equiv   1+\frac{r^{2}}{L^2}  - \frac{2 G_N M}{r}\, ,
\end{equation}
where $M$ is the mass of the black hole and $L$ is the $\mathrm{AdS}_{4}$ curvature radius.

\subsubsection{HEE for a local quench: a numerical study}
\label{subsection:HEEQuench}

Having discussed the construction of the gravity dual for the excited state, we now delve into the computation of holographic entanglement entropy (HEE) in the $\mathrm{AdS}_{4}/\mathrm{CFT}_{3}$ framework. The minimal surface, anchored to the entangling surface, is a codimension-two object. Specifically, in our setup, it is a two-dimensional surface with a boundary defined as a circle of fixed radius on the conformal boundary of AdS. In Poincaré coordinates, the conformal boundary is Minkowski and we have a circle of radius $R$ at constant Poincaré time $t$ as shown in figure \ref{fig:Penrose}. Mapping this circle to global coordinates using \eqref{eq:CoordinateTransformation} we find another circle at constant global time $\tau$ and describing a latitude of the sphere $S^2$, parametrized by $\theta$ and $\phi$ (in this case the conformal boundary is a cylinder $S^2\times \mathbb{R}$). The position of this latitude will be denoted as $\theta_\infty$. The minimal surface anchored to the circle is a parametrized by a profile $r(\theta,\phi)$, but rotational invariance kills the dependence on $\phi$ leaving us with the problem of determining the function $r=r(\theta)$ with the boundary condition $r(\theta_\infty)\rightarrow +\infty$.  

This short argument already shows that the area is a function of a single kinematical variable, \textit{i.e.}\  the boundary position of the circle $\theta_{\infty}$. This is the holographic counterpart of the CFT result that our main observable depends on a single conformal cross ratio. In particular, using \eqref{eq:CoordinateTransformation} one can show that
\begin{equation}
    \tan{\theta_\infty} = \frac{2L R \alpha}{\alpha^2 L^2 - R^2 + t^2}\,,
    \label{eq:thetainf}
\end{equation}
and remembering that $\epsilon=\alpha L$ it is easy to relate $\theta_{\infty}$ to $\xi$ defined in \eqref{eq:xicrossratiofinale}
\begin{equation}
    \xi = \sin^{2}{\theta_{\infty}}\,.
    \label{eq:crossratio2}
\end{equation}
Therefore the time evolution of the entropy will be mapped to an evolution in $\theta_{\infty}$ from $0$ to $\pi$. 






Our choice of entangling region together with the spacetime metric \eqref{eq:MetricBHAdS} implies the minimization of the area functional

\begin{equation}
    \mathrm{A} = 2\pi \int_{0}^{\theta_{\infty}} \mathrm{d}\theta \sqrt{r(\theta)^2 \sin^2\theta \left( r(\theta)^2 + \frac{L^2 r'(\theta)^2}{L^2 - \frac{2G_{N}M L^2}{r(\theta)} + r(\theta)^2} \right)}\,.
    \label{eq:inducedmetricglobalquench}
\end{equation}
This yields a second-order differential equation \eqref{eq:EuLagrM}, which is generally intractable analytically. Therefore, we proceed with a numerical approach\footnote{A similar study has been carried out in \cite{Jahn:2017xsg} with a different geometric setup for the entangling surface.}, complemented by perturbative analysis as explored in \cite{Nozaki:2013wia}.

As mentioned above, we employ global coordinates as defined in \eqref{eq:MetricBHAdS}, since their static configuration is suited to the RT surface at a constant global time slice. Moreover, the local operator introduced in \eqref{eq:ExcitedState} defines a state which is an eigenstate for the generator of global time translations \eqref{eq:eigenstate}.

Firstly, we start with the simpler case of the vacuum state, \textit{i.e.}\ $M=0$ in \eqref{eq:MetricBHAdS}. 
The minimal surface in the vacuum has been determined analytically in Poincaré coordinates  in \cite{Ryu:2006ef}, and it can be easily translated to global coordinates \eqref{eq:MetricEmptyAdS} obtaining the profile





\begin{equation}
    r_{\text{\tiny vac}}(\theta) = \frac{L}{\sqrt{-1 + \cos^2\theta \sec^2\theta_{\infty}}}\,.
    \label{eq:rthetaGlobal}
\end{equation}
By computing the functional \eqref{eq:inducedmetricglobalquench} for $M=0$ on the vacuum solution \eqref{eq:rthetaGlobal} we obtain the vacuum HEE which reads
\begin{equation}
  S_{\text{\tiny vac}} = \frac{L^2}{4G_N} \left(  2\pi  \sin{\theta_\infty} \frac{r_\infty}{L} - 2\pi  \right)\,,
\end{equation}
where $r_\infty$ represents the UV cutoff introduced to obtain a finite answer. This result will be employed below to obtain a finite variation of the time-evolution of the HEE under the local quench.

Equation \eqref{eq:rthetaGlobal} represents the starting point for the numerical and perturbative analysis for the $M \ne 0$ case, which we discuss below.
In this section we explain the main idea, leaving the details and the numerical technicalities to the appendix \ref{section:Numerical}. 

\begin{figure}
\centering
\includegraphics[scale=0.24]{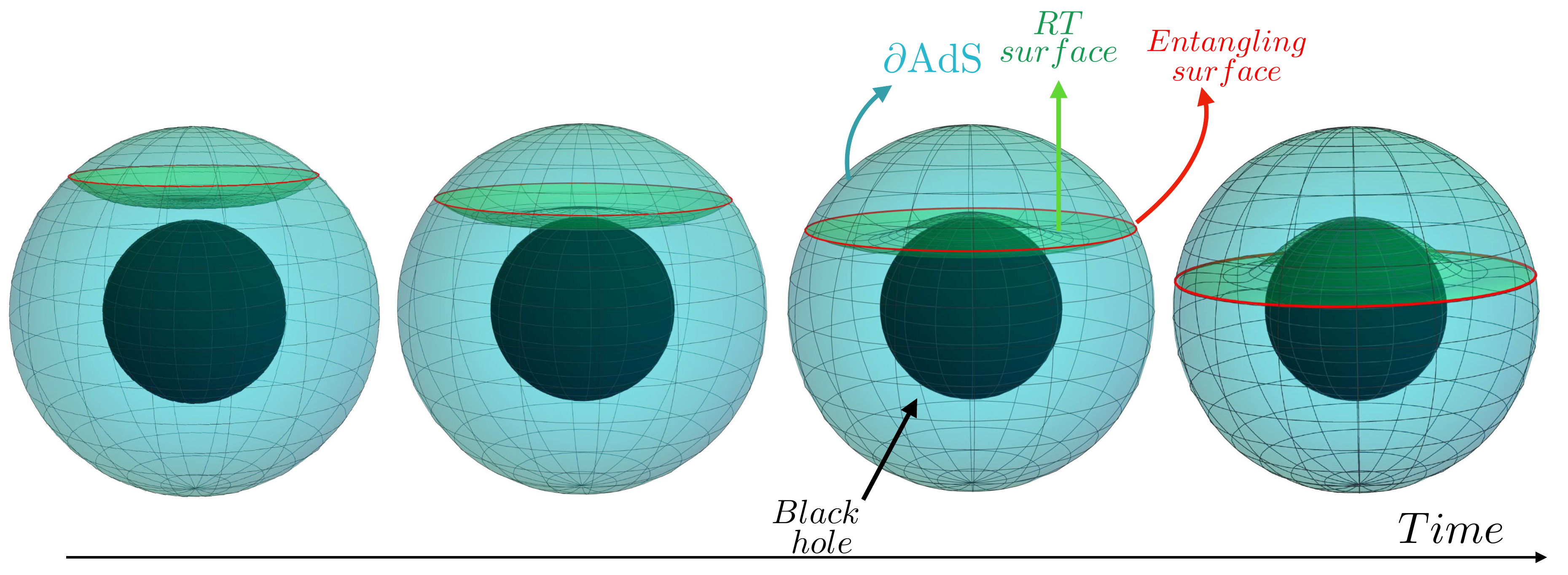}
\caption[3D Plots]
    {3D plot of the evolution for the setup with $r_h = G_N = M = L = 1$. The conformal boundary at fixed time is shown in blue, the black hole horizon in black, the entangling surface in red, and the minimal surface in green. The plot shows the time evolution from $t=0$ to the transition time, where the entangling surface is at the equator, corresponding to the blue branch in figure \ref{fig:PageCurve}. The evolution is symmetric and the orange branch is obtained when the minimal surface lies on the lower hemisphere.}
\label{fig:3dplots}
\end{figure}

We first take advantage of the analytic vacuum solution $r_{\text{\tiny vac}}(\theta)$ \eqref{eq:rthetaGlobal} in order to select the appropriate boundary conditions for the differential equation \eqref{eq:EuLagrM} with $M\ne0$. Indeed, the boundary conditions pertain to the UV property of the HEE and are independent on the value of $M$. We then obtained the profile of the minimal surface by solving numerically the ordinary differential equation. Once we got the profile, we numerically computed the area functional. To make the numerical integration easier, we found it convenient to tame the UV divergences by subtracting the UV-vacuum contribution. In particular, we computed the following integral

\begin{equation}
    S_{\text{quench}}^{\text{reg}} = \frac{2\pi}{4G_N} \int_0^{\theta_\infty} \mathrm{d}\theta \, 
    \left( 
    r(\theta) \sin{\theta} 
    \sqrt{r(\theta)^2 + \frac{L^2 r'(\theta)^2}{L^2 - \frac{2 G_N L^2 M}{r(\theta)} + r(\theta)^2}}
    - L \sin{\theta_\infty} r'(\theta)
    \right) < \infty\,,
    \label{eq:VariationHEE}
\end{equation}
for $L = M = G_N = 1$. The excess of holographic entanglement entropy is then

\begin{equation}
    \Delta S_{\textit{HEE}} = S_{\text{quench}}^{\text{reg}} - \frac{2\pi}{4G_N} L \sin{\theta_\infty} r_0 + \frac{2\pi}{4G_N} L^2\,,
    \label{eq:VariationHEE2}
\end{equation}
where $r_0$ corresponds to the value of $r$ at the turning point (or tip) $\theta=0$.
\noindent
Figure \ref{fig:3dplots} illustrates the time evolution of the minimal surface in the black hole geometry from early-time to the transition phase time, where the entangling surface lies on the equator. As time proceeds, the minimal surface lives in the lower hemisphere, reaching the south pole at late-time.

\begin{figure}[htbp!]
\centering
\includegraphics[scale=0.8]{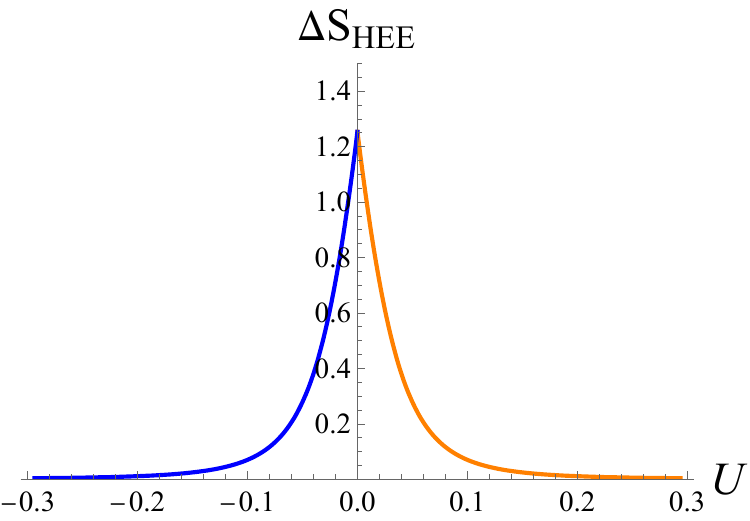}
\caption[Page Curve]
    {Evolution of $\Delta S_{\textit{HEE}}$ as a function of the lightcone coordinate $U$ \eqref{eq:lightcone coord}. Blue represents early times, while orange corresponds to late  times, symmetric about $U\rightarrow-U$.}
\label{fig:PageCurve}
\end{figure}

The final result is shown in figure \ref{fig:PageCurve}, which displays the time evolution of $\Delta S_{\textit{HEE}}$ as a function of the lightcone time $U$ introduced in \eqref{eq:lightcone coord}. The curve reveals a behavior analogous to the Page curve \cite{Page:1993wv}, even though it arises from the black hole dynamics rather than evaporation. In the dual picture, the black hole does not evaporate in a Poincaré time. Rather, it enters a Poincaré section, approaches the boundary and then disappears again behind the Poincaré horizon. This entropy can still be interpreted as the Hawking radiation of the accelerating black hole, as it is thoroughly discussed in \cite{Agon:2020fqs}, where the authors consider a two-dimensional setup analogous to the one discussed here.  

\subsubsection*{Comparison with the perturbative solution}
\label{subsubsection:perturb}

In this section, we compare our numerical results with those obtained in \cite{Nozaki:2013wia} using a perturbative treatment of the black hole, which we report below for the sake of completeness. 

To begin with, we perform a perturbative computation of the excess of the holographic entanglement entropy by expanding the square root of the determinant of the induced metric \eqref{eq:inducedmetricglobalquench} around \(G_N M = 0\)\footnote{Note that we take $M \sim 1/G_N$. Thus, even though $M$ is large, our expansion parameter can be kept small. }. Keeping terms up to the first order in \(G_N M\), we find

\begin{equation}
\sqrt{h}=\frac{\sin{\theta}\left[ r(\theta)^{3}\left( L^{2}+r(\theta)^{2}\right)^{2}+L^{2}r'(\theta)^{2}\left( G_{N}M L^{2}+L^{2}r(\theta)+r(\theta)^{3}\right)\right]}{\left(L^{2} +r(\theta)^{2}\right)^{3/2}\sqrt{L^{2}r(\theta)^{2}+r(\theta)^{4}+L^{2}r'(\theta)^{2}}} + \mathcal{O}(G_{N}^{2}M^{2})\,.
\end{equation}
At the leading perturbative order, we only need to consider the variation of the metric in terms of the mass $M$, retaining the embedding of the minimal surface to be the unperturbed vacuum solution \(r_{\text{\tiny vac}}(\theta)\) from \eqref{eq:rthetaGlobal}. This is a consequence of the fact that the first order variation of the embedding multiplies the equation of motion for \(r_{\text{\tiny vac}}(\theta)\) giving a vanishing result. Keeping this in mind, the excess of the holographic entanglement entropy in the perturbative regime is given by

\begin{equation}
\begin{split}
    \Delta S_{HEE}^{\textit{pert}} & = \frac{2\pi}{4G_{N}} \int_{0}^{\theta_{\infty}} \mathrm{d}\theta \, \left(G_{N}M L \cot{\theta_{\infty}}\sin{\theta}\tan^{2}{\theta}\right)+\mathcal{O}(G_{N}M^{2}) \\
    & = \frac{\pi L M}{2} \cot{\theta_{\infty}} \left(-2 + \cos{\theta_{\infty}} + \sec{\theta_{\infty}}\right) +\mathcal{O}(G_{N}M^{2})\,.
    \label{eq:SHEE}
    \end{split}
\end{equation}
It is convenient to express \(\Delta S_{HEE}^{\textit{pert}}\) in terms of the cross-ratio \(\xi\) defined in \eqref{eq:xicrossratiofinale} using \eqref{eq:crossratio2} 
\begin{equation}
    \Delta S_{HEE}^{\textit{pert}}=\frac{L M\pi(2-\xi-2\sqrt{1-\xi})}{2\sqrt{\xi}}\,.
    \label{eq:SHEEzeta}
\end{equation}
This result is in perfect agreement with \cite{Nozaki:2013wia} \footnote{Note that $M$ in our paper corresponds to $m$ in \cite{Nozaki:2013wia}.}.

\begin{figure}
\centering
\includegraphics[scale=0.25]{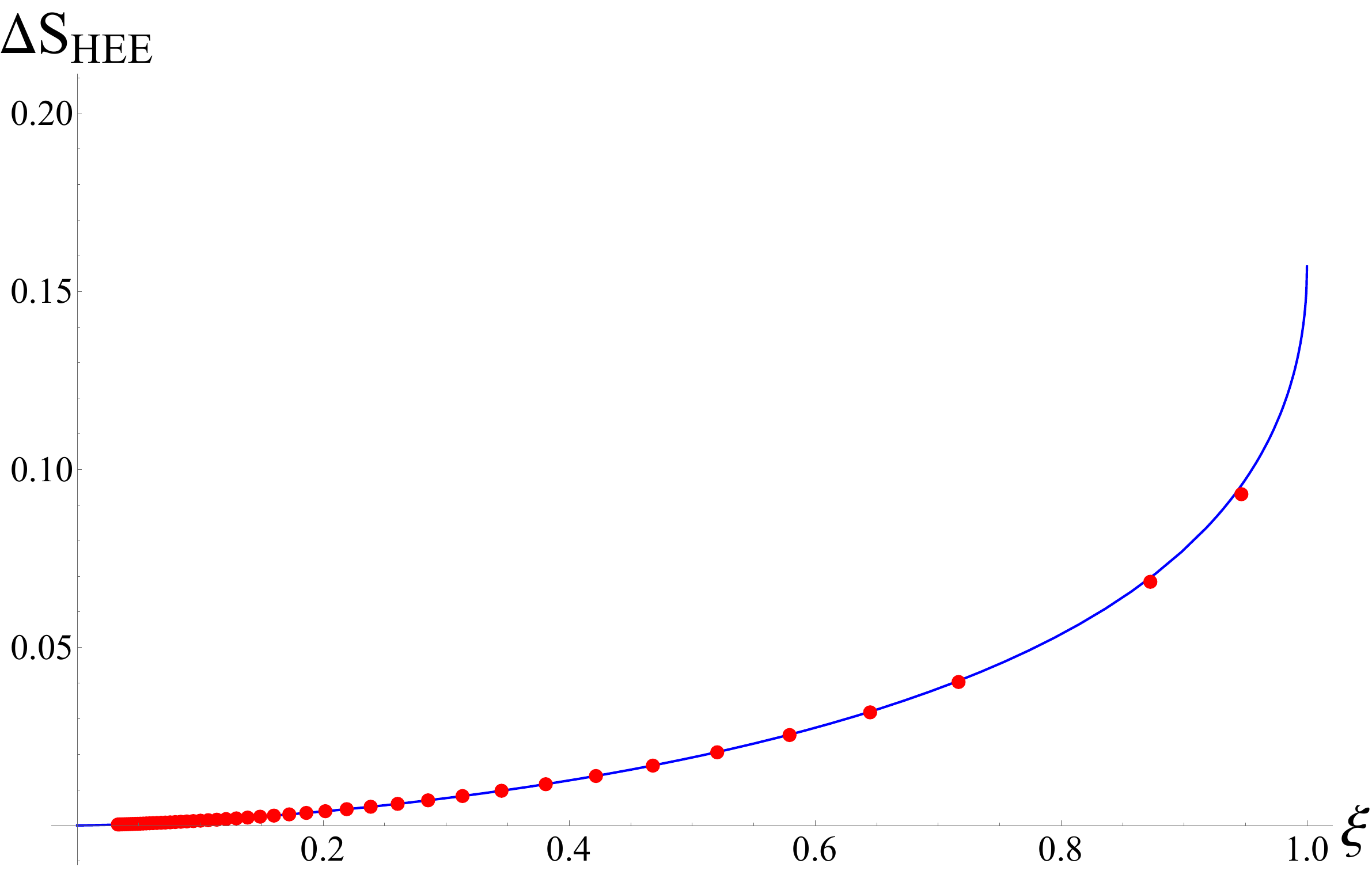}
\caption[Perturbative Cross-Ratio]
    {The blue curve represents the evolution of the excess of the holographic entanglement entropy \(\Delta S_{HEE}^{\textit{pert}}\) as a function of the cross-ratio \(\xi\), computed perturbatively in \cite{Nozaki:2013wia}. The red points correspond to our numerical results at $G_{N} M=0.1$\,.}
\label{fig:PerturbCrossRatio}
\end{figure}

We now compare our numerical results with the analytic curve obtained for the variation of the holographic entanglement entropy in the perturbative computation. This comparison is shown in figure \ref{fig:PerturbCrossRatio}, where we plot \(\Delta S_{HEE}\) as a function of the cross-ratio \(\xi\). We observe an excellent agreement between our numerical results and the perturbative computation, providing a strong validation of our approach throughout this work.

\subsubsection*{Comparison with the OPE and the bound}
\label{subsubsection:ComparisonOPEBound}

\begin{figure}[htbp!]
\centering
\includegraphics[scale=.68]{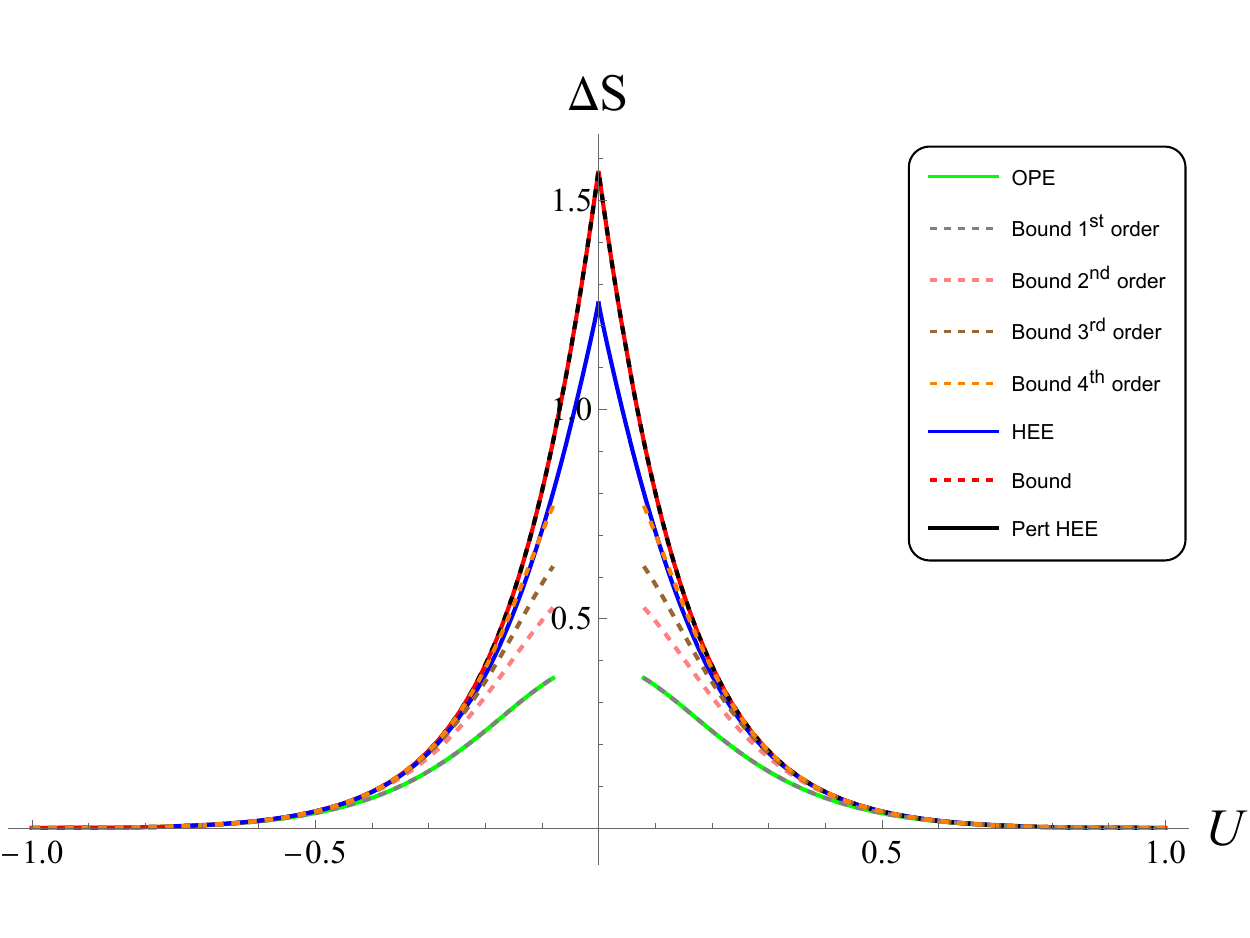}
\caption[delta SEE]
    {In $d=3$, we plot the excess of the entanglement entropy $\Delta S$ as a function of the lightcone coordinate $U$ \eqref{eq:lightcone coord}, for $G_{N}=M=L=1$ and $\epsilon=0.5$. The dashed red line represents the bound \eqref{eq:bound3M}, which is fully saturated by the black line, corresponding to the perturbative holographic result \eqref{eq:SHEEzeta}. The thick blue line shows our numerical result which lies below the bound. The lower green line represents the CFT computation \eqref{eq:DeltaSEEFInale} in the early and late time limits $\xi\rightarrow 0$, which precisely matches the expansion of the bound at the same order in $\xi$, shown by the dashed gray line.}
\label{fig:DeltaSEE}
\end{figure}

A few remarks on the previous result: at first glance, the expression \eqref{eq:SHEEzeta} appears notably familiar. Indeed, comparing it to the entanglement entropy bound for $d=3$ dimensions, we notice a similar form in \eqref{eq:bound3xi}. In the AdS/CFT correspondence, it is known \cite{Witten:1998qj} that the following relation holds

\begin{equation}
    \Delta= M L\,,
\end{equation}
where $\Delta$ is the conformal dimension of the operator in the CFT, $L$ is the radius of AdS and $M$ is the mass of the black hole.
Thus, substituting this into the bound \eqref{eq:bound3xi}, we obtain

\begin{equation}
    \Delta S_{A}^{EE} \leq \frac{\pi M L \left( 2 - \xi - 2 \sqrt{1 - \xi} \right)}{2 \sqrt{\xi}} = \Delta S_{HEE}^{\textit{pert}}\,,
    \label{eq:bound3M}
\end{equation}
which precisely matches the perturbative result found in \eqref{eq:SHEEzeta}. Since the perturbative calculation fully saturates the bound, we can conclude that, as the black hole mass increases, the holographic entanglement entropy progressively deviates below the bound. This is illustrated in figure \ref{fig:DeltaSEE}, where we can see that the holographic numerical solution falls below the bound, which coincides with the perturbative result. 

Notice that, even for high values of the mass, when the perturbative treatment is not justified, there is still good agreement between the perturbative and the numerical solution at early and late time. Indeed, if we expand the perturbative solution \eqref{eq:SHEEzeta} for $\xi \to 0$ we find
\begin{equation}
\Delta S_{HEE}^{\textit{pert}}\overset{\xi\rightarrow0}{\sim} \frac{\pi \Delta}{8} \xi^{\frac32}\,,
\end{equation}
which coincides with the late-time behavior predicted by \eqref{eq:DeltaSEEFInale}. This means that corrections in this regime become relevant at higher orders in the mass. 

\subsection{Holographic boundary local quenches}
\label{section:BCFT}

\subsubsection{Construction of the holographic dual of BCFT}
\label{subsection:BCFTconstruction}

In this section, we study the holographic dual of the boundary CFT setup introduced in section~\ref{section:BCFTsetup}. Following the bottom-up construction in \cite{Takayanagi:2011zk,Fujita:2011fp}, we aim to construct the holographic dual of a CFT defined on $\mathcal{M}$, a $d$-dimensional manifold with a conformal boundary $\partial\mathcal{M}$. The holographic dual of $\mathcal{M}$ is a $(d+1)$-dimensional asymptotically AdS spacetime, denoted by $\mathcal{N}$. The boundary of the bulk spacetime $\mathcal{N}$ is given by $\partial\mathcal{N} = \mathcal{M} \cup \mathcal{Q}$, where $\mathcal{Q}$ is a $d$-dimensional manifold homologous to $\mathcal{M}$. This additional boundary $\mathcal{Q}$ is also referred to as the End-of-the-World Brane (EoW), and its boundary coincides with $\partial\mathcal{M}$. 
As usual, we impose Dirichlet boundary conditions at the conformal boundary $\mathcal{M}$ of AdS. On $\mathcal{Q}$, however, we consider Neumann boundary conditions, allowing this boundary to be dynamical from the holographic perspective.

To correctly formulate  the variational problem, in addition to the Hilbert-Einstein action, we include the Gibbons-Hawking boundary term. The gravitational action is then expressed as

\begin{equation}
    I = \frac{1}{16\pi G_{N}} \int_{\mathrm{N}} \sqrt{-g} \left(R - 2\Lambda\right) 
    - \lambda \int_{\mathrm{Q}} \sqrt{-\hat{g}} 
    + \frac{1}{8\pi G_{N}} \int_{\mathrm{Q}} \sqrt{-\hat{g}} K\,,
    \label{eq:actionBCFT}
\end{equation}
where $g$ is the metric of the bulk AdS spacetime $\mathcal{N}$, $\hat{g}$ is the induced metric on the EoW brane $\mathcal{Q}$, and $\Lambda$ is the cosmological constant. The constant $\lambda$ represents the tension of the EoW brane. The extrinsic curvature $K = \hat{g}^{ab}K_{ab}$ is defined by

\begin{equation}
    K_{ab} = \nabla_{a} n_{b}\,, \quad a, b = 0, \dotsc, d-1\,,
    \label{eq:excurvature}
\end{equation}
where $n$ is the unit normal vector to the EoW brane $\mathrm{Q}$.
The variation of the action \eqref{eq:actionBCFT} yields the Einstein equations on $\mathcal{N}$
\begin{equation}
G_{\mu\nu}+\Lambda \, g_{\mu\nu} = 0 \, ,
\label{eq:EEq}
\end{equation}
together with the boundary term
\begin{equation}
    \delta I = \frac{1}{16\pi G_{N}} \int_{\mathrm{Q}} \sqrt{-\hat{g}} 
    \left(K_{ab} - K \hat{g}_{ab}+\lambda \hat{g}_{ab}\right) \delta \hat{g}^{ab}\,.
\end{equation}
By imposing Neumann boundary conditions, we obtain a set of equations for the extrinsic curvature of $\mathcal{Q}$
\begin{equation}
    K_{ab} - \hat{g}_{ab} K = -\lambda \hat{g}_{ab}\,,
    \label{eq:Israel}
\end{equation}

\noindent
which are also known as the Israel-Lanczos conditions \cite{Israel:1966rt}. 
To determine the location of the EoW brane, we need to solve the Israel junction conditions (\ref{eq:Israel}) together with the Einstein's equations \eqref{eq:EEq}.

\subsubsection{The holographic dual of the boundary local quench}
\label{subsection:dualboundaryquench}

In this section, we combine the analysis of the holographic dual of a local quench presented in section \ref{subsection:HEEQuench} with the prescription for a conformal boundary, whose holographic dual is described by the EoW brane. Our goal is to construct the bottom-up holographic description of a boundary local quench.

This is a formidable task as, in general, the EoW brane backreacts modifying the bulk metric. For $d>2$, analytic solutions to the equations \eqref{eq:EEq} and \eqref{eq:Israel} can be found in the case of vacuum state in AdS space-times with flat or spherical boundaries \cite{Fujita:2011fp}, while for more generic boundary shapes we need to make use of a perturbative approach \cite{Miao:2017aba,Seminara:2017hhh}. 
The situation becomes even more complicated when we deal with states different from the vacuum. Indeed, to the best of our knowledge, for $M\ne 0$ and $\lambda \ne 0$, no solution is known, even for flat or spherical boundaries. In the following, we will study in a certain detail the simplest scenario $M \ne 0$ and $\lambda = 0$, giving only some remarks of the most general $\lambda \ne 0$ situation. 

From here onward, we focus on the case of $d = 3$ dimensions, corresponding to an asymptotically $\mathrm{AdS}_4$ spacetime. In this scenario, an equivalent expression for \eqref{eq:Israel} is

\begin{equation}
    K_{ab} = \frac{\lambda}{2} \hat{g}_{ab}\,.
    \label{eq:ExCurvatureComplete}
\end{equation}
Let us consider the gravity dual of an eigenstate \eqref{eq:ExcitedState} of the conformal Hamiltonian. We work in a CFT initially defined on the Minkowski space $\mathbb{R}^{1,2}$ with flat metric $\eta_{\mu\nu} = \mathrm{diag}(-1,1,1)$ and coordinates $x^{\mu} = (t, x, y)$, where a conformal boundary is introduced at $x = 0$. For convenience, we conformally map the CFT from $\mathbb{R}^{1,2}$ to a cylinder $\mathbb{R} \times S^{2}$ with metric 

\begin{equation}
    \mathrm{d}s^{2} = R^{2} \left( -\mathrm{d}t^{2} + \mathrm{d}\theta^{2} + \sin^{2}{\theta}\,\mathrm{d}\phi^{2} \right)\,,
\end{equation}
where $R$ is the radius of the $S^{2}$ sphere, $\theta \in (0, \pi)$, and $\phi \in (0, 2\pi)$. Under this map, the boundary $x = 0$ in Minkowski space becomes the equator $\theta = \pi/2$ of the $S^{2}$ sphere, with the CFT living on the upper hemisphere. This transformation corresponds to a special conformal transformation that maps a straight line (the boundary $x = 0$) to a circle (the equator $\theta = \pi/2$ of $S^{2}$).

Next, we modify the transformations between Poincaré and global coordinates given in (\ref{eq:CoordinateTransformation}) to incorporate this setup. In particular, it is convenient to perform a rotation such that the coordinate transformations become

\begin{equation}
    \begin{split}
        & \sqrt{L^{2} + r^{2}} \cos{\tau} = \frac{\alpha L^{2} + \frac{z^{2} + x^{2} + y^{2} - t^{2}}{\alpha}}{2z}\,, \\
        & \sqrt{L^{2} + r^{2}} \sin{\tau} = \frac{L t}{z}\,, \\
        & r \sin{\theta} \cos{\phi} = \frac{\alpha L^{2} + \frac{-z^{2} - x^{2} - y^{2} + t^{2}}{\alpha}}{2z}\,, \\
        & r \cos{\theta} = \frac{L x}{z}, \\
        & r \sin{\theta} \sin{\phi} = \frac{L y}{z}\,.
    \end{split}
    \label{eq:coordinateTransformation2}
\end{equation}
We assume that the holographic dual of the boundary local quench retains the metric form of \eqref{eq:MetricBHAdS}, where the bulk spacetime describing a massive particle in global coordinates is now bounded between the conformal boundary and the EoW brane.

\subsubsection*{Tensionless \texorpdfstring{$\lambda=0$}{λ=0} limit of the EoW brane}
\label{subsection:tensionless}

In this section, we analyze the special case of a tensionless brane, $\lambda = 0$. Below, we show that in this scenario, the solution for the spacetime consists of the global AdS black hole metric with no backreaction. To start, the dynamics of the EoW brane are governed by the simplified equation

\begin{equation}
    K_{ab} = 0\,.
    \label{eq:matchcond2}
\end{equation}

Working in global AdS coordinates with time translational invariance, we can assume an ansatz given by a static EoW brane $\theta = \theta(r)$ which gives the induced metric

\begin{equation}
    \mathrm{d}\hat{s}^{2} = -L^2 f(r)\mathrm{d}\tau^{2} + \left(  \frac{1}{f(r)} + r^{2}\left(\theta'(r)\right)^2\right)\mathrm{d}r^{2} + r^{2}\sin^{2}{\theta}\,\mathrm{d}\phi^{2}\,,
    \label{eq:inducedmetric1}
\end{equation}

\noindent
where the components are obtained by pulling back the bulk metric (\ref{eq:MetricBHAdS}) and the function $f$ is defined in \eqref{eq:f(r)}. 


To reduce the matching condition \eqref{eq:ExCurvatureComplete} to a first-order differential equation, we find the unit normal vector $n^{\mu}$ to the EoW brane, which reads

\begin{equation}
    n^{\mu} = \left(0, -\frac{r f(r) \theta'(r)}{\sqrt{1+r^2 \, f(r) \left(\theta'(r)\right)^2}}, \frac{1}{\sqrt{1+r^2 \, f(r) \left(\theta'(r)\right)^2}}, 0\right)\,,
\end{equation}
where we choose the solution with the negative sign in the radial component to ensure that $n^{\mu}$ corresponds to an incoming normal vector in the radial direction.

By explicitly computing the matching condition \eqref{eq:matchcond2} for the $\tau \tau$ component, we obtain the following first-order equation 
\begin{equation}
    (G_{N} L^{2} M + r^{3})(-2G_{N} L^{2} M + L^{2} r + r^{3}) \theta'(r) = 0\,,
\end{equation}
whose solution is simply 
\begin{equation}
    \theta(r) = \text{constant}\,.
\end{equation}
At this point, the other components require that the angle $\theta$ is fixed to be
\begin{equation}
    \theta_{\mathrm{brane}} = \frac{\pi}{2}\,,
\end{equation}
corresponding to a boundary which reduces to the equatorial disk at the conformal infinity $r\rightarrow \infty$.

\begin{figure}
\centering
\includegraphics[scale=0.242]{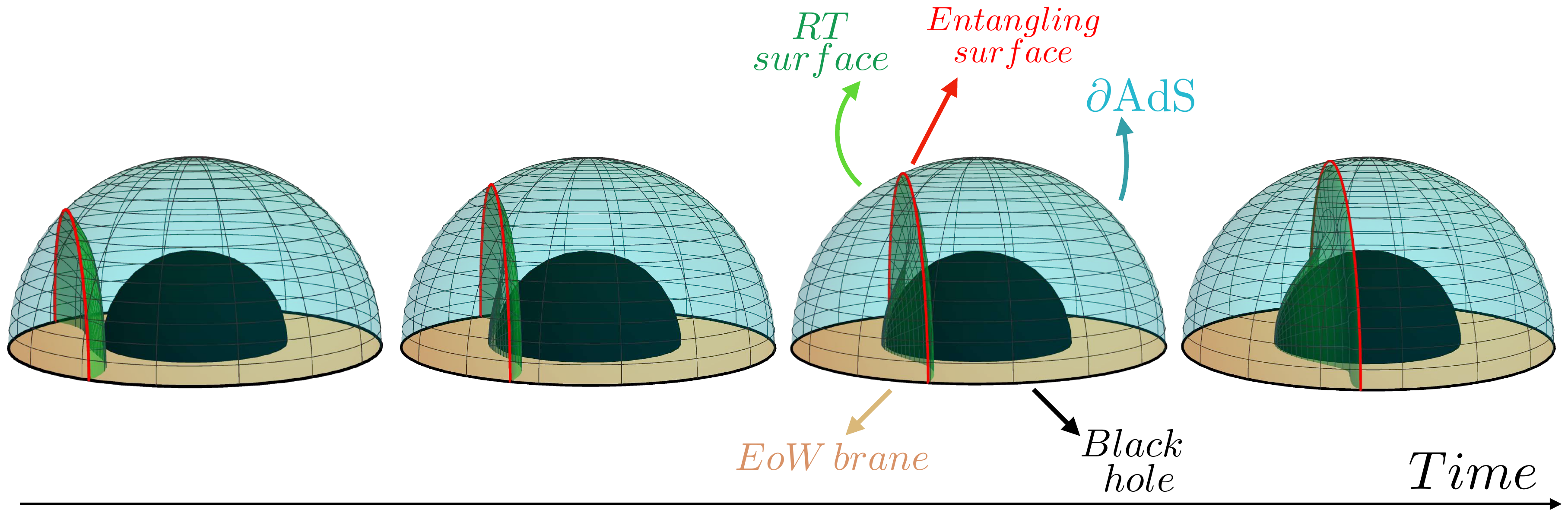}
\caption[3D Plots EoW]
    {3D plot of the evolution for the setup in with $r_h = G_N = M = L = 1$ and tension $\lambda=0$. The conformal boundary is shown in blue, the black hole horizon in black, the entangling surface in red, the minimal surface in green and the EoW brane in light brown. The plot shows the time evolution from $t=0$ to the transition time, where the entangling surface is at the equator, analogously to figure \ref{fig:3dplots}. }
\label{fig:3dplotsEoW}
\end{figure}

We aim to compute the time evolution of the holographic entanglement entropy (HEE). According to the Ryu-Takayanagi prescription, the HEE is proportional to the area of the minimal surface anchored to the conformal boundary at the entangling surface in the CFT. In our case, the entangling surface is a circle $x^{2} + y^{2} = R^{2}$ of radius $R$ in Minkowski space. Under the coordinate transformations \eqref{eq:coordinateTransformation2}, this circle maps to

\begin{equation}
    \phi(\theta) = \arccos{\left( \frac{(L^{2} - R^{2}) \csc{\theta}}{L^{2} + R^{2}} \right)}\,,
    \label{eq:EntSurfaceNew}
\end{equation}
where the entangling surface is now described by the function $\phi(\theta)$. This coordinate system emphasizes the EoW brane. Figure \ref{fig:3dplotsEoW} illustrates the setup in these coordinates. 

Since the spacetime is cut exactly in half by the EoW brane, the symmetry of the problem implies that the minimal surface is simply given by half the one we obtained in the previous section \ref{subsection:HEEQuench}, namely in the case without boundaries\footnote{In a more general setting, the minimization of the surfaces is more complicated since it is important to ensure the orthogonality condition between the minimal surface and the EoW brane. For the vacuum state, non-trivial examples have been discussed in \cite{Seminara:2017hhh,Seminara:2018pmr} }. As a result, the evolution of the holographic entanglement entropy is identical to that already derived, but rescaled by a factor of $1/2$ due to the integration being restricted to half the bulk spacetime, as depicted in figure \ref{fig:3dplotsEoW}.

\subsubsection*{Comments on the generic case $M\ne 0$ and $\lambda \ne 0$}

As discussed above, it is not possible to solve the system of differential equations in \eqref{eq:ExCurvatureComplete} by assuming the bulk metric to be \eqref{eq:MetricBHAdS}, since this leads to inconsistent solutions. Indeed, the EoW brane induces a backreaction modifying the bulk metric. This computation is highly non-trivial even if we focus on perturbation theory or numerical methods. 
Notice that this behavior differs fundamentally from the lower-dimensional cases studied in \cite{Bianchi:2022ulu, Kawamoto:2022etl}, where the simultaneous presence of the black hole and the EoW brane does not alter the bulk metric. In those cases, the bulk metric is described by a massive particle falling in an asymptotically AdS background, truncated by the EoW brane. In contrast, in our higher-dimensional setup, the backreaction is present and must be included for consistency.

We end this section mentioning an interesting case where a family of full backreacted solutions in the presence of a brane or EoW brane can be obtained analytically. This family of solution are called AdS C-metrics, a subclass of Plebiánski-Demiánski type-D metrics \cite{PLEBANSKI197698}. 

For a 4-dimensional bulk spacetime we can write the following metric
\begin{equation}
ds^2 = \frac{1}{\Omega(r,x)}\left[ - H(r)d\tau^2 +\frac{dr^2}{H(r)}+r^2 \left( \frac{dx^2}{G(x)} + G(x) d\phi^2 \right)\right] \, ,
\label{eq:C_metric}
\end{equation}
where 
\begin{equation}
 \Omega(r,x)= 1+ \mathfrak{a}_{\text{\tiny cc}}  \, r\, x \, , \qquad G(x)= 1-x^2-2 G_N M \mathfrak{a}_{\text{\tiny cc}}  \, x^3 \, , \qquad H(r) = 1+\frac{r^2}{\tilde L^2}- \frac{2 G_N M}{r} \, , 
\end{equation}
being $\tilde L = L/\sqrt{1-L^2 \mathfrak{a}_{\text{\tiny cc}} ^2}$. Here, $M$ represents again the mass of the black hole while $\mathfrak{a}_{\text{\tiny cc}} $ is the acceleration. This solution reduces to the metric \eqref{eq:MetricBHAdS} when $\mathfrak{a}_{\text{\tiny cc}} = 0$, while if $M=0$ but $ \mathfrak{a}_{\text{\tiny cc}}\ne 0$, it is possibile to find a coordinate transformation which maps \eqref{eq:C_metric} to global AdS$_4$.

These types of solutions describe a pair of accelerating black holes and have the peculiarity of admitting a timelike hypersurface  at $x=0$ which is umbilic, meaning that the extrinsic curvature is proportional to the induced metric. In our case one gets 
\begin{equation}
K_{ab}= \mathfrak{a}_{\text{\tiny cc}}  \, \hat g_{ab} \,,
\end{equation}
which indeed solves the boundary condition \eqref{eq:ExCurvatureComplete} if we take  $\mathfrak{a}_{\text{\tiny cc}} = \lambda/2$.
Choosing this umbilic surface as our EoW brane, we end up with a spacetime with a single accelerating black hole and a boundary. 
In the recent literature, the AdS C-metrics are used to construct and study lower dimensional quantum black holes living on the brane $x=0$ out of higher dimensional classical ones \cite{Tian:2024mew, Panella:2024sor}. For our purposes, the idea was to use those solutions as a possibile gravitational setup for local quenches in the presence of a boundary. 
Unfortunately, the asymptotics of such a solution does not match the expected one, as the non-trivial acceleration modifies the UV property of the theory. Indeed, we performed a double perturbative expansion around $M=0$ and $\lambda=0$ to compute the leading modification to the HEE and we found
%
%
that the area-law term acquires an additional contribution proportional to the mass of the black hole if $\lambda \ne 0$. A detailed discussion of minimal surfaces in these setups can be found in \cite{Xu:2017nut}. An extensive analysis of the HEE in such metric spaces is beyond the scope of the present work, and here we just point out that the state described by this spacetime appears to be radically different from the local quench. It is nonetheless interesting to understand the nature of boundary theory corresponding to such metrics, and we leave such an analysis to a future work.

\acknowledgments

We would like to thank Marco Billò, Elia de Sabbata, Marco Meineri and Riccardo Pozzi for useful discussions. JS was funded by the Knut and Alice Wallenberg Foundation under grant KAW 2021.0170, VR grant 2018-04438, and Olle Engkvists Stiftelse grant n. 2180108. JS thanks Nordita for their hospitality during the final stages of this work. AM was partially funded by the Grant for Internationalization of the University of Torino and would like to thank the Department of Physics at the University of Uppsala for hospitality during the initial stage of the project. We also thank the Pollica Physics Center for hosting the workshop “Defects, from condensed matter to quantum gravity”, where part of the project was carried out.

\appendix

\section{Displacement one-point function with two defects}
\label{onepoint}

In this appendix we derive the kinematical structure of the one-point function of a scalar boundary operator in the presence of two conformal defects in the configuration shown in figure \ref{fig:setupboundarytwista}. We start with the planar defects of figure \ref{fig:setupboundarytwistb} with an operator inserted on the boundary. The only dimensional quantity in this case is the orthogonal distance $|\tilde{x}^{a}|$ between the insertion and the codimension-three defect arising in the intersection between the conformal boundary and the codimension-two defect. Hence, we obtain

\begin{equation}
    \langle D(\tilde{x}^{a}) \rangle_{\mathcal{H}_{n}, \mathcal{B}} = \frac{a_{D,n}}{n}\frac{1}{|\tilde{x}^{a}|^{d/2}}\,,
    \label{eq:1pointdisplacementhalfline}
\end{equation}

\noindent
where \( \tilde{x}^{a} = (\tilde{\tau}, \tilde{y}_{1}, \ldots, \tilde{y}_{d-2}) \) are the coordinates in the flat-defect configuration and \( \mathcal{H}_n \) represents the halfplane twist operator. The coefficient $a_{D,n}$ is the one-point function coefficient and, for convenience, we introduce a factor $n$ at the denominator, analogously to \eqref{eq:1pointtwistsphericalFINAL}.
To recover the correlation function in the original space, we use the change of coordinates \eqref{eq:mappingspheretohalplane} with \( x = 0 \). Of course, we need to take into account the transformation of the displacement operator under this conformal transformation. Since the displacement is a primary operator, it transforms as
\begin{equation}
    \langle D(x^{a}) \rangle_{\sigma_{n}(R), \mathcal{B}} = \Omega^{d}(x^{a}) \langle D(\tilde{x}^{a}(x^{a})) \rangle_{\mathcal{H}_n, \mathcal{B}}\,,
\end{equation}
with the conformal factor given by
\begin{equation}
    \mathrm{d}s^{2}_{\text{hemisphere}} = \Omega^{2}(x^{a}) \, \mathrm{d}s^{2}_{\text{halfplane}}\,, \quad \Omega^{2}(x^{a}) = \frac{R^{4}}{((R - y_{d-2})^{2} + y^{\hat{a}}y_{\hat{a}}+\tau^{2})^{2}}\,,
\end{equation}
with $\hat{a}=1,\ldots,d-3$.
Using this expression along with \eqref{eq:mappingspheretohalplane} gives the required expression for the boundary one-point function after the conformal transformation

\begin{equation}
\begin{split}
     \langle D(x^{a})\rangle_{\sigma_{n}(R), \mathcal{B}}& =\frac{a_{D,n}}{n}\frac{(2R)^{d}}{\left( (R-y_{d-2})^{2}+y^{\hat{a}}y_{\hat{a}}+\tau^{2}\right)^{\frac{d}{2}}\left( (R+y_{d-2})^{2}+y^{\hat{a}}y_{\hat{a}}+\tau^{2}\right)^{\frac{d}{2}}}
     \end{split}
     \label{eq:1point2defectsA}
\end{equation}

\subsection*{Limit \texorpdfstring{$n\to 1$}{n->1} of the displacement one-point function}
\label{sec:detailsnto1}

Here, we will derive equation \eqref{eq:bsigmad} using the strategy of \cite{Smolkin:2014hba,Rosenhaus:2014woa,Rosenhaus:2014nha,Hung:2014npa} which we briefly review.
We can define a new effective twist operator \( \tilde{\sigma}_{n} \), which arises when considering the correlation function of an operator \( \chi \) inserted on a single copy of the replica CFT, say for example on the first copy, in the presence of the twist operator \( \langle \sigma_{n} \chi \rangle \). The new effective twist operator is defined as the non-local operator emerging from integrating out the \( (n-1) \) copies of the CFT without any operator insertions. This leads to the relation
\begin{equation}
    \langle \sigma_{n} \chi \rangle = \langle \tilde{\sigma}_{n} \chi \rangle_{1}\,,
    \label{eq:effectivetwist}
\end{equation}
where the correlator on the r.h.s.\ is evaluated on the first copy, and the effective twist operator \( \tilde{\sigma}_{n} \) acts within this copy of the replica. The path integral over the \( n \)-fold covering geometry can be formulated as the trace of the reduced density matrix \( \rho_{A} \), as shown in \eqref{eq:pathintegralrho}. This allows a straightforward interpretation of \( \tilde{\sigma}_{n} \) as the path integral representation of the \( (n-1) \)-th power of the reduced density matrix

\begin{equation}
    \tilde{\sigma}_{n} \equiv \langle \sigma_{n} \rangle_{\{2, \ldots, n\}} = \rho_{A}^{n-1}\,.
\end{equation}
Then, we express the reduced density matrix as the exponential of the Hermitian operator \( H_{A}^{0} \), \textit{i.e.}\ the modular Hamiltonian

\begin{equation}
    \tilde{\sigma}_{n} = e^{-(n-1)H_{A}^{0}}\,.
\end{equation}
Using this, we can rewrite the equivalence introduced earlier as 

\begin{equation}
    \underset{n \to 1}{\lim} \partial_n \langle \sigma_n \chi \rangle = \underset{n \to 1}{\lim} \partial_n \langle \tilde{\sigma}_n \chi \rangle_1 = \underset{n \to 1}{\lim} \partial_n \langle e^{-(n-1) H_A^0} \chi \rangle_1 = - \langle H_A^0 \chi \rangle_1\,,
    \label{eq:correspondencetwist}
\end{equation}
where we explicitly show how, in the \( n \to 1 \) limit, the correlation function involving an operator inserted in a single CFT copy with a twist operator corresponds to a two-point function with an extra insertion of the vacuum modular Hamiltonian $H_{A}^{0}$. 

We will apply this correspondence to the one-point function of the displacement operator \( D \) with the twist operator \( \sigma_n \) in the presence of the conformal boundary \( \mathcal{B} \), whose expression we derived earlier in \eqref{eq:1point2defectsA}

\begin{equation}
    \underset{n \to 1}{\lim} \partial_n \langle D \rangle_{\sigma_{n},\mathcal{B}} = - \langle H_A^0 D \rangle_{1,\mathcal{B}}\,.
    \label{eq:correspondenceconical}
\end{equation}
Here, both sides are evaluated in the presence of the boundary $\mathcal{B}$.
Let us emphasize that the correlation function on the l.h.s.\ of this equation is computed on a replicated manifold, while the correlation function on the r.h.s.\ is evaluated in the ordinary Minkowski (or Euclidean) space. 
For the l.h.s.\ we can use the explicit form of the correlator \eqref{eq:1point2defects}, namely

\begin{equation}
    -\lim_{n \to 1} \frac{\partial}{\partial n} \langle D(\tau=0, \vec{y}=0) \rangle_{\sigma_n, \mathcal{B}} = -\partial_n a_{D,n} \Big|_{n=1} \left( \frac{2}{R} \right)^d\,,
    \label{eq:lhsconical}
\end{equation}
where, for simplicity, we place the displacement operator $D$ at $\tau=0$ and $\vec{y}=0$ on the boundary $\mathcal{B}$.
For the r.h.s.\ we use the important fact that for a hemispherical entangling surface ending on a planar boundary the modular Hamiltonian is still a local operator and it is still given by the integral of the time-time component of the bulk stress-energy tensor, as shown in \cite{Casini:2018nym,Jensen:2013lxa}. Therefore, we use the expression for the modular Hamiltonian similar to \eqref{eq:modularHam}, integrating over the hemisphere $A$, and inserting the stress-energy tensor at a general point $(x, \vec{y})$ in the bulk at constant time slice $t=0$. For simplicity, we analytically continue to Euclidean time $\tau$ and focus on the time-time Euclidean component of the stress-energy tensor $T_{\tau\tau}$, which has an overall minus sign. Consequently, the r.h.s.\ of \eqref{eq:correspondenceconical} simplifies to

\begin{equation}
    \langle D(\tau=0, \vec{y}=0) H_A^0 \rangle_{\mathcal{B}} = -2\pi \int_A \mathrm{d}^{d-1} x \, \frac{R^2 - x^2}{2R} \langle D(\tau=0, \vec{y}=0) T_{\tau \tau} (\tau=0, x, \vec{y}) \rangle_{\mathcal{B}}\,.
    \label{eq:2pointmodham}
\end{equation}
The correlation function in the integral is a bulk-to-boundary two-point function, which is completely determined by conformal symmetry, up to the coefficient $C_D$, and takes the following form \cite{Billo:2016cpy,Bianchi:2015liz}

\begin{equation}
    \langle D(\tau=0, \vec{y}=0) T_{\tau \tau} (\tau=0, x, \vec{y}) \rangle_{\mathcal{B}} = -\frac{C_D}{d-1} \frac{1}{(x^2 + \vec{y}^2)^d}\,.
    \label{eq:2pointdisplacementstress}
\end{equation}
We perform the integral in spherical coordinates with an angle $\theta \in [0, \pi/2]$, as we are integrating over the hemisphere
\begin{equation}
\begin{split}
    \langle D(\tau, \vec{y}=0) H_A^0 \rangle_{\mathcal{B}} &= \frac{2\pi C_D \Omega_{d-3}}{d-1} \int_0^R \mathrm{d}r \int_0^{\pi/2} \mathrm{d}\theta \, r^{d-2} (\sin \theta)^{d-3} \frac{R^2 - r^2}{2R} \frac{1}{r^{2d}} \\
    &= \frac{2\pi C_D}{d-1} \frac{2\pi^{d/2-1}}{\Gamma \left( \frac{d}{2}-1 \right)} \frac{\sqrt{\pi} \Gamma \left( \frac{d}{2}-1 \right)}{2 \Gamma \left( \frac{d-1}{2} \right)} \int_0^R \mathrm{d}r \, \frac{R^2 - r^2}{2R} \frac{1}{r^{d+2}} \\
    &= \frac{C_D \pi^{(d+1)/2}}{2(d-1) \Gamma \left( \frac{d+3}{2} \right)} \frac{1}{R^d}\,.
\end{split}
\end{equation}
where the volume of $S^{d-3}$ is given by $\Omega_{d-3} = \frac{2\pi^{d/2-1}}{\Gamma\left( \frac{d}{2}-1 \right)}$.
The integral in the second line contains a divergence at $r=0$, corresponding to the point where the displacement and stress-energy tensor coincide. To evaluate this integral, we used dimensional regularization, following the approach of \cite{Hung:2014npa}.

Finally, comparing this result with the expression from \eqref{eq:lhsconical}, we obtain \eqref{eq:bsigmad} in the main text.

\section{Boundary OPE of the twist operator in BCFT}
\label{section:2OPEBCFT}

Here, we provide an alternative and equivalent derivation for the OPE performed in section \ref{subsection:OPE2defects}. The first part of the derivation is identical, up to equation \eqref{eq:OPE2boundaryoperators}. For a hemispherical twist operator of radius $R$, we can expand $\sigma_{n}(R)$ in local boundary primary operators $\hat{\mathcal{O}_{i}}$ and their descendants (labeled by $k$)\cite{Hung:2014npa}

\begin{equation}
    \sigma_{n}(R)=\langle\sigma_{n}(R)\rangle\left( 1+\sum_{i,k} R^{\hat{\Delta}_{i,k}}C^{n}_{i,k}\hat{\mathcal{O}}_{i}^{k}\right)\,,
    \label{eq:OPEtwist}
\end{equation}
where $R$ is the radius of the entangling surface, $\hat{\Delta}_{i,k}$ the conformal dimensions of the boundary operators $\hat{\mathcal{O}}_{i}^{k}$, and $C^{n}_{i,k}$ the OPE coefficients depending on the replica index $n$. Both single- and multiple-copy operators may appear, but the only non-vanishing contribution in \eqref{eq:OPE2boundaryoperators} arises from the displacement operator $\mathcal{D}$ due to conformal invariance. Using the twist operator OPE, we find

\begin{equation}
    \frac{\langle \sigma_n(R) \mathcal{B} \hat{\mathcal{O}}^{\otimes n}(x_1^a) \hat{\mathcal{O}}^{\otimes n}(x_2^a) \rangle}{\langle \sigma_n(R) \mathcal{B} \rangle \langle \hat{\mathcal{O}}(x_1^a) \hat{\mathcal{O}}(x_2^a) \rangle^n} \overset{\xi \to 0}{\sim} 1 + n R^d \frac{\tilde{a}_{\mathcal{D},n} c_{\hat{\mathcal{O}} \hat{\mathcal{O}} D}}{C_{\mathcal{D}} C_{D}} \frac{\langle \mathcal{D}(0) D(x_2^a) \rangle \langle \hat{\mathcal{O}}(x_1^a) \hat{\mathcal{O}}(x_2^a) \rangle^{n-1}}{|x_{12}^a|^{2 \hat{\Delta} - d} \langle \hat{\mathcal{O}}(x_1^a) \hat{\mathcal{O}}(x_2^a) \rangle^n}\,,
    \label{eq:OPEtwist2boundary}
\end{equation}
where $\tilde{a}_{\mathcal{D},n}$ is the OPE coefficient arising from the expansion of the twist operator, and $C_{\mathcal{D}}$ is a normalization constant. Of course, we expect that the constant $\tilde{a}_{\mathcal{D},n}$ is related to the $a_{D,n}$ in \eqref{eq:1point2defects}, but there is a proportionality constant which we are going to fix.

We now examine each term in the expression \eqref{eq:OPEtwist2boundary}. The two-boundary-point functions that appear in both the numerator and denominator simplify to a trivial result, independent of the replica index $n$.
The other correlation function contributing to the expression in \eqref{eq:OPEtwist2boundary} is the two-point function of a single-copy insertions of the displacement operator $\mathcal{D}$, symmetrized over the $n$ copies, with a displacement operator $D$ in just one copy. As noted in \eqref{eq:displacementcopies}, we are considering two boundary insertions in a single-copy of the $n$-sheeted symmetrized $\mathrm{CFT}^{\otimes n}$ and summing over all copies for both operators. This gives

\begin{equation}
\begin{split}
    \langle\mathcal{D}(0)\mathcal{D}(x_{2}^{a})\rangle &= \sum_{i,j=0}^{n-1} \langle \left(\mathbb{1}^{\otimes i} \otimes D(0) \otimes \mathbb{1}^{\otimes (n-i-1)}\right) \left( \mathbb{1}^{\otimes j} \otimes D(x_{2}^{a}) \otimes \mathbb{1}^{\otimes (n-j-1)} \right) \rangle \\
    &=n\langle \mathcal{D}(0) D(x_{2}^{a}) \rangle=n \langle D(0) D(x_{2}^{a}) \rangle\,.
\end{split}
\label{eq:2copydisplacement}
\end{equation}
This results in $n$ times the two-boundary-point function of two single-copy displacement operators. Consequently, the normalization constant $C_{\mathcal{D}}$ is related to the displacement operator normalization, or Zamolodchikov norm, $C_{D}$ by

\begin{equation}
    C_{\mathcal{D}} = n \, C_{D}\,.
    \label{eq:zamolodchikov}
\end{equation}
As we have seen, the displacement operator is a boundary operator, and for a conformal boundary, the two-point function is completely determined, yielding

\begin{equation}
    \langle D(0) D(x_{2}^{a}) \rangle = \frac{C_{D}}{|x_{2}^{a}|^{2d}}\,,
    \label{eq:2displacementfunction}
\end{equation}
up to $C_{D}$.
Finally, substituting these results into \eqref{eq:OPEtwist2boundary}, the expression simplifies to

\begin{equation}
\begin{split}
    \frac{\langle \sigma_{n}(R) \mathcal{B} \hat{\mathcal{O}}^{\otimes n}(x^{a}_{1})\hat{\mathcal{O}}^{\otimes n}(x^{a}_{2})\rangle}{\langle\sigma_{n}(R)\mathcal{B}\rangle \langle\hat{\mathcal{O}}(x^{a}_{1})\hat{\mathcal{O}}(x^{a}_{2})\rangle^{n}} & \overset{\xi\rightarrow0}{\sim}  1 + R^{d} \frac{\tilde{a}_{\mathcal{D},n} c_{\hat{\mathcal{O}}\hat{\mathcal{O}}D}}{ C_{D}} \frac{|x_{12}^{a}|^{d}}{|x_{2}^{a}|^{2d}}\,,
    \label{eq:OPEtwist2boundaryPartial}
\end{split}
\end{equation}
where \( x_{2}^{a}=( \tau_{2}-\tau, \vec{0} ) \), \( x_{1}^{a}=( \tau_{1}-\tau, \vec{0} ) \), and \( x_{12}^{a}=( \tau_{1}-\tau_{2}, \vec{0} ) \). Then, from this last expression, we recognize a power of the cross-ratio $\xi$ \eqref{eq:CrossRatioXi} in the  OPE limit \( \tau_{1} \rightarrow \tau_{2} \) and \( R \rightarrow 0 \).
Thus, \eqref{eq:OPEtwist2boundaryPartial} reduces to

\begin{equation}
    \frac{\langle \sigma_{n}(R)\hat{\mathcal{O}}^{\otimes n}(x^{a}_{1})\hat{\mathcal{O}}^{\otimes n}(x^{a}_{2})\rangle_{\mathcal{B}}}{\langle \sigma_{n}(R)\mathcal{B}\rangle \langle \hat{\mathcal{O}}(x^{a}_{1}) \hat{\mathcal{O}}(x^{a}_{2}) \rangle^{n}}\overunderset{R\rightarrow0}{\xi\rightarrow0}{\sim} 1+\frac{\tilde{a}_{\mathcal{D},n}c_{\hat{\mathcal{O}} \hat{\mathcal{O}}D}}{ C_{D}}\xi^{\frac{d}{2}}\,.
\end{equation}

\noindent
Substituting this expression into \eqref{eq:deltaSn2defectsBCFT}, we obtain the excess of the Rényi entropy, in the early and late time limit, in terms of the cross-ratio of the theory, namely

\begin{equation}
    \Delta S_{A}^{(n)} = \frac{1}{1-n} \frac{\tilde{a}_{\mathcal{D},n} c_{\hat{\mathcal{O}}\hat{\mathcal{O}}D}}{C_{D}} \xi^{\frac{d}{2}}\,.
    \label{eq:deltaSn2defectsfinal}
\end{equation}

The final term to address is \( \tilde{a}_{\mathcal{D},n} \), which is the coefficient arising in the OPE of the twist operator with a single-copy symmetrized displacement operator \( \mathcal{D} \) at the origin of the boundary. To compute this coefficient, we consider the one-point function of the single-copy displacement operator \( \mathcal{D} \) in the presence of the boundary and the spherical defect \eqref{eq:1point2defectsA} and use the OPE of the twist operator as we did in \eqref{eq:OPEtwist2boundary}

\begin{equation}
    \frac{\langle\mathcal{D}(x^{a})\sigma_{n}(R) \mathcal{B} \rangle}{\langle\sigma_{n}(R) \mathcal{B} \rangle}\overset{R\rightarrow0}{\sim}\frac{\tilde{a}_{\mathcal{D},n}}{C_{\mathcal{D}}}R^{d}\langle\mathcal{D}(x^{a})\mathcal{D}(0)\rangle=\frac{\tilde{a}_{\mathcal{D},n}R^{d}}{(\tau^{2}+\vec{y}^{2})^{d}}\,,
\end{equation}
where $x^{a}=(\tau,\vec{y})$, and we ignored the contribution of the descendants.
Then, we can compare this expression with the one we obtained in \eqref{eq:1point2defectsA} in the limit $R\rightarrow0$

\begin{equation}
\begin{split}
      \frac{\langle\mathcal{D}(x^{a})\sigma_{n}(R)\rangle_{\mathcal{B}}}{\langle\sigma_{n}(R)\rangle_{\mathcal{B}}}& =a_{D,n}\frac{(2R)^{d}}{\left( (R-y_{d-2})^{2}+y^{\hat{a}}y_{\hat{a}}+\tau^{2}\right)^{\frac{d}{2}}\left( (R+y_{d-2})^{2}+y^{\hat{a}}y_{\hat{a}}+\tau^{2}\right)^{\frac{d}{2}}}\\
      &\overset{R\rightarrow0}{\sim}\frac{a_{D,n}2^{d}R^{d}}{(\tau^{2}+\vec{y}^{2})^{d}}\,,
      \end{split}
\end{equation}
where $\hat{a}=1,\ldots,d-3$. Thus, we obtained that 

\begin{equation}
\tilde{a}_{\mathcal{D},n}=2^{d}a_{D,n}\,.    
\end{equation}
Inserting this expression in the excess of the Rényi entropy in \eqref{eq:deltaSn2defectsfinal} yields

\begin{equation}
    \Delta S_{A}^{(n)} = \frac{1}{1-n} \frac{a_{D,n} c_{\hat{\mathcal{O}}\hat{\mathcal{O}}D}}{C_{D}} 2^{d}\xi^{\frac{d}{2}}\,,
    \label{eq:deltaSn2defectsfinal2}
\end{equation}
which matches the result found in \eqref{eq:deltaSn2defectsfinal3}.

\section{Numerical analysis of the holographic entanglement entropy}
\label{section:Numerical}

In section \ref{subsection:HEEQuench}, we computed the minimal surface area to extract information about the time evolution of the excess of the holographic entanglement entropy in the presence of an excited state in the dual CFT. The area of this surface comes from the minimization of the area functional \eqref{eq:inducedmetricglobalquench}. This minimization yields a second-order differential equation in $r(\theta)$

\begin{equation}
\begin{split}
& \sin (\theta ) r(\theta )^3 \left\{\sin (\theta ) \left[-L^4 r(\theta )^2 \left(8 G M+r''(\theta )\right)-5 G L^4 M r'(\theta )^2+L^4 r(\theta ) \left(2 G M \left(4 G M \right. \right. \right. \right.\\
&\left. \left. \left. +r''(\theta )\right)+3 r'(\theta )^2\right) -L^2 r(\theta )^4 \left(8 G M+r''(\theta )\right)+4 L^2 r(\theta )^5+2 r(\theta )^3 \left(L^4+2 L^2 r'(\theta )^2\right)\right.\\
&\left.\left. +2 r(\theta )^7 \right] -L^2 \cos (\theta ) r'(\theta ) \left[-2 G L^2 M r(\theta )+L^2 r'(\theta )^2+L^2 r(\theta )^2+r(\theta )^4 \right]\right\}=0\,.
\end{split}
\label{eq:EuLagrM}
\end{equation}

We first solve the differential equation \eqref{eq:EuLagrM} numerically for the vacuum configuration with $M = 0$, whose analytic solution is given by $r_{\text{\tiny vac}}(\theta)$ in \eqref{eq:rthetaGlobal}. Setting $L = 1$ for simplicity, we impose boundary conditions for $r_{\text{\tiny vac}}(\theta_{\infty}) = r_{\infty}$ and $r'_{\text{\tiny vac}}(\theta_{tip}) = r_{tip}$, where $\theta_{tip} = 0$ corresponds to the tip of the minimal surface. To mimic a Cauchy problem, we shift all boundary conditions to $\theta_{tip}$, where the minimal surface profile exhibits a turning point. Specifically, we impose $r(\theta_{tip}) = r_0$, with $r_0 > r_h$, ensuring the minimal surface avoids entering the event horizon described by the Schwarzschild radius $r_{h}$.

For the case of a local excitation, we perform a perturbative computation of the boundary conditions at $\theta_{tip} = 0$. Expanding the solution $r(\theta)$ as a power series around $\theta = 0$, we exploit symmetry in the $\phi$-angle, yielding only even powers of $\theta$

\begin{equation}
    r(\theta) = r_0 + r_2 \theta^2 + r_4 \theta^4 + \ldots\, .
\end{equation}
Substituting this expansion into \eqref{eq:EuLagrM}, we determine the coefficients order by order in $\theta$, treating $r_0$ as an arbitrary parameter.
To locate the boundary value $\theta_{\infty}$ for a local excitation, we reformulate  the problem numerically by introducing $z(\theta)$, where $r(\theta) = \frac{1}{z(\theta)}$. This transformation simplifies the second-order differential equation \eqref{eq:EuLagrM} for $z(\theta)$.
For the vacuum solution $M = 0$, fixing $G_N$, $\epsilon$ (the UV cutoff), $L$, and using the theoretical value of $r_0$ from $r_{\text{\tiny vac}}(\theta)$ \eqref{eq:rthetaGlobal}, we compute the root of $z(\theta)$ to obtain $\theta_{\infty}$ numerically. We then solve the equation for $z(\theta)$ and confirm agreement between the theoretical and numerical solutions.
With these results, we compute the area of the minimal surface numerically for the vacuum case. 

Notably, the area is divergent due to the integration limit $\theta_{\infty}$. To regularize it, we introduce a UV cutoff $\epsilon_1$, setting the upper limit of the integral \eqref{eq:inducedmetricglobalquench} to $\theta = \theta_{\infty} - \epsilon_1^2$. Expanding in $\epsilon_1=0$ and dividing by $4G_N$, the holographic entanglement entropy becomes

\begin{equation}
    S_{\mathrm{vacuum}} = \frac{\pi L^2 \sqrt{\sin(2\theta_{\infty})}}{4 G_N \epsilon_1} - \frac{\pi L^2}{2 G_N}\,.
    \label{eq:vacuumentropyglobal}
\end{equation}
This matches the Poincaré coordinate result from \cite{Ryu:2006ef}, with $\theta_{\infty}$ replacing the dependence on the radius $R$.
Then, we express $\epsilon_1$ in terms of the IR cutoff $r_{\infty} \equiv r(\theta_{\infty} - \epsilon_1^2)$. Expanding around $\epsilon_1 = 0$ and solving for $\epsilon_1$, we find

\begin{equation}
    \epsilon_1^2 = \frac{L^2 \cot\theta_{\infty}}{2r_{\infty}^2}\,.
\end{equation}
Substituting into \eqref{eq:vacuumentropyglobal}, the entanglement entropy becomes

\begin{equation}
    S_{\mathrm{vacuum}} = \frac{\pi L \sin(\theta_{\infty}) r_{\infty}}{2 G_N} - \frac{\pi L^2}{2 G_N}\,.
    \label{eq:vacuumentropyglobal2}
\end{equation}

Having established the vacuum configuration, we now shift our focus to the local excitation scenario, where $M \neq 0$, leveraging the previous numerical results. 
To perform the same numerical analysis as in the vacuum case, we fix the parameters $M$, $G_N$, the UV cutoff $\epsilon$, $L$, and $r_0$. Solving the differential equation \eqref{eq:EuLagrM} numerically for \( z(\theta) = \frac{1}{r(\theta)} \), we extract the boundary value $\theta_\infty$. Subsequently, we numerically solve the equation for \( r(\theta) \) using appropriate boundary conditions.

As in the vacuum case, computing the area of the minimal surface \eqref{eq:inducedmetricglobalquench} reveals a divergence at $\theta_\infty$. 
Nevertheless, it is known that one can extract a finite result for the entropy variation $\Delta S = S_{\text{quench}} - S_{\text{vac}}$.

\newpage
\bibliographystyle{JHEP}
\bibliography{biblio.bib}

\end{document}